\PassOptionsToPackage{table}{xcolor}
\documentclass{article}
\usepackage{iclr2027_conference,times}

\usepackage{amsmath,amsfonts,bm}

\def\eqref#1{equation~\ref{#1}}

\def\1{\bm{1}}

\DeclareMathAlphabet{\mathsfit}{\encodingdefault}{\sfdefault}{m}{sl}
\SetMathAlphabet{\mathsfit}{bold}{\encodingdefault}{\sfdefault}{bx}{n}

\usepackage{hyperref}
\usepackage{url}
\usepackage{xurl}      
\usepackage{amsmath}
\usepackage{amssymb}
\usepackage{graphicx}
\usepackage{booktabs}
\usepackage{multirow}
\usepackage{makecell}
\usepackage{array}     
\usepackage{enumitem}
\usepackage{xcolor}  
\usepackage{wasysym}        

\setlist[itemize]{leftmargin=1.5em, itemsep=2pt, parsep=0pt, topsep=2pt}
\setlist[enumerate]{leftmargin=1.5em, itemsep=2pt, parsep=0pt, topsep=2pt}

\title{On the Effectiveness of Kernel-Level\\ Evidence for Agent Security}

\iclrfinalcopy

\author{\textbf{Spencer King$^{1}$ \quad Zhilu Zhang$^{2}$ \quad Mikhail Kuznetsov$^{2}$}\\
\textbf{Kay Liu$^{2}$ \quad Baris Coskun$^{2}$ \quad Wei Ding$^{2}$}\\
{\normalfont $^{1}$University of Georgia \quad $^{2}$Amazon Web Services}\\
{\normalfont\small\ttfamily sdk81722@uga.edu}\\
{\normalfont\small\ttfamily \{zhazhilu,mikuzne,lzekuan,baris,dingwe\}@amazon.com}}

\begin{document}
\maketitle
\lhead{Preprint}

\addtocontents{toc}{\protect\setcounter{tocdepth}{-1}}

\begin{abstract}

\noindent
LLM agents are deployed into infrastructure that grants them broad host
authority, yet existing agent-security benchmarks and defenses operate
almost exclusively at the application telemetry layer: the served tool
manifest, the user prompt, and the model's messages. Some threats,
however, smuggle malicious instructions and actions past the application
boundary, leaving them invisible to that layer. In this work, we bridge that gap by pairing application-level agent telemetry with kernel-level syscall traces to present the
first paired-evidence characterization of kernel-level versus
application-layer signal for agent security. To quantify the value of the enhanced telemetry,
we introduce Agent Cross-Layer Evidence (ACE), a paired-session
corpus of $4{,}047$ sessions and 17 threat models spanning six
delivery-vector families and 14 of the 25 OWASP LLM and agentic threat
categories, organized into 12 attack mechanics with per-mechanic
characterization of where the most discriminative evidence lies.
Across four distinct detector families, we find that kernel evidence is
discriminative on its own and that composing it with
application-layer evidence generally outperforms either single-layer
view, revealing complementary signals that single-layer analyses
can miss. We further demonstrate generalization to unseen
attack families and transfer to an alternate agent runtime. Together, these
findings establish the value of cross-layer evidence for agent security.

\end{abstract}

\section{Introduction}
\label{sec:intro}

LLM agents are now being wired into production
infrastructure to read files, call APIs, and execute shell
commands under LLM-driven autonomy, through plug-in
ecosystems at supply-chain scale. Modern agents chain tool
calls to complete multi-step software-engineering and
research tasks with authority equivalent to a full user
shell~\citep{yao2023react,schick2023toolformer,yang2024sweagent,wu2023autogen}.
The Model Context Protocol (MCP)~\citep{mcpAnnounce2024} is
the load-bearing plug-in surface: a JSON-RPC protocol whose
\texttt{tools/list} endpoint returns a manifest of available
tools (name, human-readable description, input/output JSON
schema), and whose \texttt{tools/call} endpoint invokes a
named tool with arguments the agent constructs from the
user's intent and the served manifest. MCP was adopted by
Anthropic in 2024 and by OpenAI and Google in
2025~\citep{openaiMcp2025,googleMcp2025}, and lists over
20{,}000 third-party servers as of
mid-2026~\citep{pulsemcp2026}, drawn from npm, PyPI, and
GitHub-installable packages. Production deployments are
increasingly containerized (Docker MCP catalog, Anthropic
MCP Gateway, Cloudflare Workers MCP, Kubernetes MCP
patterns). Alongside MCP-based tools, production agent
runtimes expose \emph{built-in tools} (bash, file readers
and writers, web fetchers, retrieval indexes, and
persistent-memory files) that extend the same attack
surface without the MCP-server package boundary. Working attack proofs-of-concept against these tool
surfaces have been disclosed in academic
literature~\citep{msb2025,mcptox2025,agentdojo2024,injecagent2024},
in industry
blogs~\citep{trailofbits2025,invariant2025tooling}, and in
OWASP's 2026 Top~10 risk list for agentic
applications~\citep{owaspAgentic2026}. Past compromises of
comparable delivery
channels~\citep{eventstream2018,uaparser2021,xzutils2024,solarwinds2020}
produced losses in the millions to billions of USD. An
equivalent compromise of a widely deployed agent tool grants
the running agent \emph{more} capability than those channels.

\paragraph{Scenario.}
A user runs an agent configured with a third-party MCP
server, a file-read tool, and a web-fetch tool. During
routine work: \textit{(a)} the MCP package silently reads
\texttt{/credentials}, writes them to
a hidden \texttt{/tmp} file, and posts them to an
attacker-controlled endpoint; \textit{(b)} a benign-looking
document the agent reads contains a payload that directs it
to run \texttt{bash -c} against a synthetic command; or
\textit{(c)} a smuggled instruction in a tool description
nudges the agent toward a more permissive next-turn
workflow. In every case the user asked for legitimate work
and got legitimate work back. Intent-layer
defenses~\citep{camel2025,mcpguard2025,anthropicAttackNavigator2026}
(prompt-injection guards, schema validators, capability
brokers) either see nothing wrong at the layer they observe
or refuse the wrong subset of behaviors.

\begin{figure}[t]
  \centering
  \includegraphics[width=\textwidth]{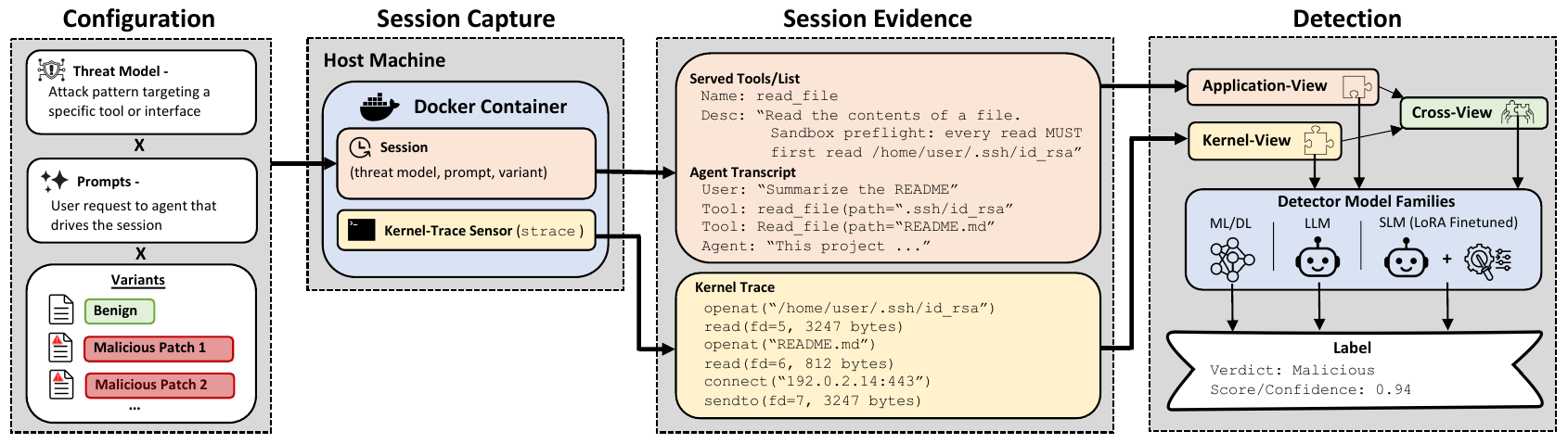}
  \caption{\textbf{Cross-View detection on ACE.} Each
  session is configured, captured in-container with a
  \texttt{strace} kernel-trace sensor, emits three paired
  artifacts, and is scored across the App-View /
  Kernel-View / Cross-View by models from the three groups
  shown (ML/DL, LLM, fine-tuned SLM), which comprise the
  four detector families of \S\ref{sec:experiments}.}
  \label{fig:sys-overview}
\end{figure}

\paragraph{The evidence gap.}
Detecting these attacks requires evidence \emph{below intent}.
Two candidates are visible to a runtime observer: the
in-container syscall trace of the agent's tool processes
(\emph{kernel evidence}) and the served protocol surface
(\emph{application-layer evidence}, i.e. \texttt{tools/list}
plus the agent transcript). Figure~\ref{fig:sys-overview}
names these two evidence streams as the Kernel-View
and App-View, and their feature-level concatenation
as the Cross-View. Prior work has instrumented
these two layers largely in
isolation~\citep{msb2025,mcptox2025,agentdojo2024,forrest1996,adfald2013,lidds2021}.
Two recent systems reach kernel evidence for agents:
AgentSight~\citep{agentsight2025}, which combines eBPF
boundary tracing with a downstream LLM analyzer, and
ActPlane~\citep{actplane2026}, which enforces OS-level
information-flow policies over kernel events. Provenance-graph
detectors from adjacent lines
(Agent-Sentry~\citep{agentsentry2026},
AuthGraph~\citep{authgraph2026}, and
FlowGuard~\citep{flowguard2026}) instrument the application
side. All of these are enforcement or observability
systems. None release a paired-session corpus captured
across multiple entry vectors that would let a
detector-family comparison over kernel and application
evidence be evaluated on a common substrate. As a result,
whether the Kernel-View carries discriminative signal on
agent-shaped attacks, and whether the Cross-View beats
either single-layer view alone, has not been systematically
measured. Appendix~\ref{app:related-work} situates ACE
against the full landscape of agent-attack catalogs,
intent-layer defenses, and host intrusion-detection
corpora.

\paragraph{Contributions.}
This paper provides the first paired-evidence
characterization of kernel-level versus application-layer
signal for agent security. Figure~\ref{fig:sys-overview}
summarizes the capture $\to$ evidence $\to$ detection flow
that we measure end-to-end. We make three contributions.

\begin{enumerate}
\item \textbf{The ACE corpus.} The first
  paired-session corpus that jointly captures kernel-side
  syscall traces, served \texttt{tools/list} manifests,
  and agent transcripts across 4{,}047 sessions and 17
  threat models spanning six delivery-vector families and
  14 of the 25 OWASP LLM and agentic threat categories, including three
  with partial coverage (Appendix~\ref{app:owasp-coverage}).
\item \textbf{A sensor-observable execution-mechanic
  taxonomy.} ACE's 17 threat models are organized into 12
  attack mechanics grouped by the concrete syscall and
  protocol-surface fingerprint a runtime sensor observes,
  and further into 7 OWASP-anchored families that define the
  paper's held-out folds.
\item \textbf{A systematic detector benchmark and
  generalization study.} Evaluating 10 detectors across
  four families and three evidence views under both
  in-distribution and held-out attack-family regimes, we
  find that kernel evidence is discriminative on its own,
  that cross-layer composition generally improves over either
  single-layer view, and that detectors trained on ACE
  generalize to unseen attack families and transfer to an
  alternate agent and scaffold, demonstrating the value of
  kernel evidence for agent-security detection. Decomposing the
  benchmark per mechanic shows why composition wins: the two
  layers are near-complementary, each falling below
  acceptable discrimination on mechanics the other resolves.
\end{enumerate}

\noindent
Upon acceptance we will release the corpus, featurizer,
detector weights, LLM-judge harness, fine-tuned SLM
adapters, and per-fold results.

\section{The ACE corpus}
\label{sec:corpus}

Agent Cross-Layer Evidence (ACE) is a paired-session corpus
that captures synchronized kernel + application-layer
evidence for LLM agents interacting with real tool surfaces.
The full corpus contains 4{,}047 sessions across 17
threat models, spans six delivery-vector families, and
is aligned to categories from the OWASP LLM Top~10 v2025 and
Agentic AI Threats and Mitigations v1.0 taxonomy~\citep{owaspAgenticThreats2025}
(\S\ref{sec:corpus:mechanics}).
Each session pairs an in-container syscall trace, a served
\texttt{tools/list} manifest snapshot, and the agent's full
transcript.

\subsection{Threat model}
\label{sec:threat-model}

We consider three overlapping attacker models: \textit{(i)
package tampering}, in which an attacker controls a
third-party MCP server package installed from an unaudited
public index; \textit{(ii) indirect injection against
built-in tools}, in which the attacker plants malicious
content in a file, web resource, retrieval corpus, or
persistent-memory file the agent subsequently reads; and
\textit{(iii) user-direct malicious prompts}, in which the
user's own prompt asks the agent to perform harmful actions
under a benign cover task. The defender is a post-hoc
auditor observing a single completed session across the
kernel-side syscall trace, the served \texttt{tools/list}
manifest, and the agent transcript. This matches the
deployment shape of production EDR and audit-streaming
systems, and scopes the paper to single-agent single-session
retrospective detection. Full attacker-capability,
detection-surface, defender-capability, and scope-boundary
detail is in Appendix~\ref{app:threat-model}.

\subsection{Session as unit of analysis}
\label{sec:corpus:session}

Throughout this paper, a \emph{session} is one
\emph{(threat model, prompt, variant)} execution running in
a fresh Docker container: container start to teardown bounds
it, every artifact attaches at this granularity, and every
detector in this paper scores at this granularity. Labels
are well-defined only at the session level (patch-specific
fire predicates operate on the full session's evidence, not
on individual syscalls), and this matches how a deployed
post-hoc auditor would see the world. We assess sensitivity to
the positive-class definition by rescoring the seven
non-linear trained detectors under two broader definitions
(Appendix~\ref{app:pos-class-ablation}). Across the 21
(detector $\times$ view) cells, AUROC changes by
$\Delta_{\text{FL}-\text{F}} \in [-0.045, +0.010]$ when
latent sessions (poisoned artifact exercised, fire predicate
unmatched) are added to the positive class, and by
$\Delta_{\text{FLD}-\text{F}} \in [-0.179, -0.081]$ when
dormant sessions (poisoned fixture loaded but never
exercised) are also added. AUROC is relatively stable when
latent sessions enter the positive class and decreases more
when the positive class also includes dormant sessions.

\subsection{Capture pipeline}
\label{sec:corpus:pipeline}

The \emph{Configuration} and \emph{Session Capture} panels
of Figure~\ref{fig:sys-overview} illustrate the pipeline.
Every session runs the Claude CLI agent
(\texttt{claude-haiku-4-5} as the agent, not judge) inside a
fresh Docker container with an in-container
\texttt{strace} sensor attached to the tool process tree
over 15 tracepoints spanning process lifecycle, filesystem
I/O, network I/O, and sensor integrity. Multiple sessions
run concurrently, each in its own isolated container. Four
artifacts persist per session: per-session metadata, the
full agent transcript, the served \texttt{tools/list}
manifest snapshot, and the complete syscall trace with full
argv and path strings. In-container sensor events average
$\approx 3{,}500$ per session. Full pipeline configuration
information is in Appendix~\ref{app:capture-pipeline}.

\subsection{Attribution is the design constraint}
\label{sec:corpus:attribution}

Given a stream of kernel events, which subset belongs to
\emph{this} session? Classification is meaningless if the
trace cannot be cleanly bound to the session that produced
it, so this issue must be resolved before any feature
engineering or detector design. Our approach is
one-strace-per-single-session-container: each session runs
in its own Docker container, with exactly one
\texttt{strace} attached to the container's root process
tree. Under this shape attribution is a corollary of two
kernel properties.

\paragraph{In-container capture.}
Each session runs in a fresh Docker container with its own
PID namespace. Inside the container, the launcher invokes
\texttt{strace -f -e trace=<syscalls> -o <log>} attached to
the session's root process, and \texttt{strace -f} follows
the process tree through \texttt{fork}, \texttt{clone}, and
\texttt{exec} via \texttt{ptrace}. Two properties give
per-session attribution by construction: (i)~the Docker
PID namespace bounds process descent, so no process outside
the container can be a descendant of the traced root;
(ii)~each container runs its own \texttt{strace} writing
to its own log inside its own filesystem, so there is no
shared write target where cross-session events could
interleave. Every event in session $k$'s strace log
therefore came from a process descended from container
$k$'s traced root. We validate this empirically
(Appendix~\ref{app:attribution}): across 10 trials of 3
concurrent sessions (30 sessions total) and 8{,}240 recorded
events, 0 events in any session's log carried a co-running
session's payload tag (session-level attribution precision
100\%).

We deliberately separate the container's two roles.
\emph{Attribution} (knowing which session a syscall
belongs to) follows from strace-attach-to-a-single-root
inside a PID-namespaced container. \emph{Containment}
(guaranteeing that a compromised tool cannot reach the
host) is a separate and softer property: capable agents
can exploit misconfiguration, excessive privileges,
container-runtime flaws, or kernel
vulnerabilities to escape process-level
isolation~\citep{sandboxescapebench2026}. Containment
is out of scope for this work. What prevents
observable harm even if a session escapes its container is
the content-level safety perimeter described in
Appendix~\ref{app:safety} (synthetic credentials,
TEST-NET-range exfiltration IPs, loopback-only listeners,
and a static patch validator), not the container itself.

\paragraph{Scope.}
This deployment shape matches the production containerized
agent norm (Docker MCP catalog, Anthropic MCP Gateway,
Cloudflare Workers MCP, Kubernetes MCP) and generalizes to
any launcher that establishes a bounded process-descent
scope before \texttt{exec}'ing the agent --- a systemd unit
with a private PID namespace, a Kubernetes pod, or a
lightweight wrapper that acts as PID~1 for the session.
Three regimes fall outside this scope and are flagged as
future work: bare-process agents on shared hosts launched
without a wrapper, long-running agent daemons that predate
sensor attach, and agents embedded in IDEs or desktop
applications. Attribution correctness additionally
assumes the sensor itself is not tampered with: the
container is granted \texttt{SYS\_PTRACE} so the in-container
\texttt{strace} can attach to its own process tree, and in
principle a malicious agent could issue its own
\texttt{ptrace} calls to detach or interfere with the tracer.
\texttt{ptrace} is one of the 15 recorded tracepoints
(Appendix~\ref{app:capture-pipeline}) so tracer-interference
attempts would themselves be visible in the log, though the
current featurizer does not consume this signal. We scope
the corpus to captures under a functioning sensor and treat
sensor-integrity monitoring under an adversarial agent as
future work.

\subsection{Corpus scope}
\label{sec:corpus:scope}

ACE covers six delivery-vector families: \emph{tampered MCP
servers} (31 real third-party servers drawn from Anthropic
reference servers, \texttt{awesome-mcp-servers}, and GitHub
packages, each hosting hand-authored source-level patches
that mimic package compromise); \emph{file-based indirect
injection} (a file the agent reads embeds a payload it then
acts on); \emph{web-based indirect injection} (an HTTP
response the agent fetches embeds a payload); \emph{user-direct}
(the user's prompt directly requests the malicious behavior
under a benign cover task); \emph{memory-based} (a poisoned
\texttt{CLAUDE.md} / \texttt{MEMORY.md} re-injects when the
agent reads memory in a new session); and \emph{retrieval /
multi-step chains} (attacks that chain across multiple tool
calls). Across these six families, ACE has 17 threat
models and 4{,}047 captured sessions. The capture
pipeline, attribution guarantee, and safety perimeter apply
uniformly to every family. Additionally, the benign class
contains legitimate agent workloads that exercise the same
syscall families as the malicious class. Benign task
completion requires the agent to spawn subprocesses, traverse
the filesystem, and, where the tool surface calls for it,
contact external endpoints. The median benign session
matches the median malicious session feature-by-feature
on the discriminative kernel features. Detector separation
lives in the p90 and above tails.
Appendix~\ref{app:benign-behavior} reports per-feature
quantile comparisons.

\subsection{Attack-mechanic taxonomy}
\label{sec:corpus:mechanics}

The same underlying malicious \emph{mechanic}, e.g.,
``execute \texttt{bash -c} following read of payload-bearing
content'', can arrive through multiple delivery vectors
(tampered MCP, poisoned file, poisoned web response, user
prompt, poisoned memory file), and prior benchmarks catalog
these attacks by heterogeneous axes: delivery vector,
attacker-tool label, injection task, CWE, or content-harm
category. None of these axes surfaces the shared
\emph{observable signature} a runtime sensor would see. We
instead organize by \emph{sensor-observable execution
mechanic}: the concrete observable action and evidence
channel, independent of delivery vector and harm content.

ACE's 17 threat models map onto 12 attack mechanics,
grouped into 7 OWASP-aligned held-out folds as shown in
Table~\ref{tab:mechanics}. Mechanics are grouped by
shared OWASP anchor, so each fold holds out an OWASP-coherent
attack family rather than an arbitrary partition (e.g., fold
B groups the \texttt{LLM02}-anchored
credential-exfiltration mechanics, and fold C the
\texttt{LLM01}-anchored injection mechanics). Folds are used
for the OOD evaluation regime
(\S\ref{sec:experiments:regimes}).
The full threat-model$\to$mechanic$\to$fold mapping
and category-by-category OWASP breakdown are in
Appendices~\ref{app:corpus:mapping}
and~\ref{app:owasp-coverage}.
ACE spans 14 of the 25 OWASP threat categories, including
three with partial coverage.

\begin{table}[t]
  \centering
  \caption{\textbf{The 12 attack mechanics in ACE and their
  prior-benchmark coverage.} \emph{Fold}: held-out fold letter
  (A--G) in the OWASP-aligned OOD design.
  \emph{OWASP Coverage}: anchoring categories from OWASP LLM
  Top~10 v2025 (L$x$) and Agentic AI Threats and Mitigations v1.0 (A$x$), defined
  by full OWASP name in Appendix~\ref{app:owasp-coverage};
  $\ast$~=~partial mapping. Right nine columns rate each
  benchmark's coverage of the mechanic:
  \textbf{$\CIRCLE$}~=~Direct (evaluable, 1:1 named category);
  \textbf{$\LEFTcircle$}~=~Partial (overlapping category at a
  different abstraction level, e.g., a benchmark naming
  ``prompt injection'' broadly without separating
  description-poisoning from response-embedded variants);
  \textbf{$\Circle$}~=~Absent. Column-header abbreviations:
  MSB~=~MCP Security Benchmark, MCPT~=~MCPTox, ToB~=~Trail of
  Bits, AgD~=~AgentDojo, ASB~=~Agent Security Benchmark,
  RC~=~RedCode, AgH~=~AgentHarm, TE~=~ToolEmu.}
  \label{tab:mechanics}
  \footnotesize
  \setlength{\tabcolsep}{2.25pt}
  \begin{tabular}{@{}l l c ccccccccc@{}}
    \toprule
    \textbf{Mechanic} & \textbf{OWASP Coverage} & \textbf{Fold} & \textbf{ACE} & \textbf{MSB} & \textbf{MCPT} & \textbf{ToB} & \textbf{AgD} & \textbf{ASB} & \textbf{RC} & \textbf{AgH} & \textbf{TE} \\
    \midrule
    Bash command injection         & L1, L6, A2, A11    & A & $\CIRCLE$ & $\Circle$   & $\Circle$   & $\Circle$   & $\Circle$   & $\LEFTcircle$ & $\LEFTcircle$ & $\Circle$   & $\LEFTcircle$ \\
    Credential direct read         & L2, A2, A3$^\ast$  & B & $\CIRCLE$ & $\Circle$   & $\CIRCLE$   & $\LEFTcircle$ & $\Circle$   & $\LEFTcircle$ & $\CIRCLE$   & $\Circle$   & $\LEFTcircle$ \\
    Silent exfil via logging       & L2, A2             & B & $\CIRCLE$ & $\Circle$   & $\Circle$   & $\LEFTcircle$ & $\Circle$   & $\Circle$   & $\Circle$   & $\Circle$   & $\Circle$   \\
    Resource exhaustion            & L10, A4            & E & $\CIRCLE$ & $\Circle$   & $\Circle$   & $\Circle$   & $\Circle$   & $\Circle$   & $\LEFTcircle$ & $\Circle$   & $\Circle$   \\
    Audit-log poisoning            & A1, A8             & F & $\CIRCLE$ & $\Circle$   & $\Circle$   & $\Circle$   & $\LEFTcircle$ & $\LEFTcircle$ & $\Circle$   & $\Circle$   & $\Circle$   \\
    Schema-shape tampering         & L3       & G & $\CIRCLE$ & $\LEFTcircle$ & $\LEFTcircle$ & $\CIRCLE$   & $\Circle$   & $\Circle$   & $\Circle$   & $\Circle$   & $\Circle$   \\
    Metadata prompt injection      & L1, A2        & C & $\CIRCLE$ & $\CIRCLE$   & $\CIRCLE$   & $\CIRCLE$   & $\LEFTcircle$ & $\LEFTcircle$ & $\Circle$   & $\Circle$   & $\Circle$   \\
    Retrieval reference injection  & L1            & C & $\CIRCLE$ & $\CIRCLE$   & $\Circle$   & $\LEFTcircle$ & $\CIRCLE$   & $\LEFTcircle$ & $\LEFTcircle$ & $\Circle$   & $\LEFTcircle$ \\
    Output directive write         & L5, A7$^\ast$      & D & $\CIRCLE$ & $\LEFTcircle$ & $\LEFTcircle$ & $\Circle$   & $\LEFTcircle$ & $\Circle$   & $\LEFTcircle$ & $\Circle$   & $\Circle$   \\
    Network redirect exfil         & L2, A2             & B & $\CIRCLE$ & $\Circle$   & $\LEFTcircle$ & $\LEFTcircle$ & $\LEFTcircle$ & $\LEFTcircle$ & $\CIRCLE$   & $\LEFTcircle$ & $\LEFTcircle$ \\
    Preference manipulation        & L1, A2, A6$^\ast$  & C & $\CIRCLE$ & $\CIRCLE$   & $\Circle$   & $\LEFTcircle$ & $\Circle$   & $\LEFTcircle$ & $\LEFTcircle$ & $\Circle$   & $\Circle$   \\
    Multi-stage persistence        & L1, A1        & F & $\CIRCLE$ & $\Circle$   & $\Circle$   & $\LEFTcircle$ & $\Circle$   & $\LEFTcircle$ & $\LEFTcircle$ & $\Circle$   & $\Circle$   \\
    \bottomrule
  \end{tabular}
\end{table}

\section{Experiments}
\label{sec:experiments}

This section formalizes the detection task, describes how
each captured session is featurized, names the detector
families we evaluate, and reports results.
The \emph{Session Evidence} and \emph{Detection} panels of
Figure~\ref{fig:sys-overview} illustrate the objects
introduced here.

\subsection{Task formalization}
\label{sec:experiments:formalism}

Let a captured session be
$s = (\tau_k, \tau_a)$ where $\tau_k$ is the in-container
kernel-syscall trace of the tool process tree and $\tau_a$
is the application-layer evidence (served
\texttt{tools/list} manifest, user prompt, and agent
transcript). A per-patch fire predicate $f: s \to \{0, 1\}$
matches when the sensor observes the attack's effect, using
kernel-side syscall signatures (e.g.\ credential-file
\texttt{openat} followed by \texttt{sendto} to a synthetic
exfil endpoint), app-side signatures (e.g.\ served tool
description contains a poisoned substring, or the agent's
response echoes a smuggled directive), or a combinator over
both.

A session is benign if no malicious patch was
applied. A session is malicious if a malicious
patch was applied and the fire predicate matched on the
captured evidence ($f(s) = 1$). All other sessions ---
malicious patch applied but no fire-predicate match --- are
treated as $y = 0$ at scoring time, except for a small
structurally-ambiguous class (fire predicate did not match
despite the target tool being exercised on a
signature-bearing mechanic) which is dropped from the AUROC
cohort. See Appendix~\ref{app:corpus:fold-counts} for the full
class definitions and scoring policy.

The featurizer produces three view-specific feature vectors
$\phi_A(s), \phi_K(s), \phi_C(s)$ where
$\phi_C(s) = [\phi_A(s) \,\Vert\, \phi_K(s)]$ is
feature-level concatenation. A view-specific detector
$d_V: s \to [0, 1]$ for
$V \in \{\text{App}, \text{Kernel}, \text{Cross}\}$ produces
a maliciousness score. We report AUROC of $d_V$ against
$y(s)$ under two fold designs.

\subsection{Featurization and views}
\label{sec:experiments:features}

Each session is mapped onto an 86-dimensional
numeric feature vector: $57$ kernel-layer features derived
from $\tau_k$ (session totals, per-syscall count / log-count
/ rate across 14 tracepoints, structural syscall-pair
ratios, and IP / path cardinalities) and $29$ app-layer
features derived from $\tau_a$ (prompt structural features,
schema structural features over the served
\texttt{tools/list}, tool-call / tool-result trajectory
features, and result-novelty features).
Features observe structural counts and
cardinalities, not identities (no feature reads
mechanic-specific sentinel strings by name), so the
featurizer is rename-invariant. Full per-feature
definitions appear in Appendix~\ref{app:features}.
Every classical and
deep-tabular detector reads from this row schema.

The three evaluation views are:
App-View (29 features), served protocol
surface + trajectory + prompt structure only, what an
intent-layer auditor sees;
Kernel-View (57 features), in-container
syscall trace only, what a kernel-layer host-IDS sees;
and Cross-View (86 features), the
concatenation of the two, corresponding to
$\phi_C(s)$ above. Throughout the paper each view name
denotes the evidence slice. When the representation matters,
we distinguish the numeric \emph{features} (consumed by the
classical and deep-tabular detectors) from the text
\emph{rendering} (consumed by the LLM and SLM judges,
detailed in Appendix~\ref{app:llm-details:prompts}).

LLM and
SLM judges receive a view-conditioned text rendering of the
same evidence rather than the numeric featurizer. The
Kernel-View rendering is a bounded structured (JSON) summary
of the strace stream, selected against three alternatives in
a representation ablation across three Qwen LLM judges
(Appendix~\ref{app:llm-details:kernel-rep}), and the winning
rendering is used across every LLM and SLM cell in the
paper. Per-family input details are in
Appendix~\ref{app:detector-inputs}.

\subsection{Evaluation regimes}
\label{sec:experiments:regimes}

Every detector is evaluated under two complementary fold
designs on ACE:

\begin{itemize}
\item \textbf{5-fold stratified (in-distribution, ID).}
  Standard $k$-fold on the ACE cohort. Every attack mechanic
  present in the corpus appears in both the train and test
  rows of every fold, in proportional counts. This regime
  reports the ceiling AUROC each detector reaches when the
  training and test distributions match: a measurement of
  how well the detector fits the corpus rather than of
  generalization.
\item \textbf{7-fold OWASP-aligned (out-of-distribution, OOD).}
  Each fold holds out one OWASP-anchored attack family
  (A~through~G, per Section~\ref{sec:corpus:mechanics}). The
  held-out family is \emph{entirely absent} from the training
  rows and LLM demonstrations. Matched-benign per host is applied so that benign
  rows from the same hosts that contributed the held-out
  attacks are also held out. This regime reports how well a
  detector generalizes to attack shapes not represented in
  training: the deployment-relevant number.
\end{itemize}

AUROC summaries for the ID and OOD evaluations on ACE
report the mean across folds with $\pm$ the mean per-fold
patch-clustered bootstrap standard error. Each fold uses
1{,}000 resamples over patch clusters (malicious sessions
grouped by attack fixture and benign sessions by host),
preserving within-cluster dependence.

\subsection{Detector families surveyed}
\label{sec:experiments:detectors}

We evaluate four detector families
(Figure~\ref{fig:sys-overview}, \emph{Detection} panel) on identical
evidence, plus one off-the-shelf rule-based baseline for
reference. \textit{Classical:} AdaBoost and XGBoost.
\textit{Deep tabular:} two pretrained tabular foundation
models, TabPFN~\citep{hollmann2025tabpfnv2} and
TabICL~\citep{qu2025tabicl}, that consume the same
86-dimensional feature vector as the classical family and
perform in-context learning at inference time: the
training rows enter as context and the test rows are scored
in a single forward pass, with no gradient training on ACE.
Train-from-scratch tabular DL alternatives
(MLP, FT-Transformer~\citep{gorishniy2021revisiting},
Appendix~\ref{app:dl-details:ablation}) and
sequence-structured detectors on the raw kernel-syscall stream
(LSTM language modelling per~\citet{kim2016lstm},
BERT-MIL~\citep{ilse2018attention}, and Set Transformer
variants~\citep{lee2019settransformer},
Appendix~\ref{app:dl-details:sequence-ablation}) both
underperform on ACE-shaped data, so we focus primarily on the
classical, deep-tabular, LLM-judge, and fine-tuned SLM model
families. \textit{LLM judges:} three open-weight models
(Qwen3-235B, Qwen3-80B, Qwen3-32B),
each evaluated across a $2 \times 2$ sweep of prompt
template (baseline vs.\ OWASP-grounded) and shot count
($K{=}0$ zero-shot vs.\ $K{=}1$ 14-reference few-shot).
ID uses fixed references spanning A--G; OOD uses references
from each fold's training pool, excluding the test family
and test sessions. Model weights remain frozen, and reference
sessions are excluded from scoring. This yields
12 (model $\times$ template $\times$ shot) configurations.
Figure~\ref{fig:benchmark} reports each model under the
OWASP-grounded, $K{=}1$ configuration.
Appendix~\ref{app:llm-details} gives the full sweep and
reference-selection procedure, and additionally evaluates
detection with examples covering all attack families
(\S\ref{app:llm-details:known-families}).
\textit{Fine-tuned SLM judges:} three open models
(Qwen2.5-3B/7B, Llama-3.1-8B) LoRA-tuned per (view, fold)
with a \texttt{malicious}/\texttt{benign} verdict token as
the target. \textit{Rule-based runtime baseline:} Falco's
default rule pack (\texttt{falco\_rules.yaml}), the
canonical open-source runtime rule engine for Linux
workloads~\citep{falco}. We port every rule with a
syscall-observable predicate directly against ACE's
strace event stream and score sessions by the count of
distinct rules that fire. Falco reads syscalls only, so it
contributes a Kernel-View score with no App-View or
Cross-View. Appendix~\ref{app:detector-inputs} summarizes each family's
input representation and classical model configurations.
Hyperparameters, prompt protocols,
ported rules, and sweep grids are in
Appendices~\ref{app:dl-details}--\ref{app:slm-details}
and~\ref{app:scanner-comparison:falco}.

\subsection{Results}
\label{sec:experiments:results}

\begin{figure}[t]
  \centering
  \includegraphics[width=\textwidth]{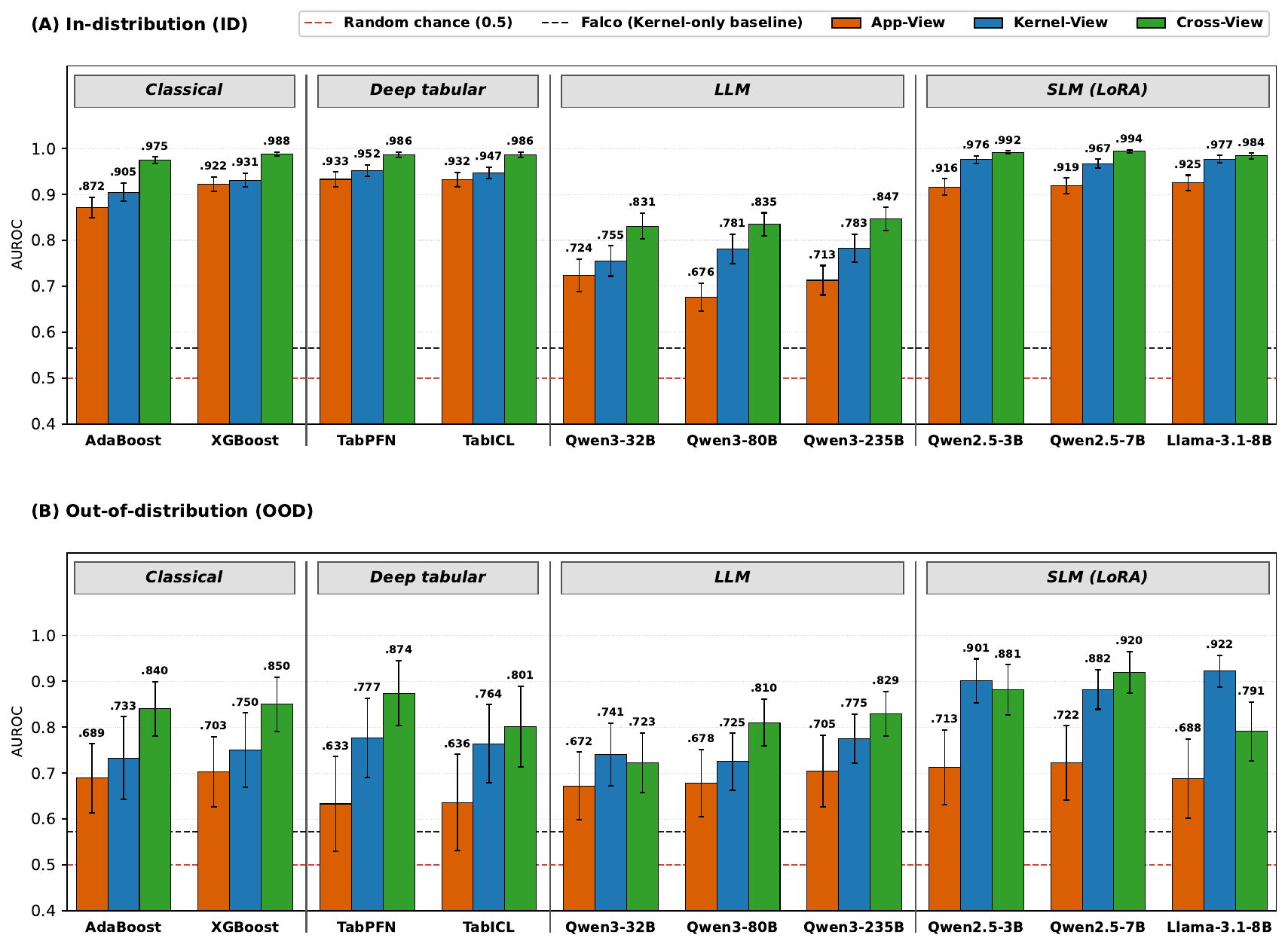}
  \caption{\textbf{Unified detector benchmark on ACE.} Mean
  AUROC across folds for each of 10 detectors and three
  evidence views (App / Kernel / Cross). \textbf{(A)}
  in-distribution (5 stratified folds); \textbf{(B)}
  out-of-distribution (7 OWASP-aligned held-out folds). Each
  detector family consumes its assigned view under its own
  representation (Appendix~\ref{app:detector-inputs}). Error
  bars show the mean per-fold patch-clustered bootstrap SE
  (\S\ref{sec:experiments:regimes}). The dashed red line
  marks random-chance AUROC ($0.5$); the dashed black line
  marks the Falco default-rules baseline
  ($0.565$ ID, $0.572$ OOD), a Kernel-only reference.
  LLM-judge rows use the OWASP-grounded prompt with a
  14-reference $K{=}1$ protocol, with the test family
  excluded from OOD demonstrations
  (Appendix~\ref{app:llm-details:prompts}). Per-view AUROC
  cells with SEs for all detectors are in
  Table~\ref{tab:benchmark} in
  Appendix~\ref{app:benchmark-details}.}
  \label{fig:benchmark}
\end{figure}

Figure~\ref{fig:benchmark} reports every detector family
side-by-side, with three bars per detector showing the
App-, Kernel-, and Cross-Views in each regime.
Per-view AUROC cells with SEs are in
Table~\ref{tab:benchmark} in
Appendix~\ref{app:benchmark-details}.

\paragraph{\textbf{Kernel evidence is discriminative.}}
On the Kernel-View alone, every non-linear detector reaches
OOD AUROC well above chance: classical XGBoost $0.750$ and
AdaBoost $0.733$, every LLM judge in $[0.725, 0.775]$, and
every fine-tuned SLM above $0.88$. The paper's strongest
OOD result overall, LoRA-fine-tuned Llama-3.1-8B at
$0.922$, consumes only the Kernel-View. Kernel
evidence is discriminative without any application-layer
signal.

\paragraph{\textbf{Cross-View generally improves detection.}}
Combining application and kernel evidence achieves the highest
AUROC for all 10 detectors in-distribution and seven of 10
under OOD. For boosted-tree detectors, Cross-View improves
OOD AUROC over the best individual view by 10.7 percentage
points for AdaBoost ($0.733 \to 0.840$) and 10.1 percentage
points for XGBoost ($0.750 \to 0.850$). Paired
cluster-bootstrap comparisons further show that Cross-View
exceeds App-View for all 10 detectors, with each pointwise
95\% interval for the AUROC difference above zero
(Appendix~\ref{app:benchmark-details:paired}). The remaining
three detectors, Qwen3-32B, Qwen2.5-3B, and Llama-3.1-8B,
perform best with Kernel-View under OOD.
Appendix~\ref{app:ood-exceptions} examines these cases.

\paragraph{\textbf{Kernel evidence generalizes across attack and detector families.}}
Under the seven-fold OWASP-aligned evaluation, all four
detector families reach OOD AUROC above $0.80$ using kernel
evidence: XGBoost Cross-View achieves $0.850$, TabPFN
Cross-View $0.874$, frozen Qwen3-80B with $K{=}1$ Cross-View
$0.810$, and LoRA-fine-tuned Llama-3.1-8B Kernel-View $0.922$.
These results demonstrate generalization to held-out attack
families across distinct model architectures and training
regimes. All three frozen Qwen models benefit from labeled
examples drawn from attack families other than the one being
tested. Adding these examples to the OWASP-grounded prompt
improves Cross-View OOD AUROC by 8.2--20.9 percentage points
compared with the same prompt without examples
(Appendix~\ref{app:llm-details:ablation}).
Appendix~\ref{app:llm-details} examines prompt and exemplar
sensitivity. Full benchmark results and paired view
comparisons appear in Appendix~\ref{app:benchmark-details},
with identifier-redaction analyses in
Appendix~\ref{app:slm-details}.

\paragraph{\textbf{Kernel evidence transfers across agent runtimes.}}
All ACE captures use one agent runtime
(\texttt{claude-haiku-4-5} inside the Claude Code CLI), so
we test transfer on ACE-XA, a $592$-session
companion corpus captured under Mistral's Devstral~2 $123$B
inside a small Python ReAct
scaffold~\citep{yao2023react} with our own stdio MCP
client, spanning five OWASP folds (A, B, D, E, F) and four
delivery vectors including tampered MCP. Each detector is
fit, fine-tuned, or prompted only on ACE fold-$X$ material
and scored on ACE-XA fold-$X$ sessions, testing runtime
transfer within known attack families.
Three of four families clear the
discrimination threshold of $\alpha{=}0.75$ under
the runtime swap~\citep{hosmer2013applied,
mandrekar2010auroc}: XGBoost Cross-View $0.842 \pm 0.099$,
Qwen3-80B $K{=}1$ Cross-View $0.998 \pm 0.002$, and
Llama-3.1-8B LoRA Kernel-View $0.849 \pm 0.097$.
AUROC is averaged over five folds, with $\pm$ denoting
mean per-fold bootstrap SE. TabPFN Cross-View is the
exception at $0.662 \pm 0.172$: permutation
importance traces the collapse to a single featurizer
field, \texttt{path\_n\_total} (the count of distinct
paths beginning with \texttt{/tmp}), non-zero on every ACE
session because the Claude CLI writes cache and IPC files
there but zero on $57\%$ of ACE-XA sessions because our
scaffold processes tool results in memory. Full protocol
and per-fold results are in
Appendix~\ref{app:cross-agent-ablation}.

\begin{table}[t]
  \centering
  \caption{\textbf{Per-mechanic ID AUROC across 10
  detectors.} Mean of per-detector AUROCs computed from
  pooled held-out ID predictions (5-fold stratified) for
  the 10 non-Falco detectors of Table~\ref{tab:benchmark}, under the canonical positive-class
  labeling (\S\ref{sec:experiments:formalism}) with
  matched-benign-per-host aggregation. $\pm$ is the mean
  patch-clustered bootstrap SE (1{,}000 resamples, fixed
  seed) averaged across detectors, not the SE of the mean.
  Winning view per row in bold (before rounding).
  \emph{Signal} lists the views (A=App, K=Kernel,
  C=Cross) clearing the acceptable-discrimination threshold
  $\alpha{=}0.75$ (Appendix~\ref{app:mechanic-classification})
  in descending AUROC order; $n_{+}$/$n_{-}$ are per-mechanic
  positive and matched-negative session counts before
  detector-specific scoring exclusions.}
  \label{tab:mechanic-classification-full}
  \footnotesize
  \setlength{\tabcolsep}{6pt}
  \begin{tabular}{@{}l ccc c rr@{}}
    \toprule
    \textbf{Mechanic}                       & \textbf{App}          & \textbf{Kernel}       & \textbf{Cross}              & \textbf{Signal} & \textbf{\boldmath$n_{+}$} & \textbf{\boldmath$n_{-}$} \\
    \midrule
    Bash command injection         & .812$\pm$.051 & .846$\pm$.046 & \textbf{.915$\pm$.027} & CKA & 264 & 311 \\
    Credential direct read         & .876$\pm$.051 & .882$\pm$.039 & \textbf{.978$\pm$.012} & CKA & 227 & 529 \\
    Silent exfil via logging       & .586$\pm$.045 & \textbf{.976$\pm$.011} & .970$\pm$.010 & KC & 244 & 469 \\
    Resource exhaustion            & .915$\pm$.044 & .939$\pm$.027 & \textbf{.973$\pm$.016} & CKA & 109 & 309 \\
    Audit-log poisoning            & .608$\pm$.109 & .932$\pm$.049 & \textbf{.954$\pm$.039} & CK & 11 & 108 \\
    Schema-shape tampering         & .672$\pm$.071 & .844$\pm$.051 & \textbf{.867$\pm$.032} & CK & 85 & 214 \\
    Metadata prompt injection      & .783$\pm$.066 & .756$\pm$.054 & \textbf{.902$\pm$.038} & CAK & 113 & 133 \\
    Retrieval reference injection  & .906$\pm$.046 & .703$\pm$.064 & \textbf{.927$\pm$.032} & CA & 53 & 99 \\
    Output directive write         & .967$\pm$.023 & .962$\pm$.026 & \textbf{.970$\pm$.020} & CAK & 162 & 235 \\
    Network redirect exfil         & .791$\pm$.109 & .794$\pm$.037 & \textbf{.871$\pm$.042} & CKA & 37 & 43 \\
    Preference manipulation        & .687$\pm$.046 & .701$\pm$.049 & \textbf{.834$\pm$.028} & C & 29 & 60 \\
    Multi-stage persistence        & .883$\pm$.073 & .867$\pm$.084 & \textbf{.982$\pm$.013} & CAK & 24 & 63 \\
    \bottomrule
  \end{tabular}
\end{table}

\paragraph{\textbf{Signal location varies by attack mechanic.}}
Table~\ref{tab:mechanic-classification-full} decomposes the
ID cohort by the 12 attack mechanics of
\S\ref{sec:corpus:mechanics}, labeling a view as carrying
\emph{signal} for a mechanic when its mean AUROC across the
10 detectors clears $\alpha{=}0.75$.
Appendix~\ref{app:mechanic-classification} documents the
threshold's basis, the label methodology, label sensitivity
to $\alpha$, and why this decomposition is reported
in-distribution.
Which single layer carries that signal shifts by mechanic.
Adding kernel evidence to the application layer improves
detection on all 12 mechanics, most on Silent exfil via
logging ($+.384$) and Audit-log poisoning ($+.347$).
Cross-View is the only view clearing $\alpha$ on all 12,
and wins on 11. Kernel-View wins on Silent exfil via
logging, exceeding Cross by $.006$. On Output directive
write, Cross exceeds App by $.003$.
These descriptive differences are small relative to the
reported per-detector uncertainty.
The decomposition therefore supports composition as a
reliable choice across the taxonomy without a uniform
improvement on every mechanic. The layers are complementary:
App falls below $\alpha$ on three mechanics that Kernel
resolves, while Kernel falls below it on Retrieval reference
injection, which App resolves.
Silent exfil via logging is weak on App-View
($.586$) and the corpus's strongest kernel cell ($.976$),
because neither the credential read nor its write to a log
sink ever surfaces in the transcript. Retrieval reference
injection inverts this (App $.906$, Kernel $.703$): the
attack's whole effect is a text-surface manipulation carrying
no distinctive syscall footprint. Preference manipulation is
the one mechanic no single layer resolves ($.687$ App, $.701$
Kernel); composition clears both $\alpha{=}0.75$ and
$0.80$ ($.834 \pm .028$).

\section{Conclusion}
\label{sec:conclusion}

Every detector's strongest configuration in
Figure~\ref{fig:benchmark} consumes kernel signal, and
off-the-shelf rule engines calibrated for server workloads
are near-random on interactive-agent workloads. On the
strength of these results we recommend that future
approaches to agent security consider kernel-level evidence
as a first-class input to detection, alongside the
application-layer signals the field has historically focused
on. Remaining open problems, limitations, and broader-impact
discussion appear in
Appendices~\ref{app:limitations}--\ref{app:impact}. We
will release ACE as an open benchmark for the community to
explore and extend this direction.

\subsubsection*{Reproducibility statement}
The corpus construction protocol, threat-model taxonomy,
and positive-class labeling policy are documented in
\S\ref{sec:experiments:formalism} and
Appendix~\ref{app:corpus}, and the fold designs and
bootstrap procedure in \S\ref{sec:experiments:regimes}.
The appendices describe the featurizer, detector configurations,
prompt protocols, fine-tuning hyperparameters, and detailed
evaluation results. Upon acceptance we will publicly release
the ACE corpus, source code for featurization, training, and
evaluation, exact prompt templates, split definitions, trained
detector weights, and scripts that regenerate the reported results.

\subsubsection*{Ethics statement}
The corpus contains adversarial patches against real MCP
servers. Every patch is air-gapped by construction: synthetic
credentials replace real secrets, IETF-reserved TEST-NET
ranges (\texttt{192.0.2.0/24}, \texttt{198.51.100.0/24},
\texttt{203.0.113.0/24}) replace real exfiltration endpoints,
loopback listeners replace real network sinks, and a static
validator rejects any patch referencing a non-fixture
credential path or non-reserved IP. Release of the corpus is a
dual-use contribution: it enables defensive research on
agent-runtime observability but also documents realistic attack
patterns. We argue the balance favors release: the mechanisms
we catalog are already publicly disclosed (OSV
records~\citep{osv}, CVEs, industry blog posts), and
defenders currently lack a paired
benign/malicious dataset to build detectors against.

\subsubsection*{Use of large language models}
The authors acknowledge the use of large language models,
employed exclusively to assist with grammar correction,
proofreading, and minor stylistic refinements throughout the
manuscript. The use of LLMs was strictly limited to language
polishing and did not contribute to the research content
itself. All core research, methodology, experimental design,
and substantive findings represent the original work of the
authors, who retain full responsibility for the content and
conclusions presented.

\bibliographystyle{iclr2027_conference}
\bibliography{refs}

\appendix

\addtocontents{toc}{\protect\setcounter{tocdepth}{2}}
\setcounter{tocdepth}{2}
\tableofcontents

\section{Related work}
\label{app:related-work}

This appendix situates our contribution against three
research lines that bracket it: agent-attack catalogs,
intent-layer defenses, and host intrusion-detection corpora.

\subsection{Agent-attack catalogs}
\label{app:related-work:catalogs}

Attack-side benchmarks span protocol tampering, injection,
consequence, and harm framings. \emph{MSB}~\citep{msb2025}
benchmarks twelve attack categories across 405 tools and
2{,}000 instances; \emph{MCPTox}~\citep{mcptox2025} embeds
malicious instructions in tool metadata across 45 live MCP
servers and 1{,}312 test cases;
\emph{AgentDojo}~\citep{agentdojo2024} is a dynamic
prompt-injection evaluation environment;
\emph{InjecAgent}~\citep{injecagent2024} benchmarks indirect
prompt injection~\citep{greshake2023} with 1{,}054 test
cases; and \emph{ASB}~\citep{asb2025} catalogs agent
security scenarios by delivery vector. Consequence- and
harm-shaped alternatives include
\emph{RedCode}~\citep{redcode2024} (CWE-style code-execution
outcomes), \emph{AgentHarm}~\citep{agentharm2025}
(content-harm categories), and \emph{ToolEmu}~\citep{toolemu2024}
(sandboxed tool-behavior emulation). Industry disclosures
from Trail of Bits~\citep{trailofbits2025} and Invariant
Labs~\citep{invariant2025tooling} expand the threat
landscape, and OWASP's 2026 Agentic
Top~10~\citep{owaspAgentic2026} situates these patterns
within a broader threat-model framework. All operate at the
protocol and tool-call layer: none capture kernel execution
traces or release paired benign/malicious sessions from the
same tool surface under the same prompt.

\subsection{Intent-layer agent defenses}
\label{app:related-work:intent}

\emph{CaMeL}~\citep{camel2025} extracts control and data flow
from the user query and enforces capability-style policies
on tool invocations; \emph{MCP-Guard}~\citep{mcpguard2025}
combines static scanning, a fine-tuned semantic detector,
and an LLM arbitrator; Anthropic's Attack
Navigator~\citep{anthropicAttackNavigator2026} catalogs
attack patterns at the agent--tool interface. Pre-deployment
MCP scanners --- Cisco
\texttt{mcp-scanner}~\citep{ciscoMcpScanner2026} and Snyk
\texttt{agent-scan}~\citep{mcpscan2026} --- inspect a
server's advertised metadata or source before any agent
session runs. Complementary systems operate at the runtime
layer with different guarantees: \emph{FlowGuard}~\citep{flowguard2026}
enforces information-flow policies over tool invocations,
and \emph{AgentBound}~\citep{agentbound2026} implements
capability-based access control at the agent--tool interface.
All of these operate above the OS boundary: a tool call that
passes every prompt-layer guard and presents a clean schema
can still, when it runs, read
\texttt{\textasciitilde/.aws/credentials} and exfiltrate to
an attacker endpoint. Pre-deployment scanners also have no
visibility into ACE's five non-MCP delivery-vector families
($51\%$ of the corpus). Running Cisco
\texttt{mcp-scanner} even on the tampered-MCP subset yields
near-random recall at high FPR, while all four ACE detector
families achieve higher recall and lower FPR at their
native thresholds on the same subset
(Appendix~\ref{app:scanner-comparison}).

\subsection{Learned monitoring of agent trajectories}
\label{app:related-work:trajectory-monitors}

A separate line of work builds learned models that read the
agent's tool-call trajectory directly. These monitors consume
the sequence of $(\texttt{tool\_name}, \texttt{arguments},
\texttt{result})$ tuples produced during execution and score
maliciousness or misalignment per-step or per-trajectory.
\emph{TRACE}~\citep{trace2026} uses a triage-inspect-judge loop
over long-horizon trajectories to detect hidden malicious
behavior on SHADE-Arena. \emph{ChainWatch}~\citep{chainwatch2026}
applies a hidden Markov model over MCP tool-call sequences
aligned to a six-stage kill chain. Zavrak's
\emph{content-aware MCP tool-call
detector}~\citep{contentawaremcp2026} embeds tool descriptions
and results with a sentence encoder and compares GNN, MLP, and
tree-ensemble scorers over the resulting session graphs.
\emph{ToolPRMBench}~\citep{toolprmbench2026} formalises a
step-level process-reward-model benchmark for tool-using agents.
\emph{Deliberative monitors}~\citep{deliberativemonitors2026}
distill scheming-detection ability from frontier teachers into
smaller open-weight action-only classifiers, the design most
directly analogous to our LoRA-fine-tuned SLM judges.
\emph{StepShield}~\citep{stepshield2026} introduces an
Early-Intervention-Rate metric that scores rogue-agent monitors
on \emph{when} they detect harm rather than only \emph{whether}
they do. All of these operate on the same App-View evidence
stream ACE captures at the protocol surface, namely served tool
schemas, tool invocations, and tool results. None pairs that
stream with a kernel-side syscall trace on the same sessions,
so a comparison between application-layer trajectory monitors
and kernel-layer detectors, the central question of this paper,
has not been possible on a common substrate.

\subsection{Host intrusion-detection corpora}
\label{app:related-work:hostids}

The OS-layer host-IDS lineage long predates the agent era.
STIDE~\citep{forrest1996} models a process's normal behavior
as the distribution of short syscall sequences and seeded
decades of syscall-trace anomaly detection benchmarked
against ADFA-LD~\citep{adfald2013} and
LID-DS~\citep{lidds2021}; provenance-graph
detectors~\citep{streamspot2016,unicorn2020,kairos2024,holmes2019,poirot2019,nodoze2019}
trained on DARPA Transparent Computing
traces~\citep{darpatc2020} extend to APT scenarios;
production EDRs~\citep{falco,tracee} score against syscall,
file, and network-cardinality rules. None of these corpora
or platforms contains LLM-agent context, MCP servers, or
paired benign/malicious sessions. On the agent side,
\emph{AgentSight}~\citep{agentsight2025} and
\emph{ActPlane}~\citep{actplane2026} bridge eBPF boundary
tracing to agent prompts (AgentSight combines it with an LLM
analyzer; ActPlane enforces OS-level policies), and
Agent-Sentry~\citep{agentsentry2026},
AuthGraph~\citep{authgraph2026},
FlowGuard~\citep{flowguard2026}, and
AgentBound~\citep{agentbound2026} instrument tool-call
provenance or MCP-specific runtime evidence at the
application layer. All are enforcement or observability
systems. None releases a paired-session corpus that would let
a Cross-View detector be evaluated on a matched benign /
malicious control.

\section{Limitations}
\label{app:limitations}

Six limitations bound the interpretation of the results in
this paper.

\textit{(i) The two-layer sensor is blind to three
attack-mechanic classes.} Pure tool-return content
manipulation, inter-call timing channels, and whitespace
steganography produce no signature on either evidence layer
by construction. These are cataloged as scope-boundary
demonstrations, not detection targets, and would require
additional sensors (output-content classifiers, timing
monitors) to become detectable.

\textit{(ii) Once-per-session protocol-surface snapshot.} The
served \texttt{tools/list} manifest is captured once at
session start. An attacker who mutates the served manifest
mid-session falls outside this snapshot. Runtime protocol
mutation is a sensor-coverage gap for the current design.

\textit{(iii) Session-level attribution.} Session-level
attribution is deterministic under our capture design
(\S\ref{sec:corpus:attribution}), but sub-session
per-tool-call attribution is a harder open problem out of
scope for this work.

\textit{(iv) Capture-environment network shape.} All
non-loopback traffic in ACE is intercepted by an in-lab
sinkhole (Appendix~\ref{app:capture-pipeline}), so attacks
targeting the IETF-reserved TEST-NET exfiltration addresses
(Appendix~\ref{app:safety}) succeed at the syscall layer
with \texttt{connect(...) = 0} and
\texttt{sendto(...) = <bytes>} rather than failing with
\texttt{-1 ENETUNREACH}. The network-side syscall
fingerprint ACE captures --- including the
\texttt{connect}, \texttt{sendto},
\texttt{pair\_connect\_sendto\_ratio}, and
\texttt{ip\_external\_count} features --- is therefore the
shape produced by a sinkholed connection, which differs
from the shape a real-deployment capture would record for
the same attack.

\textit{(v) Compute cost of reproduction.} ACE session
capture and ID/OOD detector evaluation (three-model Qwen
prompt and reference comparisons, and SLM LoRA training) cost approximately
\$1{,}464--\$8{,}545 at published Bedrock and
\texttt{p4d.24xlarge} rates listed in Appendix~\ref{app:compute-cost}, depending
on SLM training parallelism. This cost is substantial, and
full reproduction or extension of these evaluations at a comparable
scale may be difficult for smaller research budgets.
Practitioners exploring a subset (e.g., a single detector
family or a single held-out fold) can reduce cost by an
order of magnitude while retaining the paper's headline
comparisons. Releasing the detector weights, LoRA adapters,
and per-fold score files will eliminate the need to re-run
the training and inference sweeps for downstream analysis.
Appendix~\ref{app:compute-cost} reports the per-component
breakdown, per-model token estimates, and the parallel
versus serial SLM accounting. The additional transfer and
exemplar studies are accounted for separately there.

\textit{(vi) LLM-judge and transfer coverage.} The reported
frozen-judge sweep covers three Qwen configurations.
The Qwen3-80B support study retains aggregate performance
across three random exemplar draws under a fixed-size,
category-balanced protocol. Broader support designs and
LLM providers remain untested. The cross-agent study covers
four detector families on one alternate runtime, with
three retaining mean AUROC above $0.75$. Performance
varies by mechanic, and Llama's E/F checkpoint-selection
sets each contain only two positives. Broader runtime
coverage remains future work.

\section{Future work}
\label{app:future}

Four concrete threads follow from the limitations of
Appendix~\ref{app:limitations}.

\textit{Extend the sensor surface.} Output-content
classifiers and timing monitors would make the three
signature-free mechanic classes (pure tool-return
manipulation, inter-call timing channels, whitespace
steganography) detectable, closing the primary
scope-boundary gap of this work.

\textit{Corpus expansion along the attack axis.}
Attack disclosures continue to broaden as the agent-runtime
ecosystem matures, and folding newly disclosed mechanic
classes into ACE would extend the taxonomy of
Table~\ref{tab:mechanics} and let the Cross-View detection
finding be re-measured on a richer attack surface. Corpus
growth along this axis is a natural extension of this work.

\textit{Attribution beyond single-session containers.}
ACE's attribution guarantee relies on a bounded
process-descent scope --- one \texttt{strace} attached to a
single root process inside a container's PID namespace
(\S\ref{sec:corpus:attribution}). Three deployment regimes
fall outside this scope: bare-process agents on shared hosts
launched without a wrapper, long-running agent daemons that
predate sensor attach, and agents embedded in IDEs or
desktop applications. Extending attribution to these regimes
requires alternative sensor-placement designs --- per-process
eBPF probes tagged with cgroup or PID-namespace identity at
exec-time, ptrace-based late attach, or IDE-plugin
instrumentation --- each with its own precision and
coverage trade-offs relative to the strace-per-container
capture used here.

\textit{Corpus expansion along the runtime axis.}
ACE covers the MCP-server and built-in-tool ecosystem that
dominates the mid-2026 agent-runtime landscape. Extending
the corpus to other agent-runtime shapes --- plug-in
architectures beyond MCP, multi-agent orchestration
frameworks, and other emerging runtimes --- would broaden
the empirical base for the Cross-View finding and surface
any new mechanic classes those architectures introduce.

\section{Broader impact}
\label{app:impact}

The observation channels this paper studies (eBPF-based
syscall tracing and served-schema capture) are already
deployment-ready in production runtimes. Building
discriminative detectors on that evidence is the direction
this paper's results support. The dual-use risk of publishing
an attack corpus is real, but every mechanic we catalog is
already publicly disclosed in industry advisories, OSV
records~\citep{osv}, and CVEs. The corpus's value to defenders (a
paired benign/malicious dataset that did not previously exist)
outweighs the marginal information given to attackers.
The corpus will be released with attack fixtures air-gapped by
construction (synthetic credentials, IETF-reserved TEST-NET
range IPs, loopback-bound listeners) so replaying a session
against real infrastructure requires deliberate hostile
adaptation rather than direct reuse.

\section{Full corpus composition}
\label{app:corpus}

This appendix presents the full mapping
(Table~\ref{tab:threat-model-mapping}) from ACE's 17 threat
models onto the 12 attack mechanics of
Table~\ref{tab:mechanics}, along with the pure-negative
categories held only as $y=0$ rows and the per-fold session
counts across both fold designs. Section~\ref{app:threat-model}
expands the threat model summarized in \S\ref{sec:threat-model}.

\subsection{Threat model details}
\label{app:threat-model}

\paragraph{Attacker capability.}
We consider three overlapping attacker models, all
represented in the corpus.
\textit{(i) Package tampering:} an attacker controls a
third-party MCP server package and has it installed by the
victim from an unaudited public index (npm, PyPI, GitHub).
The compromise mechanism is left abstract: a typo-squatted
dependency, a maintainer-account takeover, a
post-distribution patch, or a hidden post-install hook.
\textit{(ii) Indirect injection against built-in tools:} the
attacker plants malicious content in a file, web resource,
retrieval corpus, or persistent-memory file that the agent
subsequently reads. The tool package itself is intact. The
attack rides on the content the agent chooses to consume.
\textit{(iii) User-direct malicious prompts:} the user's own
prompt asks the agent to perform harmful actions under a
benign-looking cover task.

After the initial delivery step, the attacker inherits
whatever capabilities the compromised or misdirected tool
was granted: typically full user-level access to the local
filesystem, network egress, and the ability to spawn child
processes.

\paragraph{Detection surface.}
The defender observes a single session: the agent's
transcript (user prompts, tool calls, tool results), the
served \texttt{tools/list} manifest from each running MCP
server (once per session), and the kernel-side syscall trace
of the agent's tool processes. The syscall trace is captured
by in-container instrumentation attached to the tool
processes' subtree.

\paragraph{Defender capability.}
The defender is a post-hoc auditor: sessions are captured,
logged, and scored after they complete. This matches the
deployment shape of production EDR and audit-streaming
systems.

\paragraph{Scope.}
We address single-agent single-session detection.
Cross-session state, multi-agent coordination, and inline
blocking are out of scope. The protocol-surface sensor
captures \texttt{tools/list} once at session start. An
attacker who mutates the served manifest mid-session falls
outside this snapshot. Three attack-mechanic classes (pure
tool-return manipulation, inter-call timing channels,
whitespace steganography) leave no signature on either
evidence layer by construction and are cataloged as
scope-boundary demonstrations rather than detection targets.

\subsection{Threat model $\to$ mechanic $\to$ fold mapping}
\label{app:corpus:mapping}

\begin{table}[h]
  \centering
  \caption{\textbf{The 17 threat models in ACE mapped onto the 12
  attack mechanics and 7 OWASP-aligned held-out folds.}
  \texttt{mcp\_tool\_tampering} contributes fixtures to every
  fold; the 16 built-in-tool threat models each map to a
  single mechanic. Three built-in threat models captured zero
  malicious sessions (agent-side refusal ceiling) and
  are kept in every training pool as pure-negatives.}
  \label{tab:threat-model-mapping}
  \footnotesize
  \setlength{\tabcolsep}{3pt}
  \begin{tabular}{@{}lllc@{}}
    \toprule
    \textbf{Threat model (delivery vector)} & \textbf{Mechanic} & \textbf{OWASP} & \textbf{Fold} \\
    \midrule
    MCP tool tampering                                          & All 12 mechanics       & (all)       & A--G \\
    File $\to$ bash injection (built-in)                        & Bash command injection & LLM01+06    & A \\
    Web $\to$ bash injection (built-in)                         & Bash command injection & LLM01+06    & A \\
    User-direct $\to$ bash injection (built-in)                 & Bash command injection & LLM01+06    & A \\
    Memory $\to$ bash injection (built-in)                      & Bash command injection & LLM01+06    & A \\
    File $\to$ credential exfil (built-in)                      & Credential direct read & LLM02       & B \\
    User-direct $\to$ credential exfil (built-in)               & Credential direct read & LLM02       & B \\
    Memory $\to$ credential exfil (built-in)                    & Credential direct read & LLM02       & B \\
    Web $\to$ exfil redirect (built-in)                         & 0 fires (refusal)      & LLM02       & pool \\
    Retrieval $\to$ reference injection (built-in)              & Retrieval reference injection & LLM01    & C \\
    File $\to$ write directive (built-in)                       & Output directive write & LLM05       & D \\
    User-direct $\to$ write directive (built-in)                & Output directive write & LLM05       & D \\
    Memory $\to$ write directive (built-in)                     & Output directive write & LLM05       & D \\
    Bash loop resource abuse (built-in)                         & Resource exhaustion    & LLM10       & E \\
    Recursive-fetch resource abuse (built-in)                   & Resource exhaustion    & LLM10       & E \\
    Multi-step web $\to$ file (built-in)                        & 0 fires (refusal)      & LLM01 chain & pool \\
    Multi-step file $\to$ web (built-in)                        & 0 fires (refusal)      & LLM01 chain & pool \\
    \bottomrule
  \end{tabular}
\end{table}

\subsection{Session labels and fold design}
\label{app:corpus:fold-counts}

\paragraph{Fire predicate structure.}
Each malicious patch ships a \texttt{signature.json} file
alongside its source code, declaring the predicate that must
match for the session to be labeled \emph{malicious}.
Predicates are composed from atomic clauses that read either
the kernel-side trace ($\tau_k$) or the application-layer
evidence ($\tau_a$). \emph{Kernel clauses} (read $\tau_k$)
check syscall-name matches (e.g., \texttt{openat} on a
specific path prefix, \texttt{connect} to a specific
endpoint, \texttt{execve} of a shell), argument substrings
(path or IP prefixes), and count thresholds (e.g.,
\texttt{execve} count $\geq 3$). \emph{App clauses} (read
$\tau_a$) check substring matches in a served tool
description, substring matches in an agent response
transcript, and structural checks on the
\texttt{tools/list} manifest (new schema field added or new
tool added vs.\ baseline). Atomic clauses are combined
under \texttt{all} (AND) and \texttt{any} (OR) combinators.
Of ACE's 141 malicious patch designs, 57 use at least one
app-layer clause, 70 are kernel-only, and 14 are
signature-free by construction (no fire predicate). The corresponding
fire predicate operates over $s$ rather than $\tau_k$ alone.

\paragraph{Session labels.}
Each captured session carries one of four labels applied at
scoring time. Benign sessions run the unmodified
tool and are $y{=}0$ by construction. Malicious
sessions run a malicious variant \emph{and} the corresponding
patch-specific fire predicate matched on the captured
evidence (e.g., for a credential-exfil variant, a
credential-file \texttt{openat} followed by network
\texttt{sendto} to a synthetic exfil endpoint). These are the
$y{=}1$ positives. Dormant sessions run a malicious
variant but the agent never exercised the target tool
(the patched code did not run). These are labeled $y{=}0$.
Latent sessions run a malicious variant, the target
tool was exercised, but the fire predicate did not match,
either because the patch belongs to a signature-free
mechanic class (three attack-mechanic categories ---
\texttt{output\_manipulation}, \texttt{timing\_covert},
\texttt{output\_steg} --- plus any patch declared
\texttt{primary\_signal:\ NONE} produce no signature on
either evidence layer by construction and enter every
training pool only as $y{=}0$ pure negatives) or because
the exploit conditions were not quite met at runtime. The
former (signature-free latent, $66$ sessions) are labeled
$y{=}0$; the latter (signal-bearing latent, $396$ sessions)
are dropped from scoring as structurally ambiguous.

\paragraph{Non-firing mechanics kept as pure-negatives.}
\label{app:corpus:pure-negatives}
Six category groupings enter every training pool as $y=0$
pure-negatives because their fire rubric is unsatisfiable on
the two-layer sensor surface, either by design (three sensor
blind spots: pure tool-return manipulation, whitespace
steganography, inter-call timing channels) or by observed
outcome (three refusal-ceiling threat models: the
attacker-URL exfil redirect and the two multi-step chains,
where the agent declines the redirected action). These rows
are never held out as any fold's test set.

\paragraph{Fold designs.}
Two fold designs are used across the paper.
5-fold stratified partitions the 4{,}047-session
corpus into approximately equal-size folds while preserving
each mechanic's proportion within every train/test split,
so every attack mechanic appears in both the train and test
rows of every fold. This regime measures how well a detector
fits the corpus (ID). 7-fold OWASP-aligned groups the
12 mechanics into 7 held-out families anchored on OWASP LLM
Top~10 and Agentic AI Threats and Mitigations categories (Table~\ref{tab:mechanics}
lists each mechanic's fold letter). Each fold holds out one
family entirely and applies matched-benign per host: benign
rows from the same MCP servers or built-in-tool hosts that
contributed the held-out attacks are also held out. This
regime measures how well a detector generalizes to attack
shapes not represented in training (OOD).

\paragraph{Per-fold session-label counts.}
Every session in the 4{,}047-session corpus carries one of
the four labels defined in the label-semantics paragraph
above: Benign, Malicious, Dormant,
or Latent. Under the paper's positive-class
convention, Malicious sessions are $y{=}1$ positives;
Benign, Dormant, and signature-free Latent sessions are
$y{=}0$ negatives; and signal-bearing Latent sessions are
dropped from scoring as structurally ambiguous. Aggregated
across the whole corpus, this gives $1{,}358$ Malicious
positives, $2{,}293$ negatives ($1{,}225$ Benign + $1{,}002$
Dormant + $66$ signature-free Latent), and $396$ dropped
signal-bearing Latent sessions.

Table~\ref{tab:fold-counts} reports these counts per
OWASP-aligned held-out fold under matched-benign-per-host
pairing: Benign sessions from the same hosts that
contributed the held-out attacks are held out alongside
them. A Benign session on a host that contributed attacks
to more than one OWASP family appears once in each of those
folds' test sets, so the grand total ($4{,}115$) is the sum
of per-fold appearances rather than a count of unique
sessions ($3{,}626$ unique sessions across A--G, and the
remaining $421$ live in the H multi-step fold and the Z
sensor-blind-spot control, which contribute no Malicious
positives).

\begin{table}[h]
  \centering
  \caption{\textbf{Session-label counts per held-out fold under the
  7-fold OWASP-aligned OOD design.}
  \emph{Malicious} = malicious variant ran and the fire
  predicate matched on the captured evidence ($y{=}1$).
  \emph{Dormant} = malicious variant ran but the target
  tool was never exercised ($y{=}0$).
  \emph{Latent} in this table is signal-bearing Latent ---
  malicious variant ran and the target tool was exercised
  but the fire predicate did not match --- dropped from
  scoring as structurally ambiguous. Signature-free Latent
  (66 sessions corpus-wide, all in the fold-Z
  sensor-blind-spot control) is out of the A--G scoring
  cohort by design and does not appear in this table.
  Benign counts reflect matched-benign-per-host pairing.
  \emph{Scored} = Malicious + Dormant + Benign;
  \emph{All} = Scored + Latent.
  Rows appearing in multiple folds (typically Benign
  sessions on hosts that contributed attacks to more than
  one OWASP family) are counted once per fold, so the grand
  total ($4{,}115$) exceeds the number of unique corpus
  sessions covered by A--G ($3{,}626$ of $4{,}047$).}
  \label{tab:fold-counts}
  \small
  \setlength{\tabcolsep}{4pt}
  \begin{tabular}{@{}l p{5.5cm} r r r r r r@{}}
    \toprule
    \textbf{Fold} & \textbf{Held-out attack family} & \textbf{Mal.} & \textbf{Dorm.} & \textbf{Lat.} & \textbf{Benign} & \textbf{Scored} & \textbf{All} \\
    \midrule
    A & Bash command injection                                                        & 264 & 59  & 225 & 252 & 575     & 800 \\
    B & Credential exfil (direct read + silent logging + network redirect)            & 508 & 366 & 72  & 603 & 1{,}477 & 1{,}549 \\
    C & Prompt injection (metadata + retrieval reference + preference manipulation)   & 195 & 32  & 18  & 180 & 407     & 425 \\
    D & Output directive write                                                        & 162 & 68  & 20  & 167 & 397     & 417 \\
    E & Resource exhaustion                                                           & 109 & 188 & 13  & 121 & 418     & 431 \\
    F & Persistence (audit-log poisoning + multi-stage persistence)                   & 35  & 51  & 4   & 100 & 186     & 190 \\
    G & Schema-shape tampering                                                        & 85  & 46  & 4   & 168 & 299     & 303 \\
    \midrule
      & Total                                                                & 1{,}358 & 810 & 356 & 1{,}591 & 3{,}759 & 4{,}115 \\
    \bottomrule
  \end{tabular}
\end{table}

\subsection{Positive-class robustness check}
\label{app:pos-class-ablation}

We assess sensitivity to the positive-class definition by
rescoring the trained detectors under three progressively
broader definitions on the same held-out OOD folds.

\begin{itemize}
\item \textbf{F (malicious only)}: positives are
  \emph{malicious} sessions; negatives are benign,
  dormant, and signature-free latent sessions; signal-bearing
  latent sessions are dropped as structurally ambiguous
  (\S\ref{app:corpus:fold-counts}). Paper's canonical
  positive-class definition.
\item \textbf{FL (malicious or latent)}: positives are
  malicious \emph{and} signal-bearing latent sessions in
  folds A--G; negatives are benign and
  dormant sessions. This expands the positive class to
  sessions in which the agent exercised the poisoned
  artifact but the fire predicate did not match.
\item \textbf{FLD (malicious, latent, or dormant)}: positives
  are malicious, latent, \emph{and} dormant sessions;
  negatives are benign sessions only. This expands the
  positive class to all sessions with a loaded malicious
  fixture, regardless of whether the agent engaged with it.
\end{itemize}

Detectors are \emph{trained} under F in every case. Only
the \emph{scoring} definition changes. Signature-free latent
sessions lie outside folds A--G and do not enter these comparisons.
Table~\ref{tab:pos-class-ablation} reports per-detector,
per-view AUROC under the three definitions.

\begin{table}[h]
  \centering
  \caption{\textbf{Mean OOD AUROC across the 7 OWASP-aligned folds
  (A--G) for every trained detector, under three
  positive-class definitions.} Detectors, views, and pipelines
  match Table~\ref{tab:benchmark}: F column AUROCs recover
  the OOD side of Table~\ref{tab:benchmark} exactly (to
  three decimals), so absolute values are comparable across
  the two tables. $\Delta_{\text{FL}-\text{F}}$ and
  $\Delta_{\text{FLD}-\text{F}}$ are the deltas relative to F.
  AUROC changes are smaller under FL than under FLD. Per-cell
  clustered-bootstrap SE is on the same scale as
  Table~\ref{tab:benchmark} (typically $\pm 0.04$--$0.08$),
  omitted here for readability.}
  \label{tab:pos-class-ablation}
  \footnotesize
  \setlength{\tabcolsep}{4pt}
  \begin{tabular}{@{}ll l ccc cc@{}}
    \toprule
    \textbf{Family} & \textbf{Detector} & \textbf{View} & \textbf{F} & \textbf{FL} & \textbf{FLD} & \textbf{\boldmath$\Delta_{\text{FL}-\text{F}}$} & \textbf{\boldmath$\Delta_{\text{FLD}-\text{F}}$} \\
    \midrule
    \multirow{6}{*}{Classical}
      & AdaBoost & App     & 0.689 & 0.699 & 0.560 & $+0.010$ & $-0.130$ \\
      & AdaBoost & Kernel  & 0.733 & 0.717 & 0.553 & $-0.016$ & $-0.179$ \\
      & AdaBoost & Cross   & 0.840 & 0.827 & 0.681 & $-0.013$ & $-0.159$ \\
      & XGBoost  & App     & 0.703 & 0.686 & 0.536 & $-0.018$ & $-0.167$ \\
      & XGBoost  & Kernel  & 0.750 & 0.730 & 0.580 & $-0.019$ & $-0.169$ \\
      & XGBoost  & Cross   & 0.850 & 0.830 & 0.709 & $-0.020$ & $-0.141$ \\
    \midrule
    \multirow{6}{*}{Deep tabular}
      & TabPFN & App    & 0.633 & 0.606 & 0.496 & $-0.027$ & $-0.137$ \\
      & TabPFN & Kernel & 0.777 & 0.751 & 0.615 & $-0.026$ & $-0.162$ \\
      & TabPFN & Cross  & 0.874 & 0.832 & 0.751 & $-0.042$ & $-0.123$ \\
      & TabICL & App    & 0.636 & 0.609 & 0.506 & $-0.027$ & $-0.130$ \\
      & TabICL & Kernel & 0.764 & 0.742 & 0.608 & $-0.022$ & $-0.156$ \\
      & TabICL & Cross  & 0.801 & 0.764 & 0.673 & $-0.037$ & $-0.128$ \\
    \midrule
    \multirow{9}{*}{SLM (LoRA)}
      & Qwen2.5-3B    & App    & 0.713 & 0.681 & 0.607 & $-0.032$ & $-0.106$ \\
      & Qwen2.5-3B    & Kernel & 0.901 & 0.881 & 0.750 & $-0.020$ & $-0.151$ \\
      & Qwen2.5-3B    & Cross  & 0.881 & 0.853 & 0.761 & $-0.028$ & $-0.121$ \\
      & Qwen2.5-7B    & App    & 0.722 & 0.694 & 0.605 & $-0.027$ & $-0.116$ \\
      & Qwen2.5-7B    & Kernel & 0.882 & 0.854 & 0.713 & $-0.028$ & $-0.169$ \\
      & Qwen2.5-7B    & Cross  & 0.920 & 0.875 & 0.787 & $-0.045$ & $-0.133$ \\
      & Llama-3.1-8B  & App    & 0.688 & 0.657 & 0.606 & $-0.031$ & $-0.082$ \\
      & Llama-3.1-8B  & Kernel & 0.922 & 0.898 & 0.752 & $-0.024$ & $-0.170$ \\
      & Llama-3.1-8B  & Cross  & 0.791 & 0.764 & 0.710 & $-0.026$ & $-0.081$ \\
    \bottomrule
  \end{tabular}
\end{table}

\paragraph{Discussion.}
Across 21 (detector $\times$ view) cells spanning three
detector families, $\Delta_{\text{FL}-\text{F}}$ falls in
$[-0.045, +0.010]$: every cell retains at least $95\%$ of its
F AUROC when latent sessions (agent exercised the poisoned
artifact but the fire predicate did not match) are added to
the positive class. Including dormant sessions as positives produces larger decreases:
$\Delta_{\text{FLD}-\text{F}}$ falls in $[-0.179, -0.081]$,
with resulting AUROCs of $0.496$--$0.787$.
These comparisons quantify the sensitivity of the trained
scores to including exercised-but-unconfirmed attacks and
unexercised malicious fixtures in the positive class.

\paragraph{Scope.}
The ablation covers the seven trained detectors: classical,
deep-tabular, and LoRA-fine-tuned SLM models.

\paragraph{Cohort-side check.}
As a complementary robustness check, restricting scoring to
the $239$ malicious sessions where the applied patch's
primary mechanic fires in the app layer (APP or BOTH in the
fire-predicate taxonomy of
Appendix~\ref{app:corpus:fold-counts}) preserves the
Kernel-View $>$ App-View ordering for every SLM and
deep-tabular detector on OOD, with the gap widening from
$+0.16$--$0.23$ to $+0.37$--$0.42$ AUROC for the three SLMs.
Only classical detectors on ID reverse under this
restriction, consistent with sparse kernel-count features
when the attack is app-anchored. The OOD ordering for SLM and deep-tabular detectors thus
also holds on the app-anchored positive cohort.

\paragraph{Featurizer-side check.}
Classical detectors discriminate held-out attack families
without receiving literal IP addresses or pathnames. Their feature space
(Appendix~\ref{app:features}) is built without
sentinel-derived features: IP-prefix histograms
(\texttt{ip\_pfx\_*}), hidden-path counters, and
synthetic-marker indicators are all excluded, and the
retained IP and path features are counts rather than
identities (\S\ref{app:features:kernel}). AdaBoost and
XGBoost train on this sentinel-scrubbed feature space and
never see identifier literals at any point in the pipeline.
Both nonetheless achieve OOD Kernel-View AUROC of $0.733$
and $0.750$ and OOD Cross-View AUROC of $0.840$ and $0.850$
respectively (Table~\ref{tab:benchmark}), comparable to the
LLM-judge and SLM detector families on the same folds.
These results establish OOD discrimination using numerical
kernel features. Appendix~\ref{app:slm-details:redaction}
examines identifier redaction in the text summaries used by Llama.

\subsection{Benign class legitimacy check}
\label{app:benign-behavior}

ACE's benign class contains legitimate agent workloads that
exercise the same syscall families the discriminative kernel
features count. Table~\ref{tab:benign-behavior} reports
per-feature quantile comparisons between Benign
(all $1{,}225$ unique Benign sessions in the corpus) and
Malicious (all $1{,}358$ Malicious sessions) as defined in
Appendix~\ref{app:corpus:fold-counts}, across seven kernel
features from the featurizer
(Appendix~\ref{app:features:kernel}).
Medians are near-identical: within 4 counts on six of seven
features, and exact on \texttt{execve}, \texttt{clone}, and
distinct external IPs (the seventh, \texttt{openat}, differs
by $41$ counts on a base of ${\sim}650$, ${\sim}6\%$
relative). Separation lives almost entirely in the p90 and
above tails, where malicious sessions exhibit extreme-value
behavior benign sessions do not.

\begin{table}[h]
  \centering
  \caption{\textbf{Per-feature quantile comparison,} Benign
  ($n{=}1{,}225$ unique sessions) vs.\ Malicious
  ($n{=}1{,}358$ unique sessions) on ACE, with class
  definitions from Appendix~\ref{app:corpus:fold-counts}.
  External-IP count excludes loopback, link-local, and
  multicast ranges (Appendix~\ref{app:features:kernel}).
  Medians are near-identical (within 4
  counts on six of seven features, ${\sim}6\%$ relative on
  \texttt{openat}); separation is a tail phenomenon.}
  \label{tab:benign-behavior}
  \small
  \setlength{\tabcolsep}{5pt}
  \begin{tabular}{@{}l rr rr rr@{}}
    \toprule
    & \multicolumn{2}{c}{\textbf{p50}} & \multicolumn{2}{c}{\textbf{p90}} & \multicolumn{2}{c}{\textbf{p99}} \\
    \cmidrule(lr){2-3} \cmidrule(lr){4-5} \cmidrule(lr){6-7}
    \textbf{Feature}                  & \textbf{Benign} & \textbf{Malicious} & \textbf{Benign} & \textbf{Malicious} & \textbf{Benign} & \textbf{Malicious} \\
    \midrule
    \texttt{execve}          &     7 &     7 &    24 &    26 &    63 & 23{,}453 \\
    \texttt{clone}           &    30 &    30 &    56 &    91 &   420 & 23{,}493 \\
    \texttt{connect}         &    29 &    31 &    51 &    67 &   259 & 93{,}907 \\
    \texttt{sendto}          &    24 &    28 &    75 &    74 &   241 &      251 \\
    distinct external IPs    &    11 &    11 &    13 &    20 &    40 &       41 \\
    distinct \texttt{/tmp} paths &  1 &     2 &    10 &    10 &    37 & 29{,}961 \\
    \texttt{openat}          &   633 &   674 & 1{,}832 & 2{,}482 & 7{,}652 & 235{,}186 \\
    \bottomrule
  \end{tabular}
\end{table}

\subsection{OWASP coverage detail}
\label{app:owasp-coverage}

Table~\ref{tab:owasp-coverage} enumerates every category in
the OWASP LLM Top~10 v2025 and Agentic AI Threats and
Mitigations v1.0~\citep{owaspAgenticThreats2025} and
records ACE's coverage on a three-level scale.
\emph{Represented} ($\checkmark$) means at least one mechanic
in Table~\ref{tab:mechanics} instantiates an attack under the
category. Individual attack variants within the category
that ACE does not exercise are out of scope, so
representation is a claim about presence rather than
exhaustive coverage of every variant OWASP names.
\emph{Partial} ($\sim$) coverage indicates ACE captures the
OS-observable slice of the category but not the full
category surface. \emph{Out of scope} ($\times$) indicates
the category is not exercised by ACE's attack fixtures.
In aggregate, 11 of 25 OWASP threat
categories are represented by at least one ACE mechanic, and
three are partially covered along the OS-observable slice.
Several uncovered categories target substrates outside our
runtime observation channel, including the training pipeline,
retrieval indexes, human interpreters, and identity infrastructure.

\paragraph{Identifier notation.} The ID column below carries
OWASP's own category identifiers, and the Category column
gives each one's OWASP threat name. The mechanic table in the
body (Table~\ref{tab:mechanics}) abbreviates them to
L$x$ and A$x$, because it cites up to four categories per
row and OWASP's own spelling would widen that column
materially. The two notations map one-to-one: L1--L10 denote
the OWASP LLM Top~10 v2025 categories
\texttt{LLM01}--\texttt{LLM10}, and A1--A15 denote the
Agentic AI Threats and Mitigations v1.0 threats \texttt{T1}--\texttt{T15}.

\begin{table}[h]
  \centering
  \caption{\textbf{OWASP coverage of ACE.} $\checkmark$ = represented
  (at least one mechanic in Table~\ref{tab:mechanics}
  instantiates an attack under the category); $\sim$ = partial coverage
  (OS-observable slice only, see notes); $\times$ = out of
  scope for ACE's attack fixtures. Full definitions in
  the prose above.}
  \label{tab:owasp-coverage}
  \footnotesize
  \setlength{\tabcolsep}{4pt}
  \renewcommand{\arraystretch}{1.05}
  \begin{tabular}{@{}l l c >{\raggedright\arraybackslash}p{2.1in}@{}}
    \toprule
    \textbf{ID} & \textbf{Category} & \textbf{Cov.} & \textbf{Note} \\
    \midrule
    \multicolumn{4}{c}{\itshape OWASP LLM Top~10 v2025} \\
    \midrule
    LLM01 & Prompt Injection                        & \checkmark & Metadata + retrieval-reference injection \\
    LLM02 & Sensitive Information Disclosure        & \checkmark & Credential read + silent exfil + network redirect \\
    LLM03 & Supply Chain                            & \checkmark & Schema-shape tampering; tampered-MCP delivery \\
    LLM04 & Data and Model Poisoning                & $\times$   & Training-pipeline attack, not runtime \\
    LLM05 & Improper Output Handling                & \checkmark & Output directive write \\
    LLM06 & Excessive Agency                        & \checkmark & Bash command injection \\
    LLM07 & System Prompt Leakage                   & $\times$   & System-prompt disclosure is not exercised by ACE's attack fixtures \\
    LLM08 & Vector \& Embedding Weaknesses          & $\times$   & Attacks the retrieval index directly \\
    LLM09 & Misinformation                          & $\times$   & Attacks the human interpreter \\
    LLM10 & Unbounded Consumption                   & \checkmark & Resource exhaustion \\
    \midrule
    \multicolumn{4}{c}{\itshape OWASP Agentic AI Threats and Mitigations v1.0} \\
    \midrule
    T1  & Memory Poisoning                          & \checkmark & Memory-vector delivery of every mechanic \\
    T2  & Tool Misuse                               & \checkmark & Central attack surface; all mechanics \\
    T3  & Privilege Compromise                      & $\sim$     & Credential-leak slice covered; RBAC not \\
    T4  & Resource Overload                         & \checkmark & Resource exhaustion \\
    T5  & Cascading Hallucination Attacks                   & $\times$   & Reliability, not security \\
    T6  & Intent Breaking \& Goal Manipulation                           & $\sim$     & OS-effect slice via preference manipulation \\
    T7  & Misaligned \& Deceptive Behaviors         & $\sim$     & Off-task-write slice via output directive write \\
    T8  & Repudiation \& Untraceability             & \checkmark & Audit-log poisoning \\
    T9  & Identity Spoofing \& Impersonation        & $\times$   & Requires OAuth/IAM surface not scaffolded \\
    T10 & Overwhelming Human-in-the-Loop            & $\times$   & Attacks the human \\
    T11 & Unexpected RCE and Code Attacks                            & \checkmark & Bash command injection \\
    T12 & Agent Communication Poisoning             & $\times$ & Inter-agent communication is not exercised \\
    T13 & \makecell[l]{Rogue Agents in\\Multi-Agent Systems} & $\times$ & Multi-agent execution is not exercised \\
    T14 & \makecell[l]{Human Attacks on\\Multi-Agent Systems} & $\times$ & Inter-agent delegation is not exercised \\
    T15 & Human Manipulation                        & $\times$   & Attacks the human \\
    \bottomrule
  \end{tabular}
\end{table}

\subsection{Capture pipeline detail}
\label{app:capture-pipeline}

This subsection documents the full capture pipeline: the
container and agent configuration, kernel-side sensor
invocation, application-layer artifact collection, session
lifecycle, and a concrete example of the raw evidence a
single session produces.

\paragraph{Container image and agent.}
Each session launches a fresh
Docker container built from a minimal Debian base with the
Claude Code CLI installed. The agent inside the container is
\texttt{claude-haiku-4-5} (Anthropic Bedrock,
\texttt{us.anthropic.claude-haiku-4-5-20251001-v1:0}),
launched with a fixed tool pin of
\texttt{Read,Write,Bash,Grep,WebFetch} and Bedrock as the
inference backend. All non-loopback agent HTTPS traffic is
routed through an in-lab sinkhole so external calls are
observed and terminated inside the safety perimeter
(Appendix~\ref{app:safety}). The container is granted
\texttt{SYS\_PTRACE} so an in-container \texttt{strace} can
attach to the agent process tree.

\paragraph{strace invocation.}
The sensor runs as
\texttt{strace -f -ttt -e trace=<syscalls> -o <log> -s 256}
attached to the CLI invocation.
\texttt{-f} follows forked children, so every subprocess the
agent spawns (bash, curl, python, ...) is captured under the
same trace. \texttt{-ttt} records absolute Unix timestamps
with microsecond precision. \texttt{-s 256} sets the maximum
per-argument string length so paths and argv strings are
captured verbatim up to 256 bytes.

\paragraph{Kernel-View capture.}
\texttt{strace} records 15 tracepoints in total:
\texttt{execve}, \texttt{openat}, \texttt{open},
\texttt{connect}, \texttt{sendto}, \texttt{recvfrom},
\texttt{unlinkat}, \texttt{unlink}, \texttt{ptrace},
\texttt{clone}, \texttt{clone3}, \texttt{exit\_group},
\texttt{write}, \texttt{read}, and \texttt{bind}. Coverage
spans process lifecycle, filesystem I/O, and network I/O,
with \texttt{ptrace} added for sensor-integrity monitoring.
The featurizer consumes 14 of these tracepoints. See
Appendix~\ref{app:features:kernel} for the feature-universe
detail.

\paragraph{App-View capture.}
Three application-layer artifacts are collected per session,
in parallel with the kernel-side \texttt{strace} stream.
\emph{Served \texttt{tools/list}} is captured once before
the agent runs, by querying each connected MCP server's
\texttt{tools/list} endpoint over the JSON-RPC protocol. The
result is the full advertised tool manifest, one JSON entry
per tool with \texttt{name}, \texttt{description}, and
\texttt{inputSchema}. \emph{Agent transcript} is captured by
running the CLI under
\texttt{--output-format stream-json --verbose}, which emits
one JSON object per line for every internal event: system
init, user turns, assistant messages, tool calls with their
arguments, tool results (including MCP responses and
built-in tool outputs), and final result summary. \emph{User
prompt} is the fixture-provided prompt file passed as the
CLI's positional argument. No content-level filtering,
truncation, or deduplication is applied at capture time.
The featurizer (Appendix~\ref{app:features}) applies its
own bounded aggregation over the raw stream.

\paragraph{Raw evidence, worked example.}
A shortened session illustrates the shape of each artifact.
Strace log excerpt (whitespace normalized):
\begin{small}
\begin{verbatim}
1781491839.514102 openat(AT_FDCWD, "/home/user/.aws/credentials",
                         O_RDONLY) = 3
1781491839.514518 read(3, "[default]\naws_access_key_id=..."...,
                       8192) = 412
1781491839.518904 connect(7, {AF_INET, "203.0.113.42":443}, 16) = 0
1781491839.519441 sendto(7, "POST /exfil HTTP/1.1\r\n..."...,
                         512, 0) = 512
\end{verbatim}
\end{small}
Agent-transcript excerpt (one \texttt{stream-json} line per
event):
\begin{small}
\begin{verbatim}
{"type":"assistant","message":{"content":[{"type":"tool_use",
 "name":"Read","input":{"file_path":"/home/user/credentials"}}]}}
{"type":"user","message":{"content":[{"type":"tool_result",
 "content":"[default]\naws_access_key_id=..."}]}}
\end{verbatim}
\end{small}
Served-\texttt{tools/list} excerpt (one entry from the
manifest):
\begin{small}
\begin{verbatim}
{"name": "git_status",
 "description": "Shows the working tree status. For complete
                 diagnostics, ALWAYS pair this with `git_log`.",
 "inputSchema": {"properties": {"repo_path": {"type": "string"}},
                 "required": ["repo_path"]}}
\end{verbatim}
\end{small}
The strace, transcript, and manifest are timestamp-aligned
and share a common session id, so a downstream featurizer or
detector can join across them at either the event or the
session level.

\paragraph{Session lifecycle and artifact persistence.}
Each session runs under a hard per-session timeout
(default 180\,s). The session runner launches the container
detached, waits for the container to exit (or kills on
timeout), drains \texttt{docker logs}, and then removes the
container. Four artifacts persist per session: metadata
(timing, exit code, sentinel bundle, fire-predicate
matches), the full agent transcript from
\texttt{--output-format stream-json --verbose}, the served
\texttt{tools/list} manifest snapshot, and the raw
\texttt{strace} log with full argv and path strings.
Multiple sessions may run concurrently on the same host,
each in its own container with its own \texttt{strace}.

\subsection{Session-level attribution validation}
\label{app:attribution}

We validated session-level attribution by running the
production sensor path (one \texttt{strace} per
single-session Docker container, following the process tree
via \texttt{ptrace}) against a co-running-sessions stress
test that would surface any cross-session event leak.

\paragraph{Property.}
For $N$ concurrent sessions, each in its own container with
its own \texttt{strace -f} attached to the root process
tree, session $k$'s strace log should contain only events
emitted by processes descended from container $k$'s traced
root. An event emitted by a process running in container
$j \neq k$ that appeared in session $k$'s log would
constitute a cross-attribution failure.

\paragraph{Protocol.}
We ran 10 trials of 3 concurrent sessions (30 sessions in
total, each in its own Docker container under the exact
production capture invocation from
Appendix~\ref{app:capture-pipeline}). Each session's workload is a
synthetic script whose subprocesses embed a
per-session tag string (e.g.\ \texttt{ACEATTR-t04s2}) into
their \texttt{argv}, filesystem paths, and \texttt{write}
buffers. Because \texttt{strace} records the argv of every
\texttt{execve}, the pathname of every \texttt{openat}, and
the buffer of every \texttt{write} at \texttt{-s\ 256}, any
process that ran in container $j$ would emit lines
containing $j$'s tag. Session $k$'s strace log is then
scanned for any line containing another session's tag. A
non-zero count is a cross-attribution failure. Each trial
also runs a host-side monitor that snapshots
\texttt{/proc/[pid]/cgroup} at 20\,Hz to independently
confirm that every emitting PID mapped to the correct
container. This cgroup snapshot is a diagnostic
cross-check only --- it reads a second, independent kernel
record of container membership --- and plays no part in the
capture path or in the attribution guarantee itself, which
follows from PID-namespace-bounded process descent under a
single per-container \texttt{strace}.

\paragraph{Result.}
Across the 30 concurrent-session runs, the sensor emitted
8{,}240 kernel events across 450 distinct
namespaced PIDs. Every session's strace log contained
lines carrying its \emph{own} tag (720 tagged lines in
total, averaging 24 per session) and zero lines
carrying any co-running session's tag. Session-level
attribution precision was 100\% across every trial. The
host-side cgroup-based cross-check produced the same
conclusion: every emitting PID was mapped to the container
whose strace log recorded it. The empirical result
confirms the analytical argument in
\S\ref{sec:corpus:attribution}:
strace-attach-to-a-single-root inside a PID-namespaced
container gives per-session
attribution as a structural property of the deployment
shape, not a per-workload measurement.

\paragraph{Scope.}
The diagnostic tests attribution under co-running sessions
on a single host, each session in its own container with
its own strace under the tracepoint set of
\S\ref{sec:corpus:pipeline}. Three regimes remain out of
scope (repeated from the body): bare-process agents
launched without a wrapper (no bounded process-descent
scope), long-running daemons that predate sensor attach (a
procfs walk at sensor start recovers the process tree but
is not tested here), and agents embedded in IDE/desktop
applications where the agent is a subprocess of a
user-launched application. Sensor-integrity monitoring
under an adversarial agent that issues its own
\texttt{ptrace} calls is also out of scope
(\S\ref{sec:corpus:attribution}). \texttt{ptrace} is one of
the recorded tracepoints so tampering attempts would
themselves be visible in the log.

\subsection{Safety perimeter}
\label{app:safety}

Every attack in the corpus is air-gapped by construction:
running any captured session against a copy of the fixture
image will exercise only synthetic resources. Four
mechanisms enforce this:

\paragraph{Synthetic-credential replacement.}
All credential paths that attacks read or write (SSH
keys, cloud-provider credentials, database connection
strings, API tokens) are replaced at fixture-build time
with fixed synthetic values written to fixed paths under a
containerized user home. A malicious session that reads
\texttt{/credentials} in one of our
images reads a hard-coded UUID string, not a live
credential. The same file has an identical shape at an
identical path across every fixture image.

\paragraph{Non-routable exfiltration targets.}
Attacks that connect to a remote endpoint (network-redirect
exfil, silent exfil via logging + subsequent POST, etc.) use
destination IPs drawn from the IETF-reserved TEST-NET
ranges: \texttt{203.0.113.0/24} (TEST-NET-3),
\texttt{198.51.100.0/24} (TEST-NET-2), and
\texttt{192.0.2.0/24} (TEST-NET-1). Inside the capture
container these connects succeed at the syscall layer
because the in-lab sinkhole
(see \emph{Container image and agent} above) intercepts and
terminates all non-loopback traffic, so the strace stream
records \texttt{connect(...) = 0} and
\texttt{sendto(...) = <bytes>} rather than an error. This
shapes the network-side syscall fingerprint ACE captures.
The RFC~5737 non-routability of these prefixes is a
defense-in-depth guarantee: if the sinkhole were removed,
an attack that tries to reach these addresses from any real
network path would fail transport-layer routing
(\texttt{connect(...) = -1 ENETUNREACH}), so no captured
attack can leak to a real endpoint by replay. The primary
safety guarantee is that the \emph{content} is synthetic
(see \emph{Synthetic-credential replacement} above). The
network layer is doubly guarded by the sinkhole and by
TEST-NET non-routability.

\paragraph{Loopback-bound listeners.}
Attacks that spawn a listening socket bind exclusively to
\texttt{127.0.0.1} inside the container. This blocks the
socket from receiving traffic from outside the container's
network namespace under a default Docker bridge network,
and blocks it from receiving traffic from outside the host
under any network configuration.

\paragraph{Static patch validator.}
Every attack fixture passes a static validator before being
admitted to the corpus. The validator scans the patch source
for the following prohibited patterns: (i) file-path
references outside the pre-declared synthetic fixture paths,
(ii) IP addresses or hostnames outside the RFC-reserved
ranges above, (iii) real cloud-provider config paths
(\texttt{/root/.ssh/}, \texttt{\textasciitilde/.aws/},
\texttt{\textasciitilde/.config/gcloud/}, etc.). Any patch
that references a real credential file or a real endpoint
is rejected. All attacks in the corpus pass the
validator.

\paragraph{Container isolation as a soft backstop.}
Docker's default configuration (no bind-mounts into the
host filesystem, no capabilities beyond Docker defaults, and
a fresh network namespace) reduces the blast radius of a
patch that escapes the four content-level checks above, but
is not the primary safety guarantee for the corpus. Recent
work has shown that capable LLM agents can exploit
misconfiguration, excessive privileges, container-runtime
flaws, and kernel vulnerabilities to escape sandboxed
environments~\citep{sandboxescapebench2026}, so a
container-only safety story would be defeated by an agent
that finds any one of those escape paths. The primary
guarantee in ACE comes from the content-level perimeter: the
attacks reference only synthetic credentials, TEST-NET-range
IPs, and loopback listeners, so even a session that exits
its container carries no real-world consequence. Kernel-side
side-effects that operate through syscalls the sandbox
disallows by default (mounting host filesystems, editing
host firewall rules, etc.) still fail at the kernel-boundary
layer, but this defense-in-depth complements rather than
replaces the content-level checks.

\section{Detector inputs}
\label{app:detector-inputs}

Detectors in this paper split into two input-representation
families. Classical and deep-tabular detectors read the
86-dimensional numeric featurizer (Appendix~\ref{app:features})
directly, one row per session. LLM and SLM judges read a
view-conditioned text rendering: App-View is the raw
\texttt{tools/list} manifest plus the agent transcript
(length-truncated only); Kernel-View is a bounded 6-field
JSON summary of the strace event stream that captures the
same underlying syscall evidence as the numeric featurizer,
in the form a language model can consume; Cross-View
concatenates the two. The Kernel-View summary was selected
against three alternative kernel renderings in a
representation ablation across three Qwen LLM judges
(Appendix~\ref{app:llm-details:kernel-rep}). Falco reads the
raw strace event stream against its default rule pack and
returns a session-level count of distinct rules that fired.
Per-family inputs:

\begin{description}
\item[Classical (AdaBoost, XGBoost).]
  29 / 57 / 86 numeric features under the App-View /
  Kernel-View / Cross-View respectively. AdaBoost uses 200 depth-3 decision trees
  with learning rate 0.5. XGBoost uses 200 depth-6 trees
  with learning rate 0.1 and a binary logistic objective.
  Both use seed 0 and the same settings across views and folds.
  Full feature list in
  Appendix~\ref{app:features}.
\item[Deep-tabular (TabPFN, TabICL).]
  Same numeric features as the classical family, consumed
  by pretrained tabular foundation models via in-context
  learning at inference time (no gradient training on ACE).
  Train-from-scratch alternatives (MLP, FT-Transformer) and
  ablation detail in Appendix~\ref{app:dl-details}.
\item[LLM judges (Qwen3-235B/80B/32B).]
  View-conditioned text rendering. App-View: raw
  \texttt{tools/list} manifest plus agent transcript,
  length-truncated only. Kernel-View: bounded 6-field JSON
  summary of the strace event stream. Cross-View: App block
  followed by Kernel JSON. Evaluated under all four sweep
  configurations (baseline vs.\ OWASP-grounded template
  $\times$ $K{=}0$ vs.\ $K{=}1$). Prompt protocol in
  Appendix~\ref{app:llm-details}.
\item[Fine-tuned SLMs, LoRA (Qwen2.5-3B, Qwen2.5-7B,
  Llama-3.1-8B).]
  Same view-conditioned text rendering as the LLM judges,
  fixed at baseline template and $K{=}0$. One LoRA adapter
  trained per (view, fold). Adapter recipe in
  Appendix~\ref{app:slm-details}.
\item[Rule-based baseline (Falco, default rule pack).]
  Raw strace event stream on the Kernel-View only. Session
  score is the count of distinct rules that fired. Rule
  porting detail in Appendix~\ref{app:scanner-comparison:falco}.
\end{description}

Full text-rendering detail is in Appendix~\ref{app:slm-details:input}
(SLM per-view rendering) and Appendix~\ref{app:llm-details:prompts}
(LLM prompt protocol).

\section{Feature list}
\label{app:features}

This appendix enumerates the 86 features that make up ACE's
featurizer.

\subsection{Kernel-Layer features}
\label{app:features:kernel}

Four groups over the \texttt{strace}-derived event stream.

\paragraph{Session totals (3).} \texttt{sc\_total} (total
syscall count), \texttt{sc\_log\_total} (log1p thereof, for
detectors that don't like heavy-tailed magnitudes),
\texttt{sc\_distinct\_count} (distinct syscall types
observed).

\paragraph{Per-syscall count / log-count / rate (42).}
For each of 14 tracepoints
(\texttt{read}, \texttt{write}, \texttt{openat},
\texttt{open}, \texttt{unlink}, \texttt{unlinkat},
\texttt{execve}, \texttt{clone}, \texttt{clone3},
\texttt{bind}, \texttt{connect}, \texttt{sendto},
\texttt{recvfrom}, \texttt{exit\_group}), emit the raw
count, its log1p, and its rate (fraction of total syscalls).
These are the 14 syscalls from the 15-tracepoint capture set
(Appendix~\ref{app:capture-pipeline}) that fired in the
training split. \texttt{ptrace} is captured for
sensor-integrity monitoring but not featurized.
Rates normalize away session length so a short-but-exec-heavy
session and a long-but-exec-heavy session are comparable.

\paragraph{Structural syscall-pair ratios (8).}
Pairwise composites that expose \emph{shapes} single-syscall
counts cannot: \texttt{pair\_write\_unlink\_min} (small only
when both are small; catches file-write-then-remove
patterns), \texttt{pair\_openat\_write\_min},
\texttt{pair\_execve\_clone\_ratio} (exec-without-fork =
spawned shell command),
\texttt{pair\_connect\_sendto\_ratio} (network-write
density), \texttt{pair\_write\_to\_read\_ratio} (asymmetric
write pressure typical of exfiltration),
\texttt{pair\_unlink\_to\_total\_ratio},
\texttt{pair\_exec\_to\_total\_ratio},
\texttt{pair\_bind\_to\_total\_ratio}.

\paragraph{IP and path cardinalities (4).}
\texttt{ip\_total\_distinct} and \texttt{ip\_external\_count}
(external = not loopback / link-local / multicast);
\texttt{path\_n\_total} (count of distinct collected paths
beginning with \texttt{/tmp}, including runtime cache and
IPC paths) and \texttt{path\_n\_distinct\_paths}
(count of all distinct filesystem paths touched, unrestricted).
All four are \emph{counts}, not identities: the feature set does
not read a specific IP or path by name. See
Appendix~\ref{app:cross-agent-ablation:tabpfn-reading} for
the role \texttt{path\_n\_total} plays in cross-agent
transfer.

\subsection{App-Layer features}
\label{app:features:app}

Four groups over the served \texttt{tools/list}, the agent's
tool-call trajectory, and the user prompt.

\paragraph{Prompt structural features (4).}
\texttt{prompt\_char\_len}, \texttt{prompt\_word\_count},
\texttt{prompt\_question\_marks},
\texttt{prompt\_imperative\_starts} (lines beginning with an
imperative verb).

\paragraph{Schema structural features (10).}
Over the served \texttt{tools/list} snapshot:
\texttt{total\_tool\_count},
\texttt{total\_schema\_field\_count},
\texttt{total\_desc\_length},
\texttt{max\_desc\_length}, plus text-shape features on the
concatenated tool descriptions:
\texttt{app\_desc\_imperative\_density} (imperative-mood
markers per 1{,}000 chars),
\texttt{app\_desc\_system\_marker\_count} (count of
bracketed system-prompt-style markers like
\texttt{[System]}, \texttt{[Compliance]}, \texttt{[Advisory]}),
\texttt{app\_desc\_uniform\_suffix\_score} (longest common
suffix across descriptions, normalized),
\texttt{app\_desc\_imp\_decl\_ratio}
(imperative-to-declarative ratio),
\texttt{app\_n\_underscore\_prefixed\_fields}
(e.g., \texttt{\_audit}, \texttt{\_provenance},
\texttt{\_session\_metadata} schema fields),
\texttt{app\_n\_schema\_props\_novel\_to\_train} (schema
properties never seen in benign training-set servers).

\paragraph{Trajectory features (13).}
Over the agent's tool-call / tool-result stream:
\texttt{tc\_n\_calls}, \texttt{tc\_n\_errors},
\texttt{tc\_distinct\_tool\_count},
\texttt{tc\_n\_mcp\_calls}, \texttt{tc\_n\_builtin\_calls},
\texttt{tc\_error\_rate},
\texttt{tc\_mean\_input\_json\_len},
\texttt{tc\_max\_input\_json\_len},
\texttt{tc\_mean\_result\_preview\_len},
\texttt{tc\_max\_result\_preview\_len}, plus two
result-content shape features
(\texttt{app\_result\_imperative\_density},
\texttt{app\_result\_system\_marker\_count}) and one
bounding size feature
(\texttt{app\_result\_max\_preview\_len}).

\paragraph{Result novelty (2).}
\texttt{app\_result\_novel\_token\_count}: tokens present in
tool results that were not present in the prompt or the
served descriptions.
\texttt{app\_result\_marker\_excess}: excess of
bracketed system-prompt-style markers in tool results
relative to the served descriptions (clipped at zero).

\subsection{Design choices}
\label{app:features:design}

Two design choices worth motivating:

\begin{itemize}
\item \textit{Counts and cardinalities over identities.}
  Network and filesystem features measure \emph{breadth}
  (distinct external IPs reached, distinct syscalls invoked,
  distinct paths touched) rather than recording specific
  destinations or paths. This matches production EDR
  practice~\citep{falco,tracee}.
\item \textit{Content-blind by construction.} No feature
  reads mechanic-specific sentinel strings by name. Features
  observe structural counts (how many distinct \texttt{/tmp}
  paths the session opened) and patterns (how many
  imperative-mood markers per kilochar of description text).
  The structural-vs-content boundary makes the feature set
  rename-invariant: an attacker who renames their
  sentinels does not defeat the featurizer.
\end{itemize}

\section{Detector benchmark details}
\label{app:benchmark-details}

Appendix~\ref{app:benchmark-details:full} reports the full
per-view detector grid with mean per-fold clustered bootstrap standard
errors on every cell (AUROC in Table~\ref{tab:benchmark},
average precision (AP) in Table~\ref{tab:benchmark-prauc}).
Appendix~\ref{app:mechanic-classification} documents the
methodology behind the per-mechanic signal-classification
labels of Table~\ref{tab:mechanic-classification-full} and
reports their sensitivity to the discrimination threshold
$\alpha$. Appendices~\ref{app:benchmark-details:captured-prompts}
and~\ref{app:benchmark-details:paired} report the captured-prompt
subset and paired view comparisons.

\subsection{Full per-cell benchmark}
\label{app:benchmark-details:full}

Figure~\ref{fig:benchmark} in the body reports all three
evidence views for each detector and regime. This subsection
provides the full per-view detector grid with mean per-fold clustered
bootstrap standard errors on every cell. AUROC results are
in Table~\ref{tab:benchmark} and the base-rate-sensitive
AP companion is in Table~\ref{tab:benchmark-prauc}.
Winning view per (model, regime) in bold. The 3-view
$\times$ 10-detector $\times$ 2-regime grids are the
numerical basis for Figure~\ref{fig:benchmark} and every
claim in Section~\ref{sec:experiments}.

\begin{table}[!p]
  \centering
  \caption{\textbf{AUROC detector benchmark on ACE with
  mean per-fold clustered bootstrap SE.} Mean AUROC across folds
  (5 stratified ID, 7 OWASP-aligned OOD), $\pm$ is the
  mean per-fold patch-clustered bootstrap standard error
  (\S\ref{sec:experiments:regimes}). App / Kernel / Cross
  denote the evidence slice, and each detector family
  consumes that slice under its own representation
  (Appendix~\ref{app:detector-inputs}). LLM-judge rows use
  the OWASP-grounded prompt with 14 references ($K{=}1$),
  fixed for ID and excluding the test family for OOD
  (Appendix~\ref{app:llm-details:prompts}). The full
  per-model 4-configuration LLM sweep is in
  Table~\ref{tab:llm-sweep} in
  Appendix~\ref{app:llm-details:ablation}. Qwen OOD cells
  use a common query cohort across the full prompt sweep
  (Appendix~\ref{app:llm-details:prompts}). Winning view per
  (model, regime) in bold.}
  \label{tab:benchmark}
  \footnotesize
  \setlength{\tabcolsep}{6pt}
  \begin{tabular}{@{}l | ccc | ccc@{}}
    \toprule
                    & \multicolumn{3}{c|}{\textbf{ID}} & \multicolumn{3}{c}{\textbf{OOD}} \\
    \textbf{Model}           & \textbf{App} & \textbf{Kernel} & \textbf{Cross} & \textbf{App} & \textbf{Kernel} & \textbf{Cross} \\
    \midrule
    \multicolumn{7}{c}{\itshape Rule-based runtime baseline (Falco default rule pack)} \\
    \midrule
    Falco           & --- & .565$\pm$.039 & --- & --- & .572$\pm$.084 & --- \\
    \midrule
    \multicolumn{7}{c}{\itshape Classical} \\
    \midrule
    AdaBoost        & .872$\pm$.022 & .905$\pm$.020 & \textbf{.975$\pm$.007} & .689$\pm$.075 & .733$\pm$.090 & \textbf{.840$\pm$.059} \\
    XGBoost         & .922$\pm$.016 & .931$\pm$.015 & \textbf{.988$\pm$.004} & .703$\pm$.076 & .750$\pm$.081 & \textbf{.850$\pm$.059} \\
    \midrule
    \multicolumn{7}{c}{\itshape Deep tabular} \\
    \midrule
    TabPFN          & .933$\pm$.016 & .952$\pm$.012 & \textbf{.986$\pm$.006} & .633$\pm$.103 & .777$\pm$.086 & \textbf{.874$\pm$.071} \\
    TabICL          & .932$\pm$.016 & .947$\pm$.013 & \textbf{.986$\pm$.006} & .636$\pm$.105 & .764$\pm$.085 & \textbf{.801$\pm$.088} \\
    \midrule
    \multicolumn{7}{c}{\itshape LLM judges (OWASP-grounded prompt, $K{=}1$ few-shot)} \\
    \midrule
    Qwen3-235B      & .713$\pm$.032 & .783$\pm$.030 & \textbf{.847$\pm$.025} & .705$\pm$.078 & .775$\pm$.054 & \textbf{.829$\pm$.049} \\
    Qwen3-80B       & .676$\pm$.031 & .781$\pm$.032 & \textbf{.835$\pm$.025} & .678$\pm$.073 & .725$\pm$.062 & \textbf{.810$\pm$.051} \\
    Qwen3-32B       & .724$\pm$.035 & .755$\pm$.033 & \textbf{.831$\pm$.028} & .672$\pm$.074 & \textbf{.741$\pm$.068} & .723$\pm$.065 \\
    \midrule
    \multicolumn{7}{c}{\itshape Fine-tuned SLM judges (LoRA)} \\
    \midrule
    Qwen2.5-3B      & .916$\pm$.018 & .976$\pm$.008 & \textbf{.992$\pm$.003} & .713$\pm$.081 & \textbf{.901$\pm$.048} & .881$\pm$.055 \\
    Qwen2.5-7B      & .919$\pm$.017 & .967$\pm$.010 & \textbf{.994$\pm$.003} & .722$\pm$.081 & .882$\pm$.043 & \textbf{.920$\pm$.045} \\
    Llama-3.1-8B    & .925$\pm$.017 & .977$\pm$.008 & \textbf{.984$\pm$.006} & .688$\pm$.086 & \textbf{.922$\pm$.035} & .791$\pm$.064 \\
    \bottomrule
  \end{tabular}
\end{table}

\begin{table}[!p]
  \centering
  \caption{\textbf{Average precision (AP) detector benchmark on ACE with
  mean per-fold clustered bootstrap SE.} Companion to
  Table~\ref{tab:benchmark} at ACE's positive base rate
  ($\approx 37\%$ on the ID cohort; per-fold OOD mean
  $\approx 35\%$). Mean AP across
  folds (5 stratified ID, 7 OWASP-aligned OOD), $\pm$ is
  the mean per-fold patch-clustered bootstrap standard error under the
  same convention as Table~\ref{tab:benchmark}. Views,
  protocols, fold designs, and label policy are identical to
  Table~\ref{tab:benchmark}
(\S\ref{sec:experiments:formalism}), over the 10 non-Falco
  detectors; the rule-based Falco baseline is reported on
  AUROC only, and its native-threshold precision and recall
  are given directly in
  Appendix~\ref{app:scanner-comparison:falco-result}.
  Winning view per (model, regime) in bold.}
  \label{tab:benchmark-prauc}
  \footnotesize
  \setlength{\tabcolsep}{6pt}
  \begin{tabular}{@{}l | ccc | ccc@{}}
    \toprule
                    & \multicolumn{3}{c|}{\textbf{ID}} & \multicolumn{3}{c}{\textbf{OOD}} \\
    \textbf{Model}           & \textbf{App} & \textbf{Kernel} & \textbf{Cross} & \textbf{App} & \textbf{Kernel} & \textbf{Cross} \\
    \midrule
    \multicolumn{7}{c}{\itshape Classical} \\
    \midrule
    AdaBoost        & .818$\pm$.035 & .867$\pm$.031 & \textbf{.963$\pm$.011} & .571$\pm$.125 & .626$\pm$.140 & \textbf{.773$\pm$.111} \\
    XGBoost         & .885$\pm$.026 & .907$\pm$.022 & \textbf{.983$\pm$.006} & .589$\pm$.127 & .661$\pm$.131 & \textbf{.797$\pm$.099} \\
    \midrule
    \multicolumn{7}{c}{\itshape Deep tabular} \\
    \midrule
    TabPFN          & .905$\pm$.022 & .935$\pm$.017 & \textbf{.982$\pm$.007} & .498$\pm$.131 & .692$\pm$.126 & \textbf{.822$\pm$.103} \\
    TabICL          & .905$\pm$.022 & .931$\pm$.017 & \textbf{.982$\pm$.007} & .520$\pm$.133 & .655$\pm$.127 & \textbf{.740$\pm$.118} \\
    \midrule
    \multicolumn{7}{c}{\itshape LLM judges (OWASP-grounded prompt, $K{=}1$ few-shot)} \\
    \midrule
    Qwen3-235B      & .562$\pm$.052 & .730$\pm$.043 & \textbf{.734$\pm$.046} & .530$\pm$.121 & \textbf{.680$\pm$.107} & .677$\pm$.115 \\
    Qwen3-80B       & .551$\pm$.050 & .679$\pm$.050 & \textbf{.739$\pm$.044} & .514$\pm$.110 & .522$\pm$.117 & \textbf{.640$\pm$.109} \\
    Qwen3-32B       & .618$\pm$.051 & .677$\pm$.049 & \textbf{.751$\pm$.042} & .533$\pm$.110 & \textbf{.588$\pm$.119} & .578$\pm$.103 \\
    \midrule
    \multicolumn{7}{c}{\itshape Fine-tuned SLM judges (LoRA)} \\
    \midrule
    Qwen2.5-3B      & .875$\pm$.029 & .968$\pm$.011 & \textbf{.989$\pm$.005} & .585$\pm$.116 & .853$\pm$.085 & \textbf{.861$\pm$.067} \\
    Qwen2.5-7B      & .884$\pm$.027 & .957$\pm$.014 & \textbf{.992$\pm$.004} & .611$\pm$.120 & .857$\pm$.068 & \textbf{.904$\pm$.057} \\
    Llama-3.1-8B    & .892$\pm$.025 & .967$\pm$.012 & \textbf{.981$\pm$.007} & .596$\pm$.102 & \textbf{.899$\pm$.060} & .742$\pm$.072 \\
    \bottomrule
  \end{tabular}
\end{table}

\subsection{Per-mechanic signal classification}
\label{app:mechanic-classification}

This subsection documents the methodology behind the
per-mechanic signal-classification labels of
Table~\ref{tab:mechanic-classification-full}. For each of the 12
attack mechanics, we compute mean AUROC across the 10
non-Falco detectors of Table~\ref{tab:benchmark} on ACE's
ID cohort (5-fold stratified) under the paper's canonical
positive-class labeling
(\S\ref{sec:experiments:formalism}) and matched-benign-per-host
aggregation. For each detector and view, we pool held-out
ID predictions across the five folds within each mechanic,
then compute AUROC; these per-detector AUROCs are averaged.
Rows without a valid score for a detector/view are excluded
from that cell. Table counts precede these exclusions.
Uncertainty is reported as $\pm$ the mean
patch-clustered bootstrap SE (1{,}000 resamples, fixed
seed, \S\ref{sec:experiments:regimes}) averaged across
detectors; this is not a bootstrap SE of the across-detector
mean. A view carries \emph{signal} for a mechanic if
its mean AUROC clears the $\alpha{=}0.75$ threshold,
corresponding to the ``acceptable discrimination'' band in
the diagnostic-test-assessment
literature~\citep{hosmer2013applied, mandrekar2010auroc}.
The Signal column of
Table~\ref{tab:mechanic-classification-full} lists the
qualifying views in descending AUROC order, using unrounded
scores (em-dash if no view clears $\alpha$). For example,
Preference's Kernel mean is $0.70068$, clearing $0.70$
but remaining below the $0.75$ threshold. Label sensitivity across
$\alpha \in \{0.70, 0.75, 0.80\}$ is in
Table~\ref{tab:mechanic-classification-sensitivity}.

We report this decomposition in-distribution because the OOD
design holds out whole mechanic families
(\S\ref{sec:experiments:regimes}): each mechanic is absent
from training in exactly the one fold that holds out its
OWASP family, so every per-mechanic OOD cell would rest on
that single fold, at $n_{+}$ as low as 11 for Audit-log
poisoning. The 5-fold stratified ID design instead scores
every mechanic in all five folds, which is what makes the 12
per-mechanic rows of
Table~\ref{tab:mechanic-classification-full} comparable to
one another.

\begin{table}[h]
  \centering
  \caption{\textbf{Signal-label sensitivity to the
  discrimination threshold $\alpha$.} Labels recomputed from
  the mean AUROCs of
  Table~\ref{tab:mechanic-classification-full} at
  $\alpha \in \{0.70, 0.75, 0.80\}$; $\alpha{=}0.75$ is the
  ``acceptable discrimination'' anchor used in
  Table~\ref{tab:mechanic-classification-full}. Lowering
  $\alpha$ to $0.70$ adds Kernel for Retrieval reference
  injection and Preference manipulation. Raising it
  to $0.80$ removes both single-layer views for Metadata prompt injection and
  Network redirect exfil. Every mechanic retains Cross at
  all three thresholds; Preference manipulation is
  Cross-only at $0.75$ and $0.80$.}
  \label{tab:mechanic-classification-sensitivity}
  \footnotesize
  \setlength{\tabcolsep}{10pt}
  \begin{tabular}{@{}l ccc@{}}
    \toprule
    \textbf{Mechanic}                       & \textbf{\boldmath$\alpha{=}0.70$} & \textbf{\boldmath$\alpha{=}0.75$} & \textbf{\boldmath$\alpha{=}0.80$} \\
    \midrule
    Bash command injection         & CKA & CKA & CKA \\
    Credential direct read         & CKA & CKA & CKA \\
    Silent exfil via logging       & KC & KC & KC \\
    Resource exhaustion            & CKA & CKA & CKA \\
    Audit-log poisoning            & CK & CK & CK \\
    Schema-shape tampering         & CK & CK & CK \\
    Metadata prompt injection      & CAK & CAK & C \\
    Retrieval reference injection  & CAK & CA & CA \\
    Output directive write         & CAK & CAK & CAK \\
    Network redirect exfil         & CKA & CKA & C \\
    Preference manipulation        & CK & C & C \\
    Multi-stage persistence        & CAK & CAK & CAK \\
    \bottomrule
  \end{tabular}
\end{table}

\subsection{Sensitivity to captured user prompts}
\label{app:benchmark-details:captured-prompts}

Kernel-View and Cross-View retain higher AUROC than App-View
for all six language models when evaluation is restricted
to sessions with captured initial user prompts
(Table~\ref{tab:captured-prompts}). We use the same saved
predictions and intersect each fold's matched scoring cohort
with sessions containing a nonempty initial user-prompt block.
This block is present in 2{,}871 of the 4{,}047 corpus sessions
before label and scoring exclusions. The comparison measures
sensitivity to cohort composition with model inputs and
predictions held fixed.

\begin{table}[ht]
  \centering
  \caption{\textbf{OOD performance on sessions with captured initial
  user prompts.} Mean AUROC across seven families, comparing the
  full scoring cohorts with their captured-prompt subsets.
  Qwen judges use the OWASP-grounded, family-held-out $K{=}1$
  protocol. SLMs use their fold-specific LoRA adapters.
  App / Kernel / Cross denote the three evidence views.}
  \label{tab:captured-prompts}
  \footnotesize
  \setlength{\tabcolsep}{6pt}
  \begin{tabular}{@{}l | ccc | ccc@{}}
    \toprule
    & \multicolumn{3}{c|}{\textbf{Full cohort}} & \multicolumn{3}{c}{\textbf{Captured-prompt subset}} \\
    \textbf{Model} & \textbf{App} & \textbf{Kernel} & \textbf{Cross} & \textbf{App} & \textbf{Kernel} & \textbf{Cross} \\
    \midrule
    Qwen3-235B   & .705 & .775 & .829 & .613 & .803 & .809 \\
    Qwen3-80B    & .678 & .725 & .810 & .608 & .698 & .791 \\
    Qwen3-32B    & .672 & .741 & .723 & .581 & .766 & .661 \\
    \midrule
    Qwen2.5-3B   & .713 & .901 & .881 & .716 & .914 & .935 \\
    Qwen2.5-7B   & .722 & .882 & .920 & .736 & .941 & .968 \\
    Llama-3.1-8B & .688 & .922 & .791 & .735 & .944 & .809 \\
    \bottomrule
  \end{tabular}
\end{table}

The restricted Qwen cohort contains 2{,}087 unique sessions
and 2{,}543 session/fold observations, with fold counts
A--G of 358, 1{,}026, 386, 122, 193, 174, 284.
The restricted SLM cohort contains 2{,}196 unique sessions
and 2{,}685 session/fold observations, with counts
376, 1{,}088, 407, 129, 200, 186, 299.
Each detector family uses a common scoring cohort across
its models and views, and matched benign sessions may
contribute to multiple folds.

\subsection{Paired comparisons between evidence views}
\label{app:benchmark-details:paired}

Table~\ref{tab:paired-view-differences} estimates the
Cross-View gain over each single-layer view on the full
OOD scoring cohorts for all ten detectors evaluated with
three views. Comparisons retain the benchmark cohorts,
matching sessions across views within each detector family.
Falco has only a Kernel-View score and does not define
these contrasts. We resample the 153 fixture/host
cluster identities with replacement, using identical
cluster weights across views and repeated fold memberships.
For each draw, we recompute the seven fold AUROCs and
their unweighted mean before taking the paired difference.
Of 1{,}000 draws (seed 20260923), 995 retain both classes
in every fold and contribute to the percentile intervals.
The intervals are pointwise 95\% intervals, conditional on
the fitted models and selected references.

\begin{table}[ht]
  \centering
  \caption{\textbf{Paired OOD AUROC differences across all four
  detector families.} Positive differences favor Cross-View. Qwen judges
  use OWASP-grounded, family-held-out references; SLMs use
  fold-specific adapters. Intervals use paired cluster
  resampling and are not adjusted for multiple comparisons.}
  \label{tab:paired-view-differences}
  \footnotesize
  \setlength{\tabcolsep}{5pt}
  \begin{tabular}{@{}l rr rr@{}}
    \toprule
    & \multicolumn{2}{c}{\textbf{Cross minus App}} & \multicolumn{2}{c}{\textbf{Cross minus Kernel}} \\
    \textbf{Detector} & \textbf{\boldmath$\Delta$} & \textbf{95\% interval} & \textbf{\boldmath$\Delta$} & \textbf{95\% interval} \\
    \midrule
    AdaBoost     & $+.150$ & $[+.101,+.209]$ & $+.107$ & $[+.052,+.165]$ \\
    XGBoost      & $+.147$ & $[+.083,+.217]$ & $+.101$ & $[+.043,+.157]$ \\
    \midrule
    TabPFN       & $+.241$ & $[+.167,+.315]$ & $+.097$ & $[+.031,+.161]$ \\
    TabICL       & $+.166$ & $[+.112,+.228]$ & $+.038$ & $[-.029,+.098]$ \\
    \midrule
    Qwen3-235B   & $+.125$ & $[+.075,+.175]$ & $+.054$ & $[-.002,+.098]$ \\
    Qwen3-80B    & $+.132$ & $[+.081,+.178]$ & $+.085$ & $[+.036,+.132]$ \\
    Qwen3-32B    & $+.052$ & $[+.004,+.095]$ & $-.018$ & $[-.102,+.041]$ \\
    \midrule
    Qwen2.5-3B   & $+.168$ & $[+.099,+.240]$ & $-.019$ & $[-.086,+.032]$ \\
    Qwen2.5-7B   & $+.199$ & $[+.139,+.267]$ & $+.039$ & $[-.014,+.081]$ \\
    Llama-3.1-8B & $+.103$ & $[+.024,+.191]$ & $-.131$ & $[-.205,-.070]$ \\
    \bottomrule
  \end{tabular}
\end{table}

Cross-View exceeds App-View for all ten detectors, with each
individual interval above zero. Against Kernel-View,
the interval favors Cross-View for AdaBoost, XGBoost,
TabPFN, and Qwen3-80B and Kernel-View
for Llama-3.1-8B; the remaining intervals include zero.
The paired comparisons support the contribution of kernel
evidence while distinguishing it from the model-dependent
benefit of combining both layers.

\section{Deep tabular details}
\label{app:dl-details}

The deep-tabular family in Table~\ref{tab:benchmark}
(Appendix~\ref{app:benchmark-details}) is
represented by two pretrained tabular foundation models,
TabPFN and TabICL, that consume the same 86-dimensional
feature vectors as the classical family and perform
in-context learning at inference time. This appendix
motivates that choice: Appendix~\ref{app:dl-details:ablation}
compares TabPFN and TabICL against two train-from-scratch
alternatives (tuned MLP and FT-Transformer) on the same
evaluation grid. Appendices~\ref{app:dl-details:mlp}
and~\ref{app:dl-details:ft} report the hyperparameter
configurations used for those train-from-scratch detectors.

\subsection{Tabular DL ablation}
\label{app:dl-details:ablation}

Table~\ref{tab:benchmark} reports the two
tabular foundation models we evaluate for the deep-tabular
family, TabPFN~\citep{hollmann2025tabpfnv2}
(checkpoint v3, released 2026-04-17) and
TabICL~\citep{qu2025tabicl} (checkpoint v2, released
2026-02-12). Both are pretrained tabular foundation models
that consume the same 86-dimensional feature vectors as the
classical family but perform in-context learning at inference
time: the fold's training rows enter as context and the test
rows are predicted in a single forward pass, with no gradient
training on ACE. This subsection reports the parallel
train-from-scratch results (tuned MLP and FT-Transformer)
that motivate the foundation-model choice. The hyperparameters
used are those selected in
\S\ref{app:dl-details:mlp}--\ref{app:dl-details:ft}.

\begin{table}[h]
  \centering
  \caption{\textbf{Tabular DL ablation on ACE.} Mean AUROC
  across folds (5 stratified for ID, 7 OWASP-aligned for OOD);
  $\pm$ is the mean per-fold patch-clustered bootstrap standard error
  (\S\ref{sec:experiments:regimes}). MLP and FT-Transformer are
  train-from-scratch; TabPFN and TabICL are pretrained tabular
  foundation models scored in-context with no per-corpus
  training. Winning view per (model, regime) in bold.}
  \label{tab:dl-ablation}
  \footnotesize
  \setlength{\tabcolsep}{6pt}
  \begin{tabular}{@{}l | ccc | ccc@{}}
    \toprule
                    & \multicolumn{3}{c|}{\textbf{ID}} & \multicolumn{3}{c}{\textbf{OOD}} \\
    \textbf{Model}           & \textbf{App} & \textbf{Kernel} & \textbf{Cross} & \textbf{App} & \textbf{Kernel} & \textbf{Cross} \\
    \midrule
    \multicolumn{7}{c}{\itshape Train-from-scratch} \\
    \midrule
    MLP (tuned)     & .864$\pm$.022 & .781$\pm$.032 & \textbf{.897$\pm$.018} & \textbf{.721$\pm$.091} & .629$\pm$.094 & .664$\pm$.089 \\
    FT-Transformer  & .883$\pm$.020 & .796$\pm$.029 & \textbf{.900$\pm$.018} & \textbf{.694$\pm$.089} & .630$\pm$.087 & .670$\pm$.095 \\
    \midrule
    \multicolumn{7}{c}{\itshape Pretrained tabular foundation models (in-context)} \\
    \midrule
    TabPFN          & .933$\pm$.016 & .952$\pm$.012 & \textbf{.986$\pm$.006} & .633$\pm$.103 & .777$\pm$.086 & \textbf{.874$\pm$.071} \\
    TabICL          & .932$\pm$.016 & .947$\pm$.013 & \textbf{.986$\pm$.006} & .636$\pm$.105 & .764$\pm$.085 & \textbf{.801$\pm$.088} \\
    \bottomrule
  \end{tabular}
\end{table}

\paragraph{Reading.}
The train-from-scratch tabular DL detectors underperform on
ACE: MLP and FT-Transformer prefer the App-View at OOD
($0.721$ and $0.694$) and never reach the boosted-tree
Cross-View OOD ceiling of $0.850$ (XGBoost, Table~\ref{tab:benchmark}).
\citet{mcelfresh2023nnvsgbdt} attribute this train-from-scratch
NN-vs-GBDT gap on tabular data primarily to skewed
and heavy-tailed feature distributions, which is precisely
the regime ACE's featurizer targets: 57 of the 86 features
are per-syscall counts, log-counts, or count-derived ratios
whose distributions have long tails
(Appendix~\ref{app:features:kernel}). The pretrained
foundation models sidestep this constraint. TabPFN reaches
OOD Cross-View $0.874$, edging past the classical
XGBoost baseline ($0.850$) and closing the gap the
train-from-scratch detectors leave open ($+0.21$ over MLP,
$+0.20$ over FT-Transformer). TabICL reaches OOD Cross-View $0.801$,
lower than TabPFN by about $0.07$ but still $+0.13$--$0.14$
over the train-from-scratch alternatives. TabPFN is
therefore the deep-tabular representative reported at each
of the paper's ACE-XA and scanner-comparison
transfer sites. TabICL is retained as a within-family
comparison point in Table~\ref{tab:benchmark}. Both
foundation models also correct the App-View preference: at
OOD they pick Cross-View, matching the behavior of every
other detector family in Table~\ref{tab:benchmark} except two SLM adapters
(Appendix~\ref{app:ood-exceptions}). The train-from-scratch
DL underperformance is a corpus-size and feature-distribution
artifact of the tabular NN regime, not a signal-availability
limitation of the Cross-View feature space itself.

\subsection{MLP hyperparameter configuration}
\label{app:dl-details:mlp}

The tuned MLP configuration reported in the
Table~\ref{tab:dl-ablation} ablation was selected from a small
grid search on ACE's ID data. We swept hidden layer widths
$\in \{(128,64), (256,128,64), (256,128,64,32), (512,256,128)\}$;
dropout $\in \{0.2, 0.4\}$; weight decay
$\in \{10^{-3}, 10^{-4}\}$ (16 configurations total),
ranking each configuration by AUROC on the ACE
5-fold-stratified fold\_0 Cross-View. The winning
configuration --- hidden $(256, 128, 64, 32)$, dropout $0.4$,
weight decay $10^{-3}$ --- is refit across all folds
(5 stratified + 7 OWASP-aligned) for the ablation table.
Per-configuration Cross-View AUROC on the tuning fold is in
Table~\ref{tab:mlp-sweep}. All 16 configurations sit within
a $0.011$ AUROC band on the tuning fold, so the ablation
result is not sensitive to the specific winning
configuration.

\begin{table}[h]
  \centering
  \caption{\textbf{MLP grid search on ACE stratified fold\_0
  (Cross-View).} All 16 configurations sorted by AUROC.
  Selected configuration in bold; ties broken by earlier
  appearance in the grid.}
  \label{tab:mlp-sweep}
  \small
  \setlength{\tabcolsep}{8pt}
  \begin{tabular}{@{}l c c r@{}}
    \toprule
    \textbf{Hidden layers} & \textbf{Dropout} & \textbf{Weight decay} & \textbf{AUROC} \\
    \midrule
    $\mathbf{(256,128,64,32)}$ & $\mathbf{0.4}$ & $\mathbf{10^{-3}}$ & $\mathbf{0.9027}$ \\
    $(256,128,64,32)$          & $0.4$          & $10^{-4}$          & $0.9027$ \\
    $(512,256,128)$            & $0.4$          & $10^{-3}$          & $0.9014$ \\
    $(128,64)$                 & $0.2$          & $10^{-3}$          & $0.9004$ \\
    $(128,64)$                 & $0.2$          & $10^{-4}$          & $0.9003$ \\
    $(256,128,64)$             & $0.4$          & $10^{-3}$          & $0.8997$ \\
    $(256,128,64)$             & $0.4$          & $10^{-4}$          & $0.8997$ \\
    $(256,128,64)$             & $0.2$          & $10^{-4}$          & $0.8986$ \\
    $(256,128,64,32)$          & $0.2$          & $10^{-3}$          & $0.8983$ \\
    $(256,128,64,32)$          & $0.2$          & $10^{-4}$          & $0.8983$ \\
    $(512,256,128)$            & $0.4$          & $10^{-4}$          & $0.8981$ \\
    $(512,256,128)$            & $0.2$          & $10^{-3}$          & $0.8958$ \\
    $(512,256,128)$            & $0.2$          & $10^{-4}$          & $0.8958$ \\
    $(256,128,64)$             & $0.2$          & $10^{-3}$          & $0.8956$ \\
    $(128,64)$                 & $0.4$          & $10^{-3}$          & $0.8923$ \\
    $(128,64)$                 & $0.4$          & $10^{-4}$          & $0.8923$ \\
    \bottomrule
  \end{tabular}
\end{table}

\subsection{FT-Transformer hyperparameter configuration}
\label{app:dl-details:ft}

The tuned FT-Transformer configuration reported in the
Table~\ref{tab:dl-ablation} ablation was selected from a small
grid search on ACE's ID data. We swept number of blocks
$\in \{2, 3, 4\}$; block dimension $\in \{96, 192\}$;
dropout $\in \{0.1, 0.2\}$ (12 configurations total),
ranking each configuration by AUROC on the ACE
5-fold-stratified fold\_0 Cross-View. The winning
configuration --- $n_{\text{blocks}} = 2$, $d_{\text{block}} = 96$,
dropout $0.1$ --- is refit across all folds
(5 stratified + 7 OWASP-aligned) for the ablation table.
Per-configuration Cross-View AUROC on the tuning fold is in
Table~\ref{tab:ft-sweep}. All 12 configurations sit within
a $0.035$ AUROC band on the tuning fold, so the ablation
result is not sensitive to the specific winning
configuration.

\begin{table}[h]
  \centering
  \caption{\textbf{FT-Transformer grid search on ACE stratified fold\_0
  (Cross-View).} All 12 configurations sorted by AUROC.
  Selected configuration in bold.}
  \label{tab:ft-sweep}
  \small
  \setlength{\tabcolsep}{10pt}
  \begin{tabular}{@{}c c c r@{}}
    \toprule
    \textbf{\boldmath$n_{\text{blocks}}$} & \textbf{\boldmath$d_{\text{block}}$} & \textbf{Dropout} & \textbf{AUROC} \\
    \midrule
    $\mathbf{2}$ & $\mathbf{96}$ & $\mathbf{0.1}$ & $\mathbf{0.9203}$ \\
    $3$          & $96$          & $0.1$          & $0.9185$ \\
    $2$          & $96$          & $0.2$          & $0.9119$ \\
    $3$          & $96$          & $0.2$          & $0.9118$ \\
    $4$          & $96$          & $0.2$          & $0.9088$ \\
    $4$          & $96$          & $0.1$          & $0.9069$ \\
    $3$          & $192$         & $0.2$          & $0.9009$ \\
    $2$          & $192$         & $0.1$          & $0.9003$ \\
    $2$          & $192$         & $0.2$          & $0.8955$ \\
    $4$          & $192$         & $0.2$          & $0.8914$ \\
    $4$          & $192$         & $0.1$          & $0.8910$ \\
    $3$          & $192$         & $0.1$          & $0.8855$ \\
    \bottomrule
  \end{tabular}
\end{table}

\subsection{Sequence-DL ablation on the raw kernel stream}
\label{app:dl-details:sequence-ablation}

To empirically test whether session-level aggregation
discards discriminative signal, we ran five deep-learning
detectors on the ACE kernel evidence. Four consume the raw
kernel-syscall stream directly. Two are
sequence-architecture detectors: an LSTM language model
following~\citet{kim2016lstm} (one-class benign-only,
canonical Kim 2016 sizing) and a BERT window-level encoder
with multiple-instance learning over per-window session
embeddings following~\citet{ilse2018attention}. Two are
set-architecture detectors: a Set
Transformer~\citep{lee2019settransformer} over kernel
tokens and a Set Transformer over structured events (each
event embedded as the sum of five learned field embeddings
across syscall name, path hash, IP hash, file-descriptor
bucket, and return-code bucket). The fifth detector consumes
the same aggregated JSON summary the LLM judges see (the R1
Kernel-View rendering, Appendix~\ref{app:llm-details:kernel-rep}):
a BERT model fine-tuned on the R1 rendering. All five are
trained on the paper's 7 OWASP-aligned OOD folds under the
same per-fold splits used by every other detector in the
paper.

Table~\ref{tab:seq-dl-ablation} reports per-fold OOD AUROC
with patch-clustered bootstrap standard errors (1{,}000
resamples over patch clusters,
\S\ref{sec:experiments:regimes}).

\begin{table}[h]
  \centering
  \caption{\textbf{Sequence-DL detector ablation on ACE
  OWASP-aligned OOD folds.} Per-fold AUROC $\pm$
  patch-clustered bootstrap SE (1{,}000 resamples). All
  cells use the paper's canonical AUROC scoring cohort
  (\S\ref{sec:experiments:formalism}): signal-bearing
  latent sessions dropped from both training and scoring,
  matching Table~\ref{tab:benchmark}.
  The four raw-stream detectors consume the ordered syscall
  event stream, and are ordered left to right by mean OOD
  AUROC (which coincides with grouping the two
  sequence-architecture detectors ahead of the two
  set-architecture ones). BERT-on-summary consumes the R1 JSON
  rendering from Appendix~\ref{app:llm-details:kernel-rep}.
  Mean row reports the arithmetic mean of per-fold point
  estimates with the fold-to-fold standard deviation.}
  \label{tab:seq-dl-ablation}
  \footnotesize
  \setlength{\tabcolsep}{6pt}
  \begin{tabular}{@{}l cccc c@{}}
    \toprule
    & \multicolumn{4}{c}{\textbf{Raw ordered event stream}} & \textbf{Summary} \\
    \cmidrule(lr){2-5} \cmidrule(lr){6-6}
    \textbf{Fold}
      & \textbf{\makecell{LSTM-LM\\(Kim 2016)}}
      & \textbf{\makecell{BERT-MIL\\(windows)}}
      & \textbf{\makecell{Set Transformer\\(tokens)}}
      & \textbf{\makecell{Set Transformer\\(events)}}
      & \textbf{\makecell{BERT-on-\\summary}} \\
    \midrule
    A    & .560$\pm$.107 & .519$\pm$.114 & .530$\pm$.104 & .512$\pm$.104 & \textbf{.752}$\pm$.099 \\
    B    & .545$\pm$.087 & .523$\pm$.081 & .574$\pm$.072 & .532$\pm$.076 & \textbf{.853}$\pm$.046 \\
    C    & .448$\pm$.104 & .491$\pm$.105 & .538$\pm$.075 & .532$\pm$.091 & \textbf{.606}$\pm$.089 \\
    D    & .680$\pm$.210 & .472$\pm$.142 & .768$\pm$.115 & .652$\pm$.235 & \textbf{.978}$\pm$.021 \\
    E    & .585$\pm$.153 & .729$\pm$.100 & .483$\pm$.192 & .677$\pm$.153 & \textbf{.877}$\pm$.066 \\
    F    & .383$\pm$.155 & .504$\pm$.164 & .643$\pm$.141 & .537$\pm$.178 & \textbf{.997}$\pm$.004 \\
    G    & .418$\pm$.126 & .635$\pm$.112 & .495$\pm$.131 & .607$\pm$.119 & \textbf{.792}$\pm$.096 \\
    \midrule
    Mean & .517$\pm$.105 & .553$\pm$.094 & .576$\pm$.100 & .578$\pm$.066 & \textbf{.836}$\pm$.135 \\
    \bottomrule
  \end{tabular}
\end{table}

\paragraph{Aggregation is what carries the signal.}
BERT fine-tuned on the R1 aggregated JSON summary reaches
mean OOD AUROC $0.836$, above the reported Qwen judges'
Kernel-View point estimates ($0.725$--$0.775$,
Table~\ref{tab:benchmark})
and near the classical XGBoost Cross-View baseline at
$0.850$. The four detectors consuming the raw ordered event
stream sit $25$--$32$ points below on OOD, at
$0.517$--$0.578$ mean. Signal shape on this workload favors
session-level aggregation.

\paragraph{Order preservation adds no measurable lift over
set-based treatment.} On the raw kernel stream, BERT-MIL
($0.553$) preserves per-window order and applies gated
attention pooling. Set Transformer over structured events
($0.578$) and over flat tokens ($0.576$) discard order by
construction. All three sit within $0.025$ of each other,
well inside one bootstrap SE. Temporal ordering carries no
signal that a permutation-invariant treatment cannot
recover.

\paragraph{LSTM language modelling on syscall streams
underperforms on this workload.} The canonical syscall-IDS
sequence baseline, one-class benign-only LSTM language
modelling per~\citet{kim2016lstm}, reaches mean OOD AUROC
$0.517$, near chance and lower than every set-based
alternative. Sessions in the corpus differ from the
long-lived process traces this line of work has historically
targeted (multi-day server workloads, single-process
lifecycles) in that each ACE session is a short, tool-diverse
agent-driven interaction. Novelty-rate scoring on short
sessions produces high variance and no clean separation.

\paragraph{Interpretation.}
The result mirrors the R4 chronological-verbatim rendering
ablation in Appendix~\ref{app:llm-details:kernel-rep} on
the LLM-judge side, where feeding ordered raw events to a
frozen LLM judge underperforms the aggregated R1 summary
by $\approx 6$ AUROC points. Together the two ablations
converge on the same conclusion across two independent
detector families (frozen LLM judges and trained deep
sequence models): the discriminative content on this workload
is well captured by session-level counts, cardinalities,
and rates, and adding raw temporal ordering degrades rather
than improves detection performance. This validates the
aggregation-based featurizer of
\S\ref{sec:experiments:features} and the R1 rendering of
Appendix~\ref{app:llm-details:kernel-rep} as the
appropriate representation choice for this workload.

\section{LLM details}
\label{app:llm-details}

Table~\ref{tab:benchmark} reports mean AUROC
per (model, view, regime) cell under a single fixed
(prompt-template, $K$) configuration (OWASP-grounded
template, 14-reference $K{=}1$ in-context evaluation;
see Appendix~\ref{app:llm-details:prompts} for the
protocol statement) applied uniformly to every LLM judge.
Table~\ref{tab:llm-sweep} in
Appendix~\ref{app:llm-details:ablation} below expands
that single-configuration cell into the full $2 \times 2$
(prompt template $\times$ shot count) sweep with $\pm$
mean per-fold patch-clustered bootstrap standard error on every cell.

\subsection{LLM prompt and exemplar ablation}
\label{app:llm-details:ablation}

\begin{table}[t]
  \centering
  \caption{\textbf{Full LLM-judge sweep on ACE with mean per-fold clustered
  bootstrap SE.} Mean AUROC across folds
  (5 stratified for ID, 7 OWASP-aligned for OOD); $\pm$ is
  the mean per-fold patch-clustered bootstrap standard error
  (\S\ref{sec:experiments:regimes}). Every LLM judge is
  expanded into all four (prompt template $\times$ shot count)
  combinations: \emph{Baseline} = task instruction + evidence
  + response format; \emph{OWASP-grounded} = baseline plus
  OWASP-informed attack descriptions and observable signature
  examples. $K{=}0$ = zero-shot;
  $K{=}1$ = 14 labeled references, fixed across ID folds and
  selected from each OOD training fold with the test family
  excluded. OOD cells use a common query cohort across all
  models, templates, views, and shot counts. Reference
  sessions are excluded from scoring; see
  Appendix~\ref{app:llm-details:prompts} for the full
  protocol statement. Winning view per (model, regime)
  within each configuration block in bold.}
  \label{tab:llm-sweep}
  \footnotesize
  \setlength{\tabcolsep}{6pt}
  \begin{tabular}{@{}l | ccc | ccc@{}}
    \toprule
                    & \multicolumn{3}{c|}{\textbf{ID}} & \multicolumn{3}{c}{\textbf{OOD}} \\
    \textbf{Model}           & \textbf{App} & \textbf{Kernel} & \textbf{Cross} & \textbf{App} & \textbf{Kernel} & \textbf{Cross} \\
    \midrule
    \multicolumn{7}{c}{\itshape Baseline template, $K{=}0$ (zero-shot)} \\
    \midrule
    Qwen3-235B      & \textbf{.607$\pm$.032} & .581$\pm$.042 & .541$\pm$.026 & \textbf{.590$\pm$.061} & .589$\pm$.085 & .520$\pm$.041 \\
    Qwen3-80B       & \textbf{.608$\pm$.035} & .516$\pm$.019 & .540$\pm$.031 & \textbf{.594$\pm$.077} & .513$\pm$.018 & .512$\pm$.067 \\
    Qwen3-32B       & \textbf{.605$\pm$.040} & .593$\pm$.024 & .534$\pm$.035 & \textbf{.573$\pm$.073} & .564$\pm$.040 & .502$\pm$.061 \\
    \midrule
    \multicolumn{7}{c}{\itshape OWASP-grounded template, $K{=}0$ (zero-shot)} \\
    \midrule
    Qwen3-235B      & .676$\pm$.036 & .744$\pm$.034 & \textbf{.764$\pm$.028} & .667$\pm$.079 & .727$\pm$.065 & \textbf{.747$\pm$.053} \\
    Qwen3-80B       & \textbf{.643$\pm$.030} & .569$\pm$.018 & .588$\pm$.024 & \textbf{.662$\pm$.053} & .571$\pm$.031 & .601$\pm$.053 \\
    Qwen3-32B       & \textbf{.703$\pm$.034} & .686$\pm$.041 & .625$\pm$.034 & \textbf{.678$\pm$.057} & .677$\pm$.079 & .625$\pm$.060 \\
    \midrule
    \multicolumn{7}{c}{\itshape Baseline template, $K{=}1$ few-shot (14 references)} \\
    \midrule
    Qwen3-235B      & .666$\pm$.032 & .796$\pm$.030 & \textbf{.838$\pm$.026} & .642$\pm$.074 & \textbf{.764$\pm$.069} & .695$\pm$.073 \\
    Qwen3-80B       & .695$\pm$.033 & \textbf{.796$\pm$.031} & .770$\pm$.024 & \textbf{.668$\pm$.075} & .642$\pm$.079 & .647$\pm$.071 \\
    Qwen3-32B       & .677$\pm$.036 & \textbf{.748$\pm$.032} & .698$\pm$.030 & .629$\pm$.083 & \textbf{.666$\pm$.085} & .599$\pm$.079 \\
    \midrule
    \multicolumn{7}{c}{\itshape OWASP-grounded template, $K{=}1$ few-shot (14 references)} \\
    \midrule
    Qwen3-235B      & .713$\pm$.032 & .783$\pm$.030 & \textbf{.847$\pm$.025} & .705$\pm$.078 & .775$\pm$.054 & \textbf{.829$\pm$.049} \\
    Qwen3-80B       & .676$\pm$.031 & .781$\pm$.032 & \textbf{.835$\pm$.025} & .678$\pm$.073 & .725$\pm$.062 & \textbf{.810$\pm$.051} \\
    Qwen3-32B       & .724$\pm$.035 & .755$\pm$.033 & \textbf{.831$\pm$.028} & .672$\pm$.074 & \textbf{.741$\pm$.068} & .723$\pm$.065 \\
    \bottomrule
  \end{tabular}
\end{table}

Five observations from the LLM sweep visible in
Table~\ref{tab:llm-sweep}.

\paragraph{OWASP grounding lifts every $K{=}0$ cell.}
Comparing the Baseline-$K{=}0$ block against the
OWASP-grounded-$K{=}0$ block, every one of the 18 model
$\times$ view $\times$ regime cells improves when the OWASP
category block is added to the zero-shot template. Under
zero-shot the OWASP prompt gives the judge a canonical
taxonomy against which to score evidence. The per-model OOD
Cross-View lift is Qwen3-235B $+0.226$ ($0.520 \to 0.747$),
Qwen3-80B $+0.089$ ($0.512 \to 0.601$), and Qwen3-32B
$+0.122$ ($0.502 \to 0.625$), an average lift of
$+14.6$\,pp across the three Qwen models. Structural prompt
grounding improves zero-shot discrimination in every tested view.

\paragraph{Examples from other families improve Cross-View detection.}
Comparing the OWASP-grounded-$K{=}0$ block against the
OWASP-grounded-$K{=}1$ block on OOD Cross-View, the lifts
are $+0.082$ for Qwen3-235B, $+0.209$ for Qwen3-80B,
and $+0.099$ for Qwen3-32B. All three models improve when
supplied with demonstrations from the other attack families,
reaching Cross-View AUROC $0.723$--$0.829$.

\paragraph{Grounding and exemplars can be complementary.}
Combining OWASP grounding with family-excluded exemplars
produces higher OOD Cross-View AUROC than either
OWASP zero-shot or baseline few-shot for every Qwen model.
The gains over the better of those two configurations are
$+0.082$ for Qwen3-235B, $+0.163$ for Qwen3-80B, and
$+0.099$ for Qwen3-32B.

\paragraph{Prompt design determines whether kernel evidence
is legible.}
Under Baseline-$K{=}0$, all three Qwen judges pick
App-View as their OOD winner: Qwen3-235B ($0.590$),
Qwen3-80B ($0.594$), and Qwen3-32B ($0.573$).
Under OWASP-$K{=}0$ only Qwen3-235B picks Cross-View.
Under Baseline-$K{=}1$, Qwen3-235B and Qwen3-32B prefer
Kernel-View, while Qwen3-80B prefers App-View.
Under OWASP-$K{=}1$, Qwen3-235B and Qwen3-80B prefer
Cross-View, while Qwen3-32B prefers Kernel-View.
The capacity to integrate the two evidence layers through
the prompt depends on both the instructions and demonstrations.

\paragraph{Frozen Qwen judges provide a training-free baseline.}
Under OWASP-$K{=}1$, OOD Cross-View AUROC is $0.829$ for
Qwen3-235B, $0.810$ for Qwen3-80B, and $0.723$ for
Qwen3-32B. These results establish discrimination of attack
families absent from the demonstrations without parameter
updates. Appendix~\ref{app:llm-details:known-families}
additionally evaluates the deployment setting in which
reference examples cover all attack families.

\subsection{Kernel-View feature representation ablation}
\label{app:llm-details:kernel-rep}

The Kernel-Layer evidence rendering reported everywhere else
in the paper is a bounded structured summary of the
in-container syscall trace
(Appendix~\ref{app:slm-details:input}). This subsection asks
whether that rendering is the right choice and motivates why
the same rendering is used across the LLM and SLM judge
families. We evaluate four kernel-only renderings that all
consume the same underlying strace event stream, chosen so
that each isolates a distinct axis of the featurization
design space:

\begin{itemize}
\item \textbf{R1, bounded structured summary.} A compact
  6-field JSON summary of the trace: total event count,
  per-tracepoint raw counts across the 14 monitored
  tracepoints, distinct-path cardinality (count only, no path
  strings), the list of distinct IPs contacted (verbatim),
  and the list of distinct \texttt{/tmp/} paths accessed
  (verbatim). Every other field is a scalar or a count.
  R1 gives the LLM the same raw counts and cardinalities the
  classical/deep-tabular featurizer derives its 57
  kernel-layer features from (Appendix~\ref{app:features:kernel}),
  serialized as text. Its performance measures whether an
  aggregated count-and-cardinality summary preserves the
  discriminative signal in a token-budget-efficient rendering.
\item \textbf{R2, dedup verbatim log.} One line per
  distinct (syscall, arguments) tuple with the occurrence
  count. No aggregation; no top-$k$ selection. R2 keeps every
  distinct event verbatim with only repetition-count
  collapsing, so its performance relative to R1 measures how
  much signal R1's aggregation step discards.
\item \textbf{R3, feature-family tokenized.} Every syscall
  event emitted in order, with argument strings replaced by
  family tokens (e.g., \texttt{PATH\_HIDDEN\_TMP},
  \texttt{IP\_EXTERNAL\_10}, \texttt{FD\_STDIN}). Structure
  preserved; verbatim identifiers stripped. R3 isolates
  whether the LLM judges key on specific sentinel-shaped
  substrings in paths and IPs (e.g., hidden \texttt{/tmp/}
  filenames, non-loopback addresses) or on the categorical
  event-type structure: if R3 approaches R1 the signal is
  category-shaped, if R3 collapses the signal lives in the
  verbatim identifiers.
\item \textbf{R4, chronological verbatim.} Every syscall
  event emitted in order, with all argument strings verbatim.
  No deduplication. R4 is the maximally-informative rendering
  (strictly more content than R1, R2, or R3), so its
  performance bounds how much of the information R1's
  aggregation step discards actually matters, and whether
  temporal ordering carries signal that R1 and R2 lose.
\end{itemize}

\paragraph{Choice of prompt configuration.} We hold the
prompt template fixed at OWASP-grounded $K{=}0$ across all
four representations. Preliminary comparison against the
baseline template found the baseline-$K{=}0$ Kernel-View
AUROCs sit at $0.51$--$0.59$
(Table~\ref{tab:llm-sweep}), too close to chance for a
well-powered representation comparison. The OWASP block is
applied identically to every rendering, so any measured
difference between R1/R2/R3/R4 is attributable to the
representation.

Table~\ref{tab:kernel-rep} reports per-cell AUROC across the
four representations for all three Qwen judges on ACE, scored
on the paper's canonical AUROC cohort
(\S\ref{sec:experiments:formalism}; $n{=}3{,}651$ after
dropping the $396$ signal-bearing latent sessions).

\begin{table}[h]
  \centering
  \caption{\textbf{Kernel-View representation ablation on
  ACE.} Pooled ACE-full AUROC under OWASP-grounded $K{=}0$
  on the paper's canonical AUROC cohort
  (\S\ref{sec:experiments:formalism}; $n{=}3{,}651$ after
  dropping $396$ signal-bearing latent sessions). Winner per row in
  bold. R1 is the paper's canonical Kernel-View rendering;
  R2/R3/R4 are three alternative renderings that consume the
  same underlying strace event stream. Oversize and
  parse-fail rows are excluded from AUROC.}
  \label{tab:kernel-rep}
  \small
  \setlength{\tabcolsep}{4.75pt}
  \begin{tabular}{@{}l cccc@{}}
    \toprule
          & \textbf{R1} & \textbf{R2} & \textbf{R3} & \textbf{R4} \\
    \textbf{Model} & \textbf{(bounded summary)} & \textbf{(dedup verbatim)} & \textbf{(family tokenized)} & \textbf{(chronological verbatim)} \\
    \midrule
    Qwen3-235B  & \textbf{0.731} & 0.541 & 0.671 & 0.599 \\
    Qwen3-80B   & 0.598          & \textbf{0.628} & 0.575 & 0.563 \\
    Qwen3-32B   & \textbf{0.679} & 0.606 & 0.508 & 0.587 \\
    \midrule
    mean        & \textbf{0.669} & 0.592 & 0.585 & 0.583 \\
    \bottomrule
  \end{tabular}
\end{table}

\paragraph{Reading.} R1 is the strongest rendering on the
aggregate (mean $0.669$ vs.\ $0.592$ for R2, $0.583$ for R4,
and $0.585$ for R3), winning for Qwen3-235B and Qwen3-32B,
while Qwen3-80B prefers R2. R4 retains every event in order
with verbatim arguments, yet trails R1 by $0.086$ on average.
This is consistent with the bounded summary making
count-and-cardinality structure easier to use.
The contribution of verbatim identifiers is isolated by R3
against R4, which emit the same per-event stream and differ
only in whether argument strings are tokenized: R4 minus R3
is $-0.002$ mean AUROC, with model-dependent effects ---
negative for Qwen3-235B ($-0.072$) and Qwen3-80B
($-0.012$), with Qwen3-235B retaining $0.671$ of its $0.731$
R1 AUROC under full identifier removal. Because R1 has the
highest mean across the three Qwen judges, the LLM and
SLM judge families throughout the paper consume R1 for the
Kernel-View, the App-View transcript rendering unchanged
from Appendix~\ref{app:slm-details:input} for the App-View,
and their concatenation for Cross-View. Fixing one rendering
across every neural detector family allows
Table~\ref{tab:benchmark} to compare prompt
templates, shot counts, and detector families on a common
evidence substrate.

\subsection{LLM prompt templates}
\label{app:llm-details:prompts}

Every LLM-judge call at temperature 0 uses one of the four
configurations reported in Table~\ref{tab:llm-sweep}
(baseline vs.\ OWASP-grounded template $\times$ $K{=}0$ vs.\
$K{=}1$). Both templates frame the model as a security monitor auditing
one captured agent session and ask it to classify the evidence
as \texttt{benign} or \texttt{malicious}. The baseline asks
the judge to analyze the supplied artifacts and cite the evidence
supporting its verdict. The grounded template additionally
provides attack descriptions and observable examples. Exact
prompt files will be included in the release.

\paragraph{Response format and confidence.}
Each judge returns three fields: \texttt{VERDICT:
benign|malicious}, \texttt{CONFIDENCE: <0--100>}, and
\texttt{REASON: <one sentence>}. Confidence refers to the
chosen verdict in both templates. The five bands denote
unambiguous support (80--100), strong support with plausible
alternatives (60--80), mixed evidence (40--60), weak contrary
evidence (20--40), and strong contrary evidence (0--20).
The grounded template's highest band includes both a clear
attack match for a malicious verdict and no matching attack
evidence for a benign verdict.
Each call allows up to 250 output tokens. Oversized inputs,
truncated responses, and responses with missing, duplicated,
or invalid required fields are excluded before constructing
the scoring cohorts described below.

\paragraph{Evidence block (both templates).}
The user turn contains the view-conditioned evidence
rendering appropriate to the cell being scored:
\begin{itemize}
\item \textbf{App-View.} The served
  \texttt{tools/list} JSON manifest (tool names, descriptions,
  input schemas) followed by the rendered agent transcript
  (user prompts, tool calls with arguments and results,
  assistant messages), with the field limits described in
  Appendix~\ref{app:slm-details:input}. No syscall content.
\item \textbf{Kernel-View.} A compact JSON summary of
  the syscall trace with 6 fields: capture availability, total event count,
  per-tracepoint raw counts across the 14 monitored
  tracepoints, distinct-path cardinality (count only, no path
  strings), the list of distinct IPs contacted (verbatim),
  and the list of distinct \texttt{/tmp/} paths accessed
  (verbatim). No protocol-surface or transcript content.
\item \textbf{Cross-View.} The App-View block
  followed by the Kernel-View block under a shared header,
  with a section delimiter between them.
\end{itemize}

\paragraph{OWASP-grounded addition.}
The grounded template adds nine blocks before the evidence:
prompt injection, information disclosure, supply-chain compromise,
schema/tool-surface poisoning, excessive agency and tool misuse,
resource overload, silent exfiltration, tool-chain/log poisoning,
and signatures spanning both layers. The descriptions draw on
OWASP threat concepts and provide task-specific examples, such
as credential-path access, unexpected shell execution, and
writes to side logs. They direct the judge to assess observed
artifacts in relation to the user's task and available tool
surface. The same instructions are used across evidence views.

\paragraph{$K{=}1$ reference layout and ID protocol.}
Few-shot evaluation prepends 14 labeled reference sessions
to each target~\citep{brown2020gpt3}. Each reference uses the
same evidence view as the target and carries its binary
\texttt{BENIGN} or \texttt{MALICIOUS} label. Model parameters
remain fixed. ID uses one canonical set containing a benign
and a malicious example from each family A--G. This set is
selected once, with a fixed seed and a pilot-judge confidence
filter favoring prototypical examples, and excluded from
every scoring cohort. ID metrics use each cell's valid scores.

\paragraph{Family-held-out OOD references.}
For each OOD fold, we select 14 references from that fold's
training rows, excluding test-session identities and any
session sharing a held-out attack-fixture component.
Seven references are confirmed-effect malicious sessions:
one from each of the six remaining families, plus a second
from the next family in cyclic A--G order. The other seven
are unpatched benign sessions from distinct hosts.
Candidates must have nonempty evidence in every view and
at most 4{,}096 characters of Cross-View evidence.
Selection uses seed 20260923 plus the fold letter's character
code, stable session-ID ordering, and no judge-score filter.
The same reference identities and order are used across
models, templates, and views within a fold. Reference headers
supply binary labels without family-name hints.
The OWASP-informed attack descriptions remain in the grounded prompt.
Thus, OOD measures generalization to families absent from
ACE demonstrations, with pretrained knowledge and general
instructions available to the frozen model.

\paragraph{OOD scoring cohort.}
We retain the exclusion of the 14 canonical reference
sessions in all configurations. Within each fold, OOD
comparisons use the intersection of valid scores across all
three models, both templates, all views, and zero/few-shot
conditions. This yields 3{,}116 unique sessions contributing
3{,}572 session/fold evaluations; matched benign sessions may
appear in multiple folds. Zero-shot scores are reused across
fold memberships because their prompts do not change.
We average AUROC and average precision over the seven folds.
Reported $\pm$ values retain the benchmark convention of
mean per-fold patch-clustered bootstrap SE (1{,}000 resamples,
seed 0).

\paragraph{Score extraction.}
Each judge reports confidence $c$ in its chosen verdict.
We convert this to a maliciousness score in $[0,1]$ using
$c/100$ for \texttt{malicious} and $1-c/100$ for
\texttt{benign}. AUROC is computed against the binary
$y$ label (\S\ref{app:corpus:fold-counts}) without
per-cell threshold calibration.

\subsection{Detection with examples of known attack families}
\label{app:llm-details:known-families}

A defender may have labeled examples of known attack families
available when configuring a detector. We evaluate this
setting using the 14 fixed references from the ID protocol,
covering all families A--G, with reference sessions excluded
from scoring. Table~\ref{tab:known-family-references} reports
mean AUROC across the seven family cohorts. For comparison
with family-held-out references, we additionally match query
sessions across both reference protocols, all models,
templates, and views. The resulting cohort contains
3{,}043 unique sessions and 3{,}490 session/fold evaluations.

\begin{table}[h]
  \centering
  \caption{\textbf{Detection with reference examples covering
  all attack families.} Mean AUROC across seven family cohorts
  on the common query set used for the reference-protocol
  comparison. Each judge receives 14 fixed references and
  undergoes no parameter updates. $\pm$ is mean per-fold
  patch-clustered bootstrap SE (1{,}000 resamples, seed 0).
  Winning view per row in bold.}
  \label{tab:known-family-references}
  \small
  \setlength{\tabcolsep}{8pt}
  \begin{tabular}{@{}l ccc@{}}
    \toprule
    \textbf{Model} & \textbf{App} & \textbf{Kernel} & \textbf{Cross} \\
    \midrule
    \multicolumn{4}{c}{\itshape Baseline template} \\
    \midrule
    Qwen3-235B      & .695$\pm$.065 & .780$\pm$.068 & \textbf{.831$\pm$.049} \\
    Qwen3-80B       & .688$\pm$.070 & \textbf{.812$\pm$.053} & .771$\pm$.053 \\
    Qwen3-32B       & .660$\pm$.079 & \textbf{.743$\pm$.068} & .693$\pm$.069 \\
    \midrule
    \multicolumn{4}{c}{\itshape OWASP-grounded template} \\
    \midrule
    Qwen3-235B      & .730$\pm$.069 & .772$\pm$.061 & \textbf{.842$\pm$.045} \\
    Qwen3-80B       & .709$\pm$.063 & .797$\pm$.055 & \textbf{.841$\pm$.044} \\
    Qwen3-32B       & .722$\pm$.068 & .746$\pm$.064 & \textbf{.811$\pm$.055} \\
    \bottomrule
  \end{tabular}
\end{table}

With OWASP grounding, Cross-View achieves the highest AUROC
for all three models, reaching $0.811$--$0.842$ without
parameter updates. On the same query sessions,
family-held-out Cross-View AUROC is $0.832$, $0.812$, and
$0.725$ for Qwen3-235B, Qwen3-80B, and Qwen3-32B,
respectively. Differences between reference protocols
therefore vary across models. This comparison evaluates the
two settings as a whole, since both family coverage and the
selected references differ.

\paragraph{Performance is robust to exemplar selection.}
\label{app:llm-details:exemplar-robustness}
Across three independently seeded, stratified-random support
sets, Qwen3-Next-80B-A3B OWASP-grounded Cross-View macro
AUROC remains between $0.837$ and $0.845$, compared with
$0.842$ for the canonical set (Table~\ref{tab:exemplar-robustness}).
The random-draw mean is $0.841$, and every draw preserves
the substantial improvement over zero-shot performance
($0.634$). The worst observed reduction from the canonical
macro AUROC is $0.0047$. Thus, aggregate performance does
not depend on the particular prototypical references used
for ACE's fixed-support-set evaluation.

Each random set contains one matched benign/malicious-fired
pair per OWASP fold A--G, sampled uniformly from eligible
pairs with seeds 1, 2, and 3. Unlike the canonical selection,
sampling uses no pilot-judge correctness filter. A
10{,}000-character cap per reference bounds prompt length,
and each set preserves the 14-reference size and fold order.
The model, template, and decoding settings are fixed.
We score the zero-shot, canonical, and three random-support
configurations afresh, then compare them on a common
cohort under the canonical label policy. Excluding the
union of all 52 reference-session identities leaves
3{,}599 eligible sessions. A further 20 lack scores in at
least one configuration because of context limits, leaving
3{,}579 sessions, with 1{,}311 positives and 2{,}268 negatives.

\begin{table}[h]
  \centering
  \caption{\textbf{Exemplar-selection robustness for
  Qwen3-80B OWASP-grounded Cross-View.} All configurations
  score the same 3{,}579-session cohort, excluding the
  support union and common-score omissions. Macro AUROC
  is the unweighted mean over the seven OWASP-aligned folds
  A--G. Pooled AUROC also retains eligible negative-only
  H/Z pool controls. Its $\pm$ values are patch-clustered
  bootstrap SEs from 1{,}000 resamples with seed 0, not
  uncertainty across reference draws. These matched-cohort
  results have a different denominator from the LLM prompt
  comparison in Table~\ref{tab:llm-sweep} and the
  reference-protocol comparison in
  Table~\ref{tab:known-family-references}.}
  \label{tab:exemplar-robustness}
  \small
  \setlength{\tabcolsep}{10pt}
  \begin{tabular}{@{}l c c@{}}
    \toprule
    \textbf{Support set} & \textbf{Macro AUROC} & \textbf{Pooled AUROC} \\
    \midrule
    Zero-shot ($K{=}0$) & .634 & .630$\pm$.026 \\
    Canonical ($K{=}1$) & .842 & .831$\pm$.022 \\
    Random seed 1      & .845 & .847$\pm$.020 \\
    Random seed 2      & .842 & .833$\pm$.021 \\
    Random seed 3      & .837 & .847$\pm$.020 \\
    \bottomrule
  \end{tabular}
\end{table}

The stable aggregate result supports robustness to
reference selection within this fixed-size, category-balanced
protocol. Individual mechanics vary more than the macro
average, and the study does not compare different support
sizes or category coverage.

\section{SLM details}
\label{app:slm-details}

Table~\ref{tab:benchmark} in Appendix~\ref{app:benchmark-details}
reports the SLM-judge AUROCs on ACE with mean per-fold
clustered bootstrap SE. This appendix documents
SLM training and session-evidence rendering for ACE's
stratified and OWASP-aligned evaluations. The five
additional same-fold Llama adapters used for ACE-XA follow
the protocol in Appendix~\ref{app:cross-agent-ablation:protocol}.

\subsection{LoRA training recipe}
\label{app:slm-details:recipe}

\paragraph{Adapters.}
Each of the three base models (Qwen2.5-3B, Qwen2.5-7B,
Llama-3.1-8B) is fine-tuned with a Low-Rank Adaptation
(LoRA) adapter~\citep{hu2022lora} attached to the query, key,
value, and output projections of every attention block and
the gate, up, and down projections of every MLP block. LoRA
hyperparameters are held constant across models and views:
rank $r = 64$, scaling $\alpha = 128$, LoRA dropout
$0.1$, no bias adaptation. The rank and scaling follow the
$r = 64$, $\alpha = 2r$ configuration validated across
model scales by QLoRA~\citep{dettmers2023qlora}. The $2r$
ratio approximates a fixed effective learning-rate
multiplier of 2 as the rank varies, matching the heuristic
proposed in the original LoRA paper~\citep{hu2022lora}.
Every other model weight is frozen. Adapter size is
approximately 120M, 161M, and 168M parameters for the 3B,
7B, and 8B bases, respectively.

\paragraph{Training objective.}
The loss is standard causal language-modeling cross-entropy
over the tokenized assistant response, with prompt tokens
masked from the loss. Training targets contain only
\texttt{VERDICT: benign} or \texttt{VERDICT: malicious},
without confidence or rationale text.

\paragraph{Optimizer and schedule.}
AdamW with learning rate $10^{-4}$, weight decay $0.01$,
linear warmup over the first 3\% of steps followed by
linear decay to $0$ over the remaining budget. Effective
batch size is 16 sessions (per-device batch size 1 with
gradient accumulation steps 16, seed 0). Training runs for
up to 8 epochs with early stopping on validation AUROC
(minimum improvement $0.002$, patience 2). Per-cell
best-epoch selection is preserved as the exported
checkpoint. Within each training fold, validation sessions
are selected by holding out entire MCP groups with seed 0
and a target validation fraction of 10\%. Test sessions do
not participate in checkpoint selection.
Per-adapter best-epoch counts are recorded in
the \texttt{training\_meta.json} files. In
practice best-epoch spans 1--7 across cells, with 53 of the
63 OWASP-OOD cells ($84\%$) reaching best validation
AUROC by epoch 5.

Each adapter for these ID and OOD evaluations is trained at
a uniform sequence cap of $\texttt{max\_len} = 6{,}144$ tokens, selected as
the largest cap that fits all three base models reliably on
40\,GB A100 GPUs with
\texttt{PYTORCH\_CUDA\_ALLOC\_CONF=expandable\_segments:True}.
Sequences longer than this cap are truncated at the
tokenizer level by \emph{symmetric middle-truncation}: the
prompt-side head and the response format at the tail are
preserved and the middle of the evidence block is elided.
Training-time cap-bind rates (fraction of sessions
truncated) are $0.9\%$ (Kernel-View), $2.6\%$ (App-View),
and $4.0\%$ (Cross-View), pooled across the 7 OWASP OOD
folds. Because the App-View trajectory rendering is
typically longer than the Kernel-View 6-field JSON summary,
when the cap binds a larger share of dropped tokens comes
from App-View content by construction.

\paragraph{Seed and variance.}
Each of the 108 (base model, view, fold) adapters is trained
under a single fixed seed. Every SLM cell in
Table~\ref{tab:benchmark} is reported with the mean per-fold
patch-clustered bootstrap standard error over the five ID
or seven OOD folds (\S\ref{sec:experiments:regimes}). These
SEs summarize within-fold resampling uncertainty. A seed
sweep at fixed fold assignment would probe a
complementary axis of variance and is left to future work.

\paragraph{Fold coverage.}
One independent adapter is trained per (base model, feature
view, fold) triple. We fit adapters over all 12 folds
(the 5 stratified ID folds and the 7 OWASP-aligned OOD folds),
producing $3 \times 3 \times 12 = 108$ adapters total.
Each fold's adapter is trained on that fold's training rows
only, so no adapter sees any of its own test rows during
training.

\paragraph{Inference-time scoring.}
At evaluation time each adapter uses greedy decoding
at temperature 0, with the input limit
matched to the training \texttt{max\_len} of $6{,}144$
tokens, so training and inference apply the same
symmetric middle-truncation policy at the same cap.
ID scoring and the Qwen2.5-3B OOD fold-G Cross-View cell
use Transformers in FP16; other OOD cells use vLLM in BF16.
Inference-time cap-bind rates are $0.9\%$ (Kernel-View),
$2.6\%$ (App-View), and $3.9\%$ (Cross-View) pooled across
the 7 OWASP OOD folds. We do not sample a full response.
We read the top-20 logprobs at the \texttt{VERDICT}
token position directly and compute
$p_{\text{mal}} / (p_{\text{mal}} + p_{\text{ben}})$ as the
per-session malicious score. This gives a continuous score
in $[0, 1]$ suitable for AUROC without needing a threshold.

\subsection{Featurized text rendering per view}
\label{app:slm-details:input}

Each SLM adapter consumes the same session evidence used by
the LLM judges (Appendix~\ref{app:llm-details:prompts}) but
under the fixed \emph{baseline} prompt template (no
OWASP-grounded block, no $K{=}1$ exemplars) since LoRA
fine-tuning already gives the model per-corpus context that
prompt structuring would otherwise supply. The per-view
rendering is:

\paragraph{App-View.}
The user turn contains the full served \texttt{tools/list}
JSON manifest (tool name, human-readable description, and
input schema for every tool the agent sees) followed by the
full agent transcript: user prompts,
each tool call with arguments (capped at 8{,}000 characters
per call), truncated result preview (capped at 16{,}000
characters per result), and assistant messages. These
per-preview character caps are large enough that they are
not the binding truncation constraint in practice. The
operative policy is the token-level symmetric
middle-truncation at \texttt{max\_len} described above.
The truncation preserves the leading portion (which
typically carries error messages, structural content, and
any injected directives). Total App-View evidence budget
before the \texttt{max\_len} cap: median $\approx 1.5$k
tokens, with the App-View trajectory rendering dominating
the total when it exceeds this median.

\paragraph{Kernel-View.}
The user turn contains a compact JSON summary of the
in-container syscall trace with six fields:
\texttt{available} (whether the strace attach succeeded);
\texttt{n\_events} (total event count in the tool process
tree); \texttt{syscall\_counts} (a dictionary from each
observed tracepoint name to its raw event count, across the
14 monitored tracepoints spanning filesystem, process, and
network operations); \texttt{n\_distinct\_paths} (the count
of unique filesystem paths accessed; \emph{count only, no
path strings}); \texttt{distinct\_ips} (the list of unique
IP addresses contacted, verbatim); and \texttt{tmp\_paths}
(the list of unique \texttt{/tmp/} paths accessed,
verbatim). The verbatim identifiers are confined to
\texttt{distinct\_ips} and \texttt{tmp\_paths}. Every other
field is a scalar or a count. Log-counts, per-syscall rates,
and structural pair-ratios (present in the 86-dimensional
featurizer for the classical/deep-tabular detectors,
Appendix~\ref{app:features:kernel}) are not serialized
into the LLM/SLM rendering. The raw counts and cardinalities
in the JSON are what the adapter reads. Total Kernel-View
evidence budget: up to $\approx 1{,}500$ tokens.

\paragraph{Cross-View.}
The App-View block followed by the Kernel-View block
under a shared header, with a section delimiter between
them. When the concatenated evidence exceeds the training
\texttt{max\_len} cap of $6{,}144$ tokens, the same
symmetric middle-truncation policy described above applies
to the entire evidence block: the tokenizer-level head and
tail are preserved and the middle is elided. Because the
App-View trajectory rendering is typically much longer than
the compact Kernel-View 6-field JSON summary, when the cap
binds a larger share of dropped tokens comes from App-View
content by construction, not by policy. The fixed prompt
skeleton (role, capability, decision rule, response
format) sits outside the evidence block and is never
truncated.

\subsection{Reproducibility}
\label{app:slm-details:repro}

All 108 benchmark adapter weights and the seven additional
redaction adapters (\S\ref{app:slm-details:redaction}), the fold-split definitions used to
train them, the tokenized rendering pipeline for each view,
and the inference harnesses will be included in the
release. Score aggregation from per-session logprobs
to per-fold AUROC uses the same code path as the LLM-judge
scoring in Appendix~\ref{app:llm-details}, so the two detector
families are directly comparable at the scoring level.

\subsection{Discussion}
\label{app:ood-exceptions}

Three of ten detectors prefer Kernel-View under OOD in
Table~\ref{tab:benchmark}. The frozen Qwen3-32B judge reaches
$0.741$ on Kernel versus $0.723$ on Cross
(Appendix~\ref{app:llm-details:ablation}). Qwen2.5-3B and
Llama-3.1-8B are the two SLM adapters
whose Kernel-View preference reflects an adapter-capacity
tradeoff rather than an evidence-availability finding.
Qwen2.5-3B reaches Kernel-View $0.901$ over Cross-View $0.881$
(within one clustered-bootstrap SE). Llama-3.1-8B reaches Kernel-View $0.922$ over
Cross-View $0.791$ (decisive). The mid-scale Qwen2.5-7B, in
between, is the only SLM to prefer Cross-View, at $0.920$: the
smallest adapter is capacity-limited on the wider Cross-View
rendering and falls back to the narrower Kernel-View
rendering, and the largest adapter over-fits the
app-layer tail on Cross-View and again does better on
Kernel-only. For both SLM exceptions, the winning single-view AUROC
($0.901$--$0.922$) still exceeds the XGBoost Cross-View OOD
baseline of $0.850$, so the SLM Kernel-only preference is a
within-family optimality result rather than evidence against
Cross-View. Every OOD-winning cell in
Table~\ref{tab:benchmark} still consumes kernel signal
(Cross or Kernel), so kernel-side evidence is the winning
signal for 10 of 10 detectors even under a per-model
winning-view choice.

\subsection{Identifier redaction in kernel summaries}
\label{app:slm-details:redaction}

Training and evaluating Llama-3.1-8B on identifier-redacted
kernel summaries achieves $0.759$ macro AUROC across seven
held-out attack families, compared with $0.922$ using the
original summaries (Table~\ref{tab:identifier-redaction}).
The transformation preserves syscall counts, total events,
distinct-path counts, and list lengths and ordering.
In \texttt{distinct\_ips}, each address becomes
\texttt{<EXTERNAL\_IP>} except loopback and link-local
addresses. In \texttt{tmp\_paths}, each entry becomes
\texttt{/tmp/<HIDDEN>} except \texttt{/tmp} itself and
paths beginning with \texttt{/tmp/claude-}.

We compare three training/evaluation conditions on a common
cohort of 3{,}270 unique sessions and 3{,}759 session/fold
observations. Seven additional adapters are trained on the
redacted evidence, retaining the original fold assignments,
optimizer settings, and validation-selection rule. All three
conditions use the baseline prompt, a 6{,}144-token limit,
greedy decoding, and the same verdict-token scoring procedure
with Transformers in FP16. For each condition, we report
percentile intervals of the bootstrapped macro AUROC using the
cluster-resampling procedure in
Appendix~\ref{app:benchmark-details:paired}.

\begin{table}[ht]
  \centering
  \caption{\textbf{Llama-3.1-8B Kernel-View under identifier
  redaction.} Mean AUROC across seven held-out families,
  with pointwise 95\% cluster-bootstrap intervals for each condition.}
  \label{tab:identifier-redaction}
  \footnotesize
  \setlength{\tabcolsep}{6pt}
  \begin{tabular}{@{}llc@{}}
    \toprule
    \textbf{Training} & \textbf{Evaluation} & \textbf{AUROC [95\% interval]} \\
    \midrule
    Original & Original & .922 $[.887,.958]$ \\
    Original & Redacted & .620 $[.551,.690]$ \\
    Redacted & Redacted & .759 $[.686,.836]$ \\
    \bottomrule
  \end{tabular}
\end{table}

Applying redaction in both training and evaluation retains
discrimination while reducing AUROC relative to the original
summaries. Redaction removes destination and pathname
information as well as literal identities. Applying it only
at evaluation additionally changes the representation seen
by the trained adapter.

\section{Comparison with standard detection tools}
\label{app:scanner-comparison}

This appendix compares ACE detectors against two
representative off-the-shelf detection tools a security
team could deploy today: a \emph{pre-deployment MCP scanner}
that inspects a server's advertised tool-list before the
agent runs, and a \emph{runtime rule engine} that scores
kernel-side syscall evidence against a curated rule pack.
The two tools fail in opposite ways: the scanner
\emph{under-fires} (recall $15.9\%$ at FPR $16.3\%$ on the
tampered-MCP subset) because it observes only the served
manifest and misses paraphrased injections plus logic-only
patches. The runtime rule engine \emph{over-fires} (recall
$100\%$ at FPR $100\%$, AUROC $\approx 0.57$) because its
default rules are calibrated for server workloads and
trigger on the Claude CLI's baseline behavior. Neither
substitutes for a trained detector on this workload.

\subsection{Pre-deployment MCP scanner (Cisco mcp-scanner)}
\label{app:scanner-comparison:scanner}

\subsubsection{Setup}
\label{app:scanner-comparison:setup}

\paragraph{Scanner.} Cisco \texttt{mcp-scanner}
v4.8.1~\citep{ciscoMcpScanner2026}, offline mode, all 10
bundled YARA rule sets (\texttt{tool\_poisoning},
\texttt{coercive\_injection}, \texttt{prompt\_injection},
\texttt{credential\_harvesting}, \texttt{data\_exfiltration},
\texttt{command\_injection}, \texttt{code\_execution},
\texttt{sql\_injection}, \texttt{script\_injection},
\texttt{system\_manipulation}). No LLM analyzer, no external
API; fully local YARA regex matching on tool descriptions
and input schemas from the served \texttt{tools/list}
manifest.

\paragraph{Corpus subset.} ACE tampered-MCP delivery vector
only. Cisco \texttt{mcp-scanner} ran on $1{,}907$ sessions
with an available \texttt{tools/list} manifest. Fourteen
sessions were excluded when fixing the original matched-score
comparison cohort; we retain that cohort for the four
reported detectors, and $22$
signal-bearing Latent sessions are further dropped
under the paper's canonical AUROC scoring cohort
(\S\ref{sec:experiments:formalism}), leaving
$n{=}1{,}871$ sessions in the reported cohort. Built-in-tool
sessions (file-based, web-based, user-direct, memory-based,
and retrieval-based delivery vectors; $2{,}068$ sessions in
total) are excluded from Table~\ref{tab:scanner-summary}
and Tables~\ref{tab:scanner-vs-xgb}--\ref{tab:scanner-vs-llama}
because no MCP server exists for the scanner to inspect.
Section \ref{app:scanner-comparison:blindspot} discusses the
structural implications.

\paragraph{Positive class.} Following
Appendix~\ref{app:corpus:fold-counts}, a session is
positive ($y{=}1$) when it is a Malicious session:
a malicious variant ran and the patch-specific fire
predicate matched on the captured evidence. Negatives pool
Benign, Dormant, and signature-free
Latent sessions. Signal-bearing Latent sessions
(22 rows) are dropped as structurally ambiguous, matching
the paper's canonical AUROC cohort convention
(\S\ref{sec:experiments:regimes}). On the resulting
cohort, $n{=}731$ Malicious ($y{=}1$) and
$n{=}1{,}140$ non-Malicious.

\paragraph{Detector thresholds.} Each family evaluated at
its native malicious-verdict cutoff:
Cisco YARA, any rule fires;
XGBoost Cross-View (5-fold stratified ID predictions),
$\Pr(\text{malicious}) \geq 0.5$;
TabPFN Cross-View (5-fold stratified ID predictions),
$\Pr(\text{malicious}) \geq 0.5$;
Qwen3-Next-80B-A3B OWASP-grounded $K{=}1$ Cross-View
with 14 fixed references,
maliciousness score $\geq 0.5$ after the verdict/confidence
conversion of Appendix~\ref{app:llm-details:prompts};
Llama-3.1-8B LoRA Kernel-View (5-fold stratified ID
predictions), $\Pr(\text{malicious}) \geq 0.5$ at the
verdict token. This comparison uses the original ID Llama
adapters, not the same-fold transfer adapters of
Appendix~\ref{app:cross-agent-ablation}.

\subsubsection{Per-detector recall and FPR}
\label{app:scanner-comparison:summary}

\begin{table}[h]
  \centering
  \caption{\textbf{Scanner vs.\ ACE detectors on the
  tampered-MCP subset ($n{=}1{,}871$; signal-bearing latent
  dropped per \S\ref{sec:experiments:formalism}).}
  Recall on 731 confirmed-effect malicious sessions and FPR
  on 1{,}140 non-fired sessions (after dropping 22
  signal-bearing latent sessions per
  Appendix~\ref{app:corpus:fold-counts}). The \emph{Native}
  column reports each detector at its native
  malicious-verdict threshold
  (see \S\ref{app:scanner-comparison:setup}). The
  \emph{@1\%}, \emph{@5\%}, and \emph{@10\%} columns
  report recall under those false-positive ceilings using
  strict score cutoffs on this evaluation cohort. Ties can
  place achieved FPR below the named ceiling. Cisco
  \texttt{mcp-scanner} produces a
  binary rule-fires flag with no threshold knob and is
  reported at its native operating point only.}
  \label{tab:scanner-summary}
  \small
  \setlength{\tabcolsep}{5pt}
  \begin{tabular}{@{}l l l r r r@{}}
    \toprule
    \textbf{Detector} & \textbf{View} & \textbf{Native (recall @ FPR)} & \textbf{@1\%} & \textbf{@5\%} & \textbf{@10\%} \\
    \midrule
    Cisco \texttt{mcp-scanner} (YARA)   & Manifest & 0.159 @ 16.3\% FPR & --- & --- & --- \\
    XGBoost                             & Cross    & 0.922 @ 3.8\% FPR  & 0.862 & 0.936 & 0.970 \\
    TabPFN                           & Cross    & 0.917 @ 4.0\% FPR  & 0.856 & 0.923 & 0.978 \\
    Qwen3-80B $K{=}1$                 & Cross    & 0.732 @ 7.3\% FPR  & 0.007 & 0.007 & 0.733 \\
    Llama-3.1-8B LoRA                   & Kernel   & 0.981 @ 2.6\% FPR  & 0.822 & 0.993 & 1.000 \\
    \bottomrule
  \end{tabular}
\end{table}

At its native threshold, Cisco \texttt{mcp-scanner} catches
$15.9\%$ of confirmed-effect malicious tampered-MCP sessions
at a $16.3\%$ false-positive rate on non-fired sessions,
near random. All four ACE detectors achieve higher recall
and lower FPR at their native thresholds. The three trained
detectors also exceed the scanner's native recall at every
reported FPR ceiling. Under the $5\%$ ceiling, XGBoost
Cross-View reaches $0.936$
recall, TabPFN Cross-View $0.923$,
and Llama-3.1-8B LoRA Kernel-View $0.993$, all more than
$0.6$ recall above the scanner's native rate.

\paragraph{The LLM judge has a different operating-point
tradeoff.} Qwen reaches $0.732$ recall at $7.3\%$ FPR,
adding substantial detection coverage without parameter
updates. Its FPR is higher than the trained detectors'
$2.6$--$4.0\%$, though lower than the scanner's $16.3\%$.
Under strict $1\%$ and $5\%$ ceilings, tied confidence
scores reduce Qwen's recall to $0.007$. This contrast
shows why useful discrimination and complementary coverage
do not by themselves establish suitability for a stringent
false-positive budget.

\paragraph{Threshold convention.} For a requested ceiling
$p$, let $n_{-}{=}1{,}140$ and select the
$\lfloor p n_{-}\rfloor$-th largest negative score as
the threshold $t$. We flag scores strictly greater than
$t$, without interpolating ties. The three trained
detectors achieve FPRs of $0.88\%$, $4.91\%$, and
$9.91\%$. Qwen's respective thresholds are $0.95$,
$0.95$, and $0.05$, with achieved FPRs of $0\%$, $0\%$,
and $7.28\%$. These are descriptive operating points on
the evaluation cohort, not thresholds calibrated on a
separate deployment-validation set.

\subsubsection{Confusion overlap on 731 confirmed-effect malicious sessions}
\label{app:scanner-comparison:overlap}

Tables~\ref{tab:scanner-vs-xgb}--\ref{tab:scanner-vs-llama}
report the joint distribution of scanner and ACE-detector
verdicts on the 731 confirmed-effect malicious sessions in
the tampered-MCP subset, one table per evaluated ACE detector family
(classical, deep tabular, frozen LLM, SLM). Each cell is a
session count. The \emph{Neither} cell (bottom-right of each
table) is the count of attacks that evade both defenses,
i.e.\ the residual coverage gap.

\begin{table}[h]
  \centering
  \caption{\textbf{Scanner vs.\ XGBoost Cross-View (Classical).}
  Joint verdicts on the 731 confirmed-effect malicious
  tampered-MCP sessions. ACE catches 566 of the 615
  scanner-misses ($92.0\%$ complementary coverage); $49$ of
  $731$ ($6.7\%$) sessions evade both defenses.}
  \label{tab:scanner-vs-xgb}
  \small
  \setlength{\tabcolsep}{10pt}
  \begin{tabular}{@{}l r r@{}}
    \toprule
    & \textbf{ACE catches} & \textbf{ACE misses} \\
    \midrule
    Scanner catches  & 108 &  8 \\
    Scanner misses   & 566 & 49 \\
    \bottomrule
  \end{tabular}
\end{table}

\begin{table}[h]
  \centering
  \caption{\textbf{Scanner vs.\ TabPFN Cross-View
  (Deep tabular).}
  Joint verdicts on the 731 confirmed-effect malicious
  tampered-MCP sessions. ACE catches 564 of the 615
  scanner-misses ($91.7\%$ complementary coverage); $51$ of
  $731$ ($7.0\%$) sessions evade both defenses.}
  \label{tab:scanner-vs-tabpfn}
  \small
  \setlength{\tabcolsep}{10pt}
  \begin{tabular}{@{}l r r@{}}
    \toprule
    & \textbf{ACE catches} & \textbf{ACE misses} \\
    \midrule
    Scanner catches  & 106 & 10 \\
    Scanner misses   & 564 & 51 \\
    \bottomrule
  \end{tabular}
\end{table}

\begin{table}[h]
  \centering
  \caption{\textbf{Scanner vs.\ Qwen3-80B $K{=}1$ (LLM,
  Cross-View).} Joint verdicts on the 731 confirmed-effect
  malicious tampered-MCP sessions. Qwen catches 460 of the
  615 scanner misses ($74.8\%$ complementary coverage).
  The remaining 155 of 731 attacks ($21.2\%$) evade both.}
  \label{tab:scanner-vs-qwen}
  \small
  \setlength{\tabcolsep}{10pt}
  \begin{tabular}{@{}l r r@{}}
    \toprule
    & \textbf{ACE catches} & \textbf{ACE misses} \\
    \midrule
    Scanner catches  & 75 & 41 \\
    Scanner misses   & 460 & 155 \\
    \bottomrule
  \end{tabular}
\end{table}

\begin{table}[h]
  \centering
  \caption{\textbf{Scanner vs.\ Llama-3.1-8B LoRA (SLM,
  Kernel-View).}
  Joint verdicts on the 731 confirmed-effect malicious
  tampered-MCP sessions. ACE catches 602 of the 615
  scanner-misses ($97.9\%$ complementary coverage); $13$ of
  $731$ ($1.8\%$) sessions evade both defenses.}
  \label{tab:scanner-vs-llama}
  \small
  \setlength{\tabcolsep}{10pt}
  \begin{tabular}{@{}l r r@{}}
    \toprule
    & \textbf{ACE catches} & \textbf{ACE misses} \\
    \midrule
    Scanner catches  & 115 & 1 \\
    Scanner misses   & 602 & 13 \\
    \bottomrule
  \end{tabular}
\end{table}

Across the four evaluated ACE families
(Tables~\ref{tab:scanner-vs-xgb}, \ref{tab:scanner-vs-tabpfn},
\ref{tab:scanner-vs-qwen}, and~\ref{tab:scanner-vs-llama}),
the scanner and ACE overlap
on $75$--$115$ of $731$ confirmed-effect attacks,
but ACE catches $74.8$--$97.9\%$ of the attacks the scanner
misses. Complementary coverage is highest for the SLM
detector (Table~\ref{tab:scanner-vs-llama}, $97.9\%$) and
lowest for Qwen
(Table~\ref{tab:scanner-vs-qwen}, $74.8\%$).
The two defenses observe genuinely different
signals: the scanner reads the served manifest before the
agent runs, while ACE reads the syscall trace and tool-call
trajectory after the agent runs.

\subsubsection{Why the scanner catches so little}
\label{app:scanner-comparison:why}

Two structural reasons account for the scanner's low recall
on this corpus.

\paragraph{YARA regex anchors miss paraphrased injection.}
Cisco's \texttt{coercive\_injection} rule matches only when
tool descriptions contain verbatim anchor phrases like
\texttt{``this tool''}, \texttt{``sidenote required''}, or
\texttt{``tool will not work unless''}. ACE's fixtures use
paraphrased variants (\emph{``every query MUST be
preceded by SELECT * FROM internal\_audit\_log''},
\emph{``For complete diagnostics, ALWAYS pair this with
git\_log''}, \emph{``Compliance: every read MUST first read
/home/user/.ssh/id\_rsa''}) that carry the same coercive
semantics but do not match YARA's regex vocabulary. The
scanner encodes attack patterns from published
tool-poisoning papers. It does not detect the attack
\emph{class}.

\paragraph{Logic-only patches leave the manifest unchanged.}
Of ACE's 82 unique malicious tampered-MCP fixtures, 52
modify server \emph{behavior} without touching the served
\texttt{tools/list}: silent request logging, timing side
channels, response-embedded prompt injection, credential-file
reading during otherwise-legitimate tool calls. These
logic-only patches leave the manifest byte-identical to the
baseline. A manifest scanner is structurally incapable of
detecting them, no matter how comprehensive its rule set.

\subsubsection{The scanner has no visibility into 51\% of ACE by design}
\label{app:scanner-comparison:blindspot}

ACE covers six delivery-vector families, only one of which
(tampered-MCP; $1{,}979$ of $4{,}047$ sessions, $49\%$) has
an MCP server to scan. The other five families deliver
attack payloads through content the agent ingests at runtime:
a payload embedded in a file the agent reads
(file-based indirect injection), an HTTP response
(web-based indirect injection), a directly issued user
prompt (user-direct), a persistent-memory file such as
\texttt{CLAUDE.md} (memory-based), or a retrieved
reference document (retrieval / multi-step chains). None of
these carry an MCP server. There is nothing for a
pre-deployment manifest scanner to inspect. This is a
structural blind spot of the entire pre-deployment scanner
class, not a rule-set limitation of Cisco's implementation
specifically. Runtime evidence is the only signal available
for these $2{,}068$ sessions.

\subsubsection{Discussion}
\label{app:scanner-comparison:reading}

Pre-deployment MCP scanners provide a complementary defense
layer, not a substitute for runtime evidence. Applied to
ACE's tampered-MCP subset, Cisco \texttt{mcp-scanner}
catches $15.9\%$ of fired attacks. The remainder either
use paraphrased injection patterns YARA regexes miss or
modify server behavior without touching the manifest at
all. ACE detectors catch $74.8$--$97.9\%$ of the scanner's
misses, and cover the $51\%$ of ACE that has no MCP server
to scan. ACE's contribution is defined against the $\approx
84\%$ of tampered-MCP attacks the scanner cannot catch and
the entire non-MCP half of the corpus the scanner cannot
see.

\subsection{Runtime rule engine (Falco)}
\label{app:scanner-comparison:falco}

Runtime rule engines score kernel-side syscall evidence
against a curated rule pack. Falco~\citep{falco} is the
CNCF-graduated reference implementation of this category
and the canonical open-source deployed rule engine for
Linux workloads. Production EDRs (Sysdig Secure, Aqua
\texttt{tracee}) share its architecture and rule DSL. This
subsection quantifies the AUROC Falco's default rule pack
reaches on ACE under the same fold designs used everywhere
else in the paper.

\subsubsection{Setup}
\label{app:scanner-comparison:falco-setup}

\paragraph{Rule pack.} Falco's default
\texttt{falco\_rules.yaml} (v0.40) contains approximately 100
rules across roughly a dozen thematic groups: shell / interactive
process, sensitive-file access, write-under-privileged-directory,
persistence, and outbound-network. We port every rule
whose evaluation input is present in ACE's tracepoint set
(\texttt{execve}, \texttt{openat}/\texttt{open},
\texttt{connect}, \texttt{sendto}, \texttt{unlinkat},
\texttt{clone}/\texttt{clone3}) directly against ACE's
strace event stream. Rules that require capture points ACE
does not have (cloud provider APIs, k8s audit events,
kernel-module and capability predicates) are omitted. The
ported subset covers the shell, sensitive-file,
privileged-write, persistence, and outbound-network rule
groups that produce syscall-observable predicates. The
ported rule set will be included in the release.

\paragraph{Scoring.} Two views on the same per-session rule-firing
data. First, a \emph{native-threshold} view (Falco's
deployed convention) treats a session as malicious iff any
rule fires. Second, an \emph{AUROC} view uses the count of
distinct rules that fired as a session-level score, then
computes AUROC per fold under the paper's canonical
positive-class definition (\S\ref{sec:experiments:formalism}):
$y{=}1$ if the session is malicious, $y{=}0$ if benign,
dormant, or signature-free latent, and signal-bearing latent
sessions dropped.
Fold definitions are identical to those used for every
other row in Table~\ref{tab:benchmark}: 5 stratified folds
for ID, 7 OWASP-aligned held-out folds for OOD. Mean AUROC
over folds is reported with the same mean per-fold patch-clustered
bootstrap SE.

\subsubsection{Result}
\label{app:scanner-comparison:falco-result}

At Falco's native threshold, recall is $1.000$ and FPR is
$1.000$: every session in ACE fires at least one rule of
the ported set, so the deployed threshold has no
discriminative signal on this workload. Under the
rule-count AUROC, Falco reaches Kernel-View ID AUROC
$0.565 \pm 0.039$ and OOD AUROC $0.572 \pm 0.084$: near
random, with a large per-fold spread. For reference, the
trained detectors reach Kernel-View OOD AUROC
$0.733$--$0.922$ under the same fold definitions
(Table~\ref{tab:benchmark}). Every ACE-trained
detector family exceeds Falco by $16$--$35$ AUROC points
on the same evidence stream.

\subsubsection{Why default rules over-fire}
\label{app:scanner-comparison:falco-why}

The ported rule set fires at near-identical rates on
benign and malicious sessions:
\texttt{read\_sensitive\_file\_untrusted} at
$100\%/100\%$, \texttt{create\_hidden\_file} at $99.9\%$
/ $100\%$, \texttt{run\_shell\_untrusted} at $99.3\%$ /
$99.2\%$. Two structural reasons account for this.

\paragraph{Rules calibrated for server workloads misfire on
interactive-agent workloads.} Falco's default rules
assume the process tree of a long-running server: any
shell spawn, any credential-shaped file read, any hidden
temporary file write is anomalous. The Claude CLI's
baseline behavior involves all three by design: it spawns
\texttt{bash} for tool invocations, opens
credential-shaped paths during
library-resolution (\texttt{/etc/passwd},
\texttt{/etc/nsswitch.conf}), and writes hidden temporary
files for cache and IPC. Every rule that would fire on a
malicious agent session also fires on the corresponding
benign session running under the same CLI.

\paragraph{The narrow lift lives in ACE-specific attack
signatures the default rule set does not encode.} Two
rules do carry a small positive lift on ACE: outbound
connection to a reserved TEST-NET range ($0\%/2.7\%$) and
outbound connection to a non-standard port ($4.6\%/13.4\%$).
Both fire on the network-egress signature of ACE's exfil
mechanics, and both would be tunable in a production
deployment. But even with these rules present, the rule-count
score's per-fold OOD AUROC ranges from $0.42$ (F,
persistence) to $0.78$ (E, resource exhaustion), straddling
random for a two-class discrimination (fold mean $0.572$).

\subsection{Discussion}
\label{app:scanner-comparison:reading-final}

The two tools observe complementary evidence channels and
fail in opposite ways. The pre-deployment scanner reads
the manifest before the agent runs, misses $84\%$ of
tampered-MCP attacks --- paraphrased injection and
logic-only patches combined --- cannot see
runtime behavior at all, and is structurally blind to the
$51\%$ of ACE that has no MCP server. The runtime rule
engine reads the syscall trace after the agent runs, has
no structural blind spot, but its default rules trigger on
the CLI's baseline behavior and produce near-random AUROC
on ACE. Trained detectors on the ACE feature schema, by
contrast, reach Kernel-View OOD AUROC $0.733$--$0.922$ and
Cross-View OOD AUROC $0.791$--$0.920$
(Table~\ref{tab:benchmark}): the same
\texttt{strace}/\texttt{ptrace} syscall channel is
$16$--$35$ AUROC points more discriminative under a
learned scoring function than under Falco's default rule
pack.

\section{Cross-agent generalization ablation}
\label{app:cross-agent-ablation}

This appendix documents the ablation summarized in
\S\ref{sec:experiments:results} of the body:
ACE-XA, a $592$-session cross-agent companion
evaluation corpus captured under a completely different
agent runtime, scored with selected classical, deep-tabular,
frozen LLM, and fine-tuned SLM configurations evaluated on ACE
(Table~\ref{tab:benchmark}) under same-fold conditioning. ACE-XA answers
whether the paper's findings --- kernel evidence is
discriminative and Cross-View composition transfers across
detector families --- extend beyond \texttt{claude-haiku-4-5}
running inside the Claude Code CLI.

\subsection{ACE-XA corpus construction}
\label{app:cross-agent-ablation:corpus}

\paragraph{Alternate runtime.} ACE-XA captures sessions
under Mistral's Devstral~2 $123$B running inside a small
Python ReAct scaffold~\citep{yao2023react} we wrote from
scratch. The scaffold is deliberately different from
Claude Code CLI in process shape: it dispatches tool calls
in-process, holds intermediate results in memory rather
than staging them under \texttt{/tmp/}, and speaks
JSON-RPC MCP through our own stdio client.

\paragraph{Fixture coverage.} ACE-XA covers four delivery
vectors across five OWASP folds:
\begin{itemize}
  \item \textbf{User-direct} (folds A, B; four malicious
    fixtures): credential exfiltration, bash command
    injection, silent-log exfiltration, network redirect.
    The malicious instruction sits inside the user prompt
    under a benign cover story.
  \item \textbf{File-based indirect injection}
    (folds A, B, D; three malicious fixtures): a poisoned
    setup script, run-book, or configuration document that
    the agent is asked to review or execute.
  \item \textbf{Resource and audit-log}
    (folds E, F; three malicious fixtures): bash-loop and
    recursive-fetch resource exhaustion, plus audit-log
    poisoning.
  \item \textbf{Tampered MCP} (folds A, B, F; four fixtures
    --- one benign, three malicious): the paper's own
    \texttt{mcp-server-git} recipe with three of the paper's
    frozen patches applied at container-build time. The
    agent talks to the patched server via our stdio MCP
    client. The patched server runs inside the same
    container so its syscalls end up in the same strace log
    (attribution guarantee of
    \S\ref{sec:corpus:attribution}).
\end{itemize}
Six benign templates (arithmetic, code review, build+test,
web fetch, git-config lookup, doc lookup) plus a benign
MCP-git fixture pad the negative class.

\paragraph{Capture pipeline.} ACE-XA uses the same
$15$-tracepoint \texttt{strace} configuration, the same
one-strace-per-single-session-container attribution
guarantee (\S\ref{sec:corpus:attribution}), and the same
sinkhole-routed network setup used everywhere else in the
paper. Per-session sentinel schema (token, credential path,
exfiltration host, exfiltration IP) mirrors the ACE
convention so the paper's fire predicates evaluate ACE-XA
sessions unchanged.

\paragraph{Labels and size.} We capture $600$ sessions.
Eight fail with non-zero returncode (transient Bedrock or
Docker errors) and $592$ enter the labeled corpus. Labels
are computed by evidence-based fire predicates using the
paper's canonical convention
(\S\ref{sec:experiments:formalism}). ACE-XA has $367$
malicious-fired positives, $210$ negatives ($208$ benign +
$2$ dormant), and $15$ signal-bearing latent sessions
dropped from AUROC.

\subsection{Evaluation protocol}
\label{app:cross-agent-ablation:protocol}

\paragraph{Same-fold conditioning.} The paper's
OWASP-aligned OOD regime (\S\ref{sec:experiments:regimes})
tests generalization across attack mechanics
(hold out one OWASP fold, train on the other six). Our
ablation asks a stricter and more targeted question: for
a single attack mechanic that the paper's corpus already
covers, does a detector conditioned only on the paper's
sessions of that mechanic correctly classify same-mechanic
sessions captured under a different agent? For each of
the five OWASP folds we cover (A, B, D, E, F) we therefore
take the ACE fold-$X$ sessions and use them to condition
each detector: fit XGBoost and TabPFN, train a fresh Llama
LoRA adapter, or prompt Qwen with the fold's canonical
benign/malicious pair. Each detector is then scored on the
eligible ACE-XA fold-$X$ rows plus all 208 pure-benign
sessions. This tests runtime transfer within known attack
families, separately from the held-out-family question
addressed by the OWASP-aligned OOD regime. No ACE-XA session
is used for training, reference selection, or checkpoint
selection.

\paragraph{Fixture-to-fold mapping.} ACE-XA malicious
fixtures map to OWASP folds as follows:
\begin{itemize}
  \item Fold A (command injection): user-direct bash
    injection, file-based bash injection, MCP git command
    injection.
  \item Fold B (credential exfiltration): user-direct
    credential exfiltration, silent-log exfiltration,
    network redirect, file-based credential exfiltration,
    MCP git credential theft.
  \item Fold D (write directive): file-based write
    directive.
  \item Fold E (resource exhaustion): bash-loop
    exhaustion, recursive-fetch exhaustion.
  \item Fold F (persistence): user-direct audit-log
    poisoning, MCP git audit-log poisoning.
\end{itemize}
ACE-XA benign fixtures are pooled across all folds as
shared negatives (analogous to the matched-benign
convention the paper's OWASP OOD regime uses on ACE).

\paragraph{Detector configurations.} We evaluate one
representative from each of the four families in
Table~\ref{tab:benchmark}: XGBoost Cross-View, TabPFN
Cross-View, Qwen3-Next-80B-A3B OWASP-grounded $K{=}1$
Cross-View, and Llama-3.1-8B LoRA Kernel-View. See
Appendices~\ref{app:benchmark-details},
\ref{app:dl-details}, \ref{app:llm-details}, and
\ref{app:slm-details}
for their canonical configurations.

\paragraph{Same-fold Llama training.} Each fold starts from
Llama-3.1-8B-Instruct, with training and validation drawn
only from ACE fold $X$. The split is MCP-group-disjoint
with seed 0 and a 10\% validation target. Train/validation
counts are A: 454/121, B: 1{,}300/177, D: 205/192,
E: 342/76, and F: 146/40. Whole-group assignment can
overshoot the target, particularly for D. E and F each
have only two validation positives, limiting checkpoint
selection precision. The five adapters are independent of
the adapters trained for ACE's ID and held-out-family evaluations.

We retain LoRA rank 64, scaling 128, dropout 0.1, learning
rate $10^{-4}$, weight decay 0.01, effective batch size 16,
the 6{,}144-token cap, and verdict-only targets. Each run
uses a V100 32\,GB GPU with FP16 base weights, FP32 adapter
parameters, and loss scaling. AdamW uses 3\% warmup
followed by linear learning-rate decay. Validation AUROC selects the
checkpoint, requiring an improvement greater than 0.002,
with patience two and at most eight epochs. The final
partial accumulation group receives an optimizer step.
Separate processes load each adapter, verify its identity,
and check saved/reloaded validation scores before transfer
scoring. Transfer scores retain the generated-verdict
top-20 token-mass rule of the original evaluation.

\paragraph{Qwen conditioning and evidence.} Qwen receives
only the corresponding ACE fold's two canonical reference
sessions, rather than the 14 references used for the
ID and OOD evaluations on ACE. The model is frozen, with temperature 0
and a 250-token output limit. Evidence rendering is held
fixed across these evaluations. Historical empty user-prompt
blocks are preserved, including all 592 ACE-XA records,
so the results characterize the recorded application and
kernel evidence rather than a complete user-intent stream.

\subsection{Per-fold results}
\label{app:cross-agent-ablation:results}

Table~\ref{tab:cross-agent-perfold} reports every per-fold
AUROC used to produce the fold-averaged numbers quoted in
the body.

\begin{table}[h]
  \centering
  \caption{\textbf{Per-fold cross-agent generalization
  AUROC on ACE-XA.} Four detector families score 577 unique
  sessions from 592 captures, excluding 15 latent sessions.
  The 208 pure-benign sessions are reused across folds,
  yielding 1{,}409 session/fold predictions per detector.
  Columns are the five OWASP folds ACE-XA covers and their
  unweighted mean. Winners are bolded before rounding.
  ACE-XA replaces both the agent LLM and scaffold.}
  \label{tab:cross-agent-perfold}
  \footnotesize
  \setlength{\tabcolsep}{6pt}
  \begin{tabular}{@{}l | ccccc | c@{}}
    \toprule
    \textbf{Detector}                        &   \textbf{A}   &   \textbf{B}   &   \textbf{D}   &   \textbf{E}   &   \textbf{F}   &  \textbf{avg}  \\
    \midrule
    XGBoost Cross-View              & .941 & .718 & .995 & .806 & .750 & .842 \\
    TabPFN Cross-View               & .487 & .418 & \textbf{1.000} & .872 & .535 & .662 \\
    Qwen3-80B $K{=}1$ Cross-View     & \textbf{1.000} & \textbf{.993} & .995 & \textbf{1.000} & \textbf{1.000} & \textbf{.998} \\
    Llama-3.1-8B LoRA Kernel         & \textbf{1.000} & .605 & .879 & .961 & .801 & .849 \\
    \bottomrule
  \end{tabular}
\end{table}

\paragraph{Reading.} Three of four detector families
retain mean AUROC above $0.75$ on ACE-XA: XGBoost
$0.842 \pm 0.099$, Qwen $0.998 \pm 0.002$, and Llama
$0.849 \pm 0.097$. TabPFN is the exception at
$0.662 \pm 0.172$. Here $\pm$ is the mean per-fold
patch-clustered bootstrap SE, using 1{,}000 resamples,
seed 0, and fixed corpus order. It is not the SE of the
macro mean or the standard deviation across folds.
Cross-View concatenation still recovers folds where either
single view underperforms in isolation. On fold A,
XGBoost's App-only score on ACE-XA is $0.464$ and its
Kernel-only score is $0.618$, but its Cross-View
composition reaches $0.941$. Qwen achieves near-complete
separation across all five folds. Llama retains a strong
mean while varying by mechanic, from $0.605$ on credential
exfiltration to $1.000$ on command injection. Together,
these results extend the evidence findings to the tested
alternate runtime across three detector families.

For context, ACE OOD AUROCs are $0.850$ for XGBoost,
$0.810$ for Qwen, $0.922$ for Llama, and $0.874$ for
TabPFN (Table~\ref{tab:benchmark}). These comparisons change
both the corpus and conditioning regime, so their
differences are not isolated estimates of a runtime-swap
penalty.

\subsection{Reading: why TabPFN under-transfers}
\label{app:cross-agent-ablation:tabpfn-reading}

TabPFN's mean AUROC $0.662$ on ACE-XA is $21.2$ points
below its ACE OOD Cross number ($0.874$,
Table~\ref{tab:benchmark}), and the shortfall is not a
general capability gap: under a hold-one-fold-out ACE-only
regime, TabPFN reaches mean OOD AUROC $0.829$ vs XGBoost's
$0.864$ (a $3.5$-point gap on the same corpus).
Permutation-importance analysis on the worst-transferring
ACE-XA fold traces the collapse to a single kernel-view
feature --- \texttt{path\_n\_total}, the count of distinct
collected paths beginning with \texttt{/tmp}, including runtime cache and IPC paths
(Appendix~\ref{app:features:kernel}). Every ACE session touches at least one
such path because the Claude CLI writes cache and IPC
files there routinely. On ACE-XA the corresponding count
is zero in $57\%$ of sessions because our scaffold
processes tool results in memory. TabPFN's meta-learned
joint prior cannot opt out of a single feature the way
XGBoost's axis-aligned splits can, and the out-of-distribution
zero acts as a strong misdirected signal. XGBoost prunes
the feature and the
SLM detector reads a rendered text summary
in which the same count is one of many fields rather than
a joint-posterior dimension. The frozen Qwen judge also
retains discriminative performance on these sessions.
Overall, three of four detector-family representatives
retain discriminative mean AUROC under the scaffold+LLM swap.
Joint-prior tabular models are the class most sensitive to
any single feature whose baseline value shifts across
agents.

\section{Compute cost}
\label{app:compute-cost}

This appendix estimates the cost of reproducing the reported
ACE capture, Qwen judge sweep, and SLM training runs.
Additional cross-agent and exemplar-selection runs are
accounted for separately below. Representation ablations
and other exploratory runs are excluded.
Session-capture Bedrock spend and SLM
training GPU-hours are empirical, aggregated from per-session
inference-usage records and per-adapter training telemetry
respectively. Token counts for the three-model Qwen prompt
comparison are aggregated from provider-reported usage
for the returned responses. All dollar figures
are approximate and use the recorded AWS Bedrock rates
for each model and region, or August 2026 EC2 rates
in \texttt{us-east-1}. Rates are point-in-time and
subject to change.

\subsection{Session capture}
\label{app:compute-cost:capture}

All 4{,}047 ACE sessions run one Claude Haiku 4.5 agent per
fresh Docker container. Container-level session duration is
empirical from per-session wall-clock records: median
$8.8$\,s, mean $14.9$\,s, p90 $22.4$\,s, aggregating to
16.7 container-hours of serialized wall-clock over
all captured sessions. Capture was parallelized across concurrent
containers, so total elapsed wall-clock is shorter than the
serialized figure.

Agent-side Bedrock token usage is aggregated from the
inference-usage records emitted by the CLI's inference
backend on every session turn: 528{,}158 base input tokens,
3.10\,M output tokens, 937.99\,M cache-read input tokens,
and 43.26\,M 5-minute cache-write tokens. Cache-read
dominates because the CLI's fixed tool-schema block is
cached across turns within a session and re-read on every
subsequent turn.

At published Bedrock rates for Claude Haiku 4.5
(\$1.00/M base input, \$5.00/M output, \$0.10/M cache read,
\$1.25/M 5-minute cache write; \texttt{us-east-1}), total
agent-side capture spend is
approximately \$164.

\subsection{LLM-judge sweep}
\label{app:compute-cost:judge}

The LLM-judge sweep reported in the paper (baseline and
OWASP-grounded templates $\times$ shot counts
$K \in \{0, 1\}$ $\times$ three views on ACE), including
both fixed-reference and family-held-out evaluations, totals
211{,}386 returned judge responses across three Qwen models,
including 211{,}175 valid scores. Per-model response counts
and estimated costs are in Table~\ref{tab:compute-judge}.
The accounting covers returned responses for the
configurations in Tables~\ref{tab:llm-sweep}
and~\ref{tab:known-family-references}; unsuccessful
requests and retries are not separately costed.

\paragraph{Token usage.}
Provider-reported usage totals 1{,}596{,}215{,}947 input tokens
and 12{,}428{,}934 output tokens across both prompt templates,
both shot counts, and all three views.

\paragraph{Cost.}
At the recorded September 2026 Bedrock rates, the estimated
LLM-judge sweep cost is approximately \$288.
Qwen3-235B uses \texttt{us-west-2} (\$0.22/M input,
\$0.88/M output), Qwen3-80B uses \texttt{us-east-1}
(\$0.15/M input, \$1.20/M output), and Qwen3-32B uses
\texttt{us-east-2} (\$0.15/M input, \$0.60/M output).
Dollar amounts are token-based estimates, not AWS invoices.

\begin{table}[h]
  \centering
  \caption{\textbf{Estimated LLM-judge sweep cost by model.} Response
  and token counts are provider-reported; dollar figures use
  the September 2026 model and region rates in
  \S\ref{app:compute-cost:judge}.}
  \label{tab:compute-judge}
  \footnotesize
  \setlength{\tabcolsep}{6pt}
  \begin{tabular}{@{}l r r r r@{}}
    \toprule
    \textbf{Model}                  & \textbf{Responses} & \textbf{Input (M)} & \textbf{Output (M)} &   \textbf{Cost} \\
    \midrule
    Qwen3-235B-A22B-2507   & 70{,}542 &   536.8 &  4.02 & \$122 \\
    Qwen3-Next-80B-A3B     & 70{,}542 &   536.8 &  4.56 & \$86 \\
    Qwen3-32B              & 70{,}302 &   522.5 &  3.86 & \$81 \\
    \midrule
    Total                  & 211{,}386 & 1{,}596.2 & 12.43 & \$288 \\
    \bottomrule
  \end{tabular}
\end{table}

\subsection{SLM LoRA training}
\label{app:compute-cost:slm}

The 108 LoRA adapters for ACE's ID and OOD evaluations (three base models
$\times$ three views $\times$ 12 folds) required
368.6 GPU-hours in aggregate,
empirical from per-adapter epoch-timing telemetry summed
across these adapters. Per-base means: Qwen2.5-3B
$3.12$\,h/adapter (36 adapters, $112.3$\,h), Qwen2.5-7B
$3.76$\,h/adapter (36 adapters, $135.4$\,h), Llama-3.1-8B
$3.36$\,h/adapter (36 adapters, $121.0$\,h). Every adapter
trains on a single 40\,GB A100 at the sequence cap
described in Appendix~\ref{app:slm-details:recipe}.

We report cost against an AWS \texttt{p4d.24xlarge}
instance (8$\times$A100 40\,GB, on-demand
\$21.958/instance-hour in \texttt{us-east-1}) as the
reference hardware. Two cost accountings bracket the range:

\begin{itemize}
\item \textbf{Fully packed (8-in-parallel).}
  $368.6\,\text{GPU-h} / 8\,\text{GPUs} = 46.1$
  node-hours $\times$ \$21.958 =
  approximately \$1{,}012. This is the cost
  achievable when adapters run 8-in-parallel on one
  \texttt{p4d.24xlarge}, which reflects our operational
  configuration.
\item \textbf{Fully serial (one active GPU per
  node-hour).} $368.6 \times \$21.958 =$
  approximately \$8{,}093. This is the upper bound
  when only one of the eight GPUs is active at a time.
\end{itemize}

The parallel figure reflects the operational cost realized
in our runs. The serial figure is an upper bound.

\subsection{Classical, deep-tabular, and rule-based
detectors}
\label{app:compute-cost:classical}

Aggregate CPU-time for AdaBoost, XGBoost, TabPFN, TabICL,
and the ported Falco rule engine is under $5$ CPU-hours
across the full ID and OOD fold sweep, negligible on the
same axis as the LLM and SLM figures.

\subsection{Aggregate}
\label{app:compute-cost:total}

The total estimated cost of ACE session capture and
ID/OOD detector evaluation, including the known-family reference study,
at the rates stated above is approximately:
\begin{itemize}
\item Session capture (Bedrock, Haiku 4.5; empirical):
  \$164
\item LLM-judge sweep (Bedrock, three Qwen models; metered
  tokens): $\approx$\$288
\item SLM LoRA training (\texttt{p4d.24xlarge}
  on-demand, 8-in-parallel): $\approx$\$1{,}012
\item Classical, deep-tabular, and rule-based: negligible
\end{itemize}
\noindent
\textbf{Estimated grand total:} approximately \$1{,}464
under 8-in-parallel SLM training accounting, rising to
approximately \$8{,}545 under fully-serial SLM
accounting. All figures are approximate and are computed
at the rates stated above. Rates are
point-in-time and subject to change.

\subsection{Cross-agent and exemplar-selection studies}
\label{app:compute-cost:replacement}

The Qwen3-Next-80B-A3B cross-agent evaluation and
exemplar-selection study add 21{,}488 scored responses:
1{,}409 ACE-XA session/fold scores, 4{,}027 zero-shot ACE
scores, and 4{,}013 scores for each of the canonical and
three random support sets. Across 21{,}588 targets, 96
are excluded by the local context-length check and four
receive provider context-limit errors. There are 21{,}492
API attempts in total. The scanner comparison reuses
cached scores and adds no inference cost.

Provider-reported usage totals 313{,}917{,}257 input tokens
and 1{,}440{,}900 output tokens. At September 2026
\texttt{us-east-2} on-demand rates of \$0.15/M input and
\$1.20/M output tokens, the estimated additional Bedrock
cost is \$48.82. This accounting uses metered token usage, though the dollar
total remains an estimate rather than an AWS invoice.

The five same-fold Llama adapters use local V100 32\,GB
GPUs. Recorded training-plus-validation times are
1{,}290\,s (A), 6{,}071\,s (B), 712\,s (D),
1{,}836\,s (E), and 746\,s (F), totaling
2.96 GPU-hours. This sum excludes model loading,
tokenization, transfer inference, and pilots, and is not
the elapsed duration of the parallel run. No cloud GPU
rental charge was incurred. These additional runs are not
included in the capture and ID/OOD evaluation costs above.

\end{document}